\documentclass[11pt]{article}
\usepackage{amsfonts,amsmath,amsxtra}
\usepackage{latexsym}
\usepackage{amssymb}

\def\hybrid{\topmargin 0pt      \oddsidemargin 0pt
        \headheight 0pt \headsep 0pt
        \textwidth 16.5cm
        \textheight 23cm
        \marginparwidth 0.0in
        \parskip 5pt plus 1pt   \jot = 1.5ex}
\catcode`\@=11
\def\marginnote#1{}
\newcount\hour
\newcount\minute
\newtoks\amorpm
\hour=\time\divide\hour by60 \minute=\time{\multiply\hour by60
\global\advance\minute by-\hour}
\edef\standardtime{{\ifnum\hour<12 \global\amorpm={am}%
        \else\global\amorpm={pm}\advance\hour by-12 \fi
        \ifnum\hour=0 \hour=12 \fi
      \number\hour:\ifnum\minute<10 0\fi\number\minute\the\amorpm}}
\edef\militarytime{\number\hour:\ifnum\minute<10 0\fi\number\minute}

\def\draftlabel#1{{\@bsphack\if@filesw {\let\thepage\relax
   \xdef\@gtempa{\write\@auxout{\string
      \newlabel{#1}{{\@currentlabel}{\thepage}}}}}\@gtempa
   \if@nobreak \ifvmode\nobreak\fi\fi\fi\@esphack}
        \gdef\@eqnlabel{#1}}
\def\@eqnlabel{}
\def\@vacuum{}
\def\draftmarginnote#1{\marginpar{\raggedright\scriptsize\tt#1}}

\def\draft{\oddsidemargin -0.1truein
        \def\@oddfoot{\sl preliminary draft \hfil
        \rm\thepage\hfil\sl\today\quad\militarytime}
        \let\@evenfoot\@oddfoot \overfullrule 3pt
        \let\label=\draftlabel
        \let\marginnote=\draftmarginnote
\def\@eqnnum{{\rm (\theequation)}
\rlap{\kern\marginparsep\tt\@eqnlabel}%
\global\let\@eqnlabel\@vacuum}  }

\newfont{\Bbbb}{msbm7 scaled 1\@ptsize00}
\newcommand{\zs}{\raise-1pt\hbox{$\mbox{\Bbbb Z}$}}

scaled 1\@ptsize00

\font\sevenmsa=msam6 
\newfam\msafam
\textfont\msafam=\sevenmsa
\def\hexnumber@#1{\ifnum#1<10 \number#1\else
\ifnum#1=10 A\else\ifnum#1=11 B\else\ifnum#1=12 C\else \ifnum#1=13
D\else\ifnum#1=14 E\else\ifnum#1=15 F\fi\fi\fi\fi\fi\fi\fi}
\def\msa@{\hexnumber@\msafam}
\def\llcorner{\delimiter"4\msa@78\msa@78 }
\def\lrcorner{\delimiter"5\msa@79\msa@79 }
\mathchardef\blacktriangleright="3\msa@49
\mathchardef\blacktriangleleft="3\msa@4A \font\tenmsb=msbm10 scaled
1\@ptsize00
\newfam\msbfam
\textfont\msbfam=\tenmsb \scriptfont\msbfam=\tenmsb

\newdimen\Squaresize \Squaresize=14pt
\newdimen\Thickness \Thickness=0.5pt

\def\Square#1{\hbox{\vrule width \Thickness
   \vbox to \Squaresize{\hrule height \Thickness\vss
      \hbox to \Squaresize{\hss#1\hss}
   \vss\hrule height\Thickness}
\unskip\vrule width \Thickness} \kern-\Thickness}

\def\Vsquare#1{\vbox{\Square{$#1$}}\kern-\Thickness}

\def\numberbysection{\@addtoreset{equation}{section}
        \def\theequation{\thesection.\arabic{equation}}}
\numberbysection

\renewcommand{\theequation}{\thesection.\arabic{equation}}
\def\titlepage{\@restonecolfalse\if@twocolumn\@restonecoltrue\onecolumn
     \else \newpage \fi \thispagestyle{empty}\c@page\z@
        \def\thefootnote{\fnsymbol{footnote}} }

\def\endtitlepage{\if@restonecol\twocolumn \else  \fi
        \def\thefootnote{\arabic{footnote}}
        \setcounter{footnote}{0}}  
\relax

\hybrid
\makeatletter
\newdimen\normalarrayskip            
\newdimen\minarrayskip               
\normalarrayskip\baselineskip \minarrayskip\jot
\newif\ifold             \oldtrue            \def\new{\oldfalse}
\def\arraymode{\ifold\relax\else\displaystyle\fi}
\def\eqnumphantom{\phantom{(\theequation)}} 
\def\@arrayskip{\ifold\baselineskip\z@\lineskip\z@
     \else
     \baselineskip\minarrayskip\lineskip1\baselineskip\fi}

\def\@arrayclassz{\ifcase \@lastchclass \@acolampacol \or
\@ampacol \or \or \or \@addamp \or
   \@acolampacol \or \@firstampfalse \@acol \fi
\edef\@preamble{\@preamble
  \ifcase \@chnum
     \hfil$\relax\arraymode\@sharp$\hfil
     \or $\relax\arraymode\@sharp$\hfil
     \or \hfil$\relax\arraymode\@sharp$\fi}}

\def\@array[#1]#2{\setbox\@arstrutbox=\hbox{\vrule
     height\arraystretch \ht\strutbox
     depth\arraystretch \dp\strutbox
width\z@}\@mkpream{#2}\edef\@preamble{\halign \noexpand\@halignto
\bgroup \tabskip\z@ \@arstrut \@preamble \tabskip\z@ \cr}%
\let\@startpbox\@@startpbox \let\@endpbox\@@endpbox
    \if #1t\vtop \else \if#1b\vbox \else \vcenter \fi\fi
  \bgroup \let\par\relax
  \let\@sharp##\let\protect\relax
  \@arrayskip\@preamble}
\def\eqnarray{\stepcounter{equation}%
              \let\@currentlabel=\theequation
              \global\@eqnswtrue
              \global\@eqcnt\z@
              \tabskip\@centering              
              \let\\=\@eqncr
              $$%
            \halign to \displaywidth  \bgroup
             \eqnumphantom \@eqnsel
      \hskip\@centering                               
    $\displaystyle  \tabskip\z@ {##}$%
    &\global\@eqcnt\@ne \hskip 2\arraycolsep
         $ \displaystyle  \arraymode{##}$\hfil
    &\global\@eqcnt\tw@ \hskip 2\arraycolsep
         $\displaystyle\tabskip\z@{##}$\hfil
         \tabskip\@centering
    &{##}\tabskip\z@\cr}
\makeatother
\newcommand{\RR}{{\mathbb{R}}}
\newcommand{\CC}{{\mathbb{C}}}

\def\IC{\mathbb{C}}

\def\IP{\mathbb{P}}

\def\IR{\mathbb{R}}
\def\IZ{\mathbb{Z}}
\def\CA {\mathcal{A}}

\def\CC {\mathcal{C}}

\def\CE {\mathcal{E}}
\def\CF {\mathcal{F}}
\def\CG {\mathcal{G}}
\def\CH {\mathcal{H}}

\def\CL {\mathcal{L}}
\def\CM {\mathcal{M}}
\def\CN {\mathcal{N}}
\def\CO {\mathcal{O}}

\def\CQ {\mathcal{Q}}

\def\CU {\mathcal{U}}
\def\CV {\mathcal{V}}

\def\CZ {\mathcal{Z}}

\def\s {{\sigma}}

\def\la{\lambda}

\def\e{\epsilon}
\def\pr {\partial}
\def\apr {\overline {\partial }}

\def\jb{\bar{j}}

\def\wb {\bar{w}}
\def\zb {\bar{z}}

\def\bvarphi{\bar{\varphi}}

\def\nn{\nonumber}

\def\frak{\mathfrak}

\newtheorem{te}{Theorem}[section]
\newtheorem{de}{Definition}[section]
\newtheorem{prop}{Proposition}[section]
\newtheorem{cor}{Corollary}[section]
\newtheorem{lem}{Lemma}[section]
\newtheorem{ex}{Example}[section]
\newtheorem{rem}{Remark}[section]

\newcommand{\proof}{\noindent {\it Proof}.\,\,}
\newcommand\bqa{\begin{eqnarray}}
\newcommand\eqa{\end{eqnarray}}
\def\be{\begin{eqnarray}\new\begin{array}{cc}}
\def\ee{\end{array}\end{eqnarray}}
\def\nn{\nonumber}

\def\beq{\begin{equation}}
\def\eeq{\end{equation}}
\def\bse{\begin{subequations}}                
\def\ese{\end{subequations}}
\def\bp{\begin{pmatrix}}
\def\ep{\end{pmatrix}}

\def\h{\hbar}
\def\i{\imath}

\newcommand\rk{\operatorname{rank}}

\def\stack#1#2{\raise0.7pt\hbox{$\mathrel{\mathop{#2}\limits^{#1}}$}}
\def\tr{\triangleright}
\def\tl{\triangleleft}
\def\sem{\mathsurround=0pt \raise1pt
\hbox{$\scriptscriptstyle>\!\!$}\:\!\!\tl}
\def\mes{\mathsurround=0pt \tr\!\:\!\raise0.8pt
\hbox{$\scriptscriptstyle\!\!<$}\,}
\def\]{\mathsurround=0pt ]\raise-2pt\hbox{$_\ast$}}

\def\<{\langle}
\def\>{\rangle}

\def\CQ{{\cal Q}}
\def\frak{\mathfrak}

\def\CO{{\cal O}}
\def\CU{{\cal U}}
\def\CZ{{\cal Z}}

\def\CH{\mathcal{H}}

\def\we{\raise-1pt\hbox{$\,\stackrel{\wedge}{,}\,$}}
\def\tr{{\rm tr}\,}

\def\pr {\partial}

\def\vp{\varphi}
\newcounter{pac}[section]

\newcounter{pacc}[subsection]

0

\title{\bf Parabolic Whittaker Functions and \\
Topological Field Theories I}
\begin{document}
\author{Anton Gerasimov, Dimitri Lebedev and Sergey Oblezin}
\date{}
\maketitle

\renewcommand{\abstractname}{}

\begin{abstract}
\noindent {\bf Abstract}.
First, we define a generalization of the standard quantum Toda chain
inspired by a construction of
quantum cohomology of partial flags spaces $GL_{\ell+1}/P$, $P$
a  parabolic subgroup. Common eigenfunctions of the
parabolic quantum Toda chains are generalized Whittaker functions
given by matrix elements of infinite-dimensional representations of
$\mathfrak{gl}_{\ell+1}$. For maximal parabolic subgroups (i.e. for
$P$ such that $GL_{\ell+1}/P=\IP^{\ell}$) we construct two different representations
of the corresponding  parabolic Whittaker functions as correlation functions
in topological  quantum field theories on a two-dimensional disk. In
one case the parabolic Whittaker function is given by a correlation function
in a type $A$ equivariant topological sigma model with the target
space $\IP^{\ell}$. In the other case the same Whittaker function
appears as a correlation function in a type $B$ equivariant
topological Landau-Ginzburg model related with the type $A$ model
by  mirror symmetry. This note is a continuation of our
project of establishing a relation between two-dimensional
topological field theories (and more generally topological string theories)
and Archimedean ($\infty$-adic)  geometry.
From this perspective the existence of
two, mirror dual,   topological field theory
representations  of the parabolic Whittaker functions
provide  a quantum field theory realization of the
local Archimedean  Langlands duality for  Whittaker functions.
The established relation between the Archimedean  Langlands duality
and mirror symmetry in two-dimensional topological quantum field
theories should be considered as a main result of this note.

\end{abstract}
\vspace{5 mm}

\section*{Introduction}

In  \cite{GLO1}, \cite{GLO2} we propose two-dimensional topological
field theories as a proper framework for a description  of the
Archimedean completion of arithmetic schemes ($\infty$-adic geometry
according to \cite{MaP}). In particular, we give a representation of
local Archimedean $L$-factors (we include local epsilon-factor
in the definition of the $L$-factors) in terms of two-dimensional
topological  field theories.  It is well-known that local
$L$-factors allow two types of constructions - ``arithmetic''
construction based on  representation theory  of the Weil-Deligne
group of the local  field  and ``automorphic'' construction
relying on representation theory of reductive groups over local
field (see e.g. \cite{Bu}, \cite{L}, \cite{ABV}). The equivalence of
these constructions for various types of $L$-factors is a subject of
the local Langlands duality. In an interpretation suggested in
\cite{GLO1}, \cite{GLO2} the ``arithmetic'' construction of local
Archimedean $L$-factors is naturally identified with a type $A$
topological field theory description \cite{GLO1} in terms of
equivariant volumes of  spaces of holomorphic maps of a disk into
complex vector spaces. The ``automorphic'' construction of the same
local $L$-factors  is realized using  a type $B$ topological field
theory via  periods of holomorphic forms \cite{GLO2}. The Archimedean Langlands
duality between these two constructions of the local Archimedean
$L$-factors appears as  a  mirror duality between underlying type
$A$ and type $B$ topological sigma models.

The duality between two constructions of $L$-factors
can be extended to a duality between two constructions of the Whittaker
functions. In the  non-Archimedean case this duality manifests in
the existence of the Shintani-Casselman-Shalika (SCS) formula \cite{Sh},
\cite{CS} for the non-Archimedean Whittaker functions
along with a standard realization of the  Whittaker functions as matrix
elements  of representations of reductive groups $G$ over local non-Archimedean
fields. According to \cite{Sh}, \cite{CS} the
 non-Archimedean Whittaker functions can
be identified with  characters of finite-dimensional representations
of Langlands dual groups ${}^LG$. In \cite{GLO4} we propose a
$q$-version of the classical SCS formula
providing  a $q$-version of the Langlands duality pattern for
the $q$-deformed Whittaker functions. In appropriate limit the
$q$-version of SCS formula reduces to the non-Archimedean  one. The
limiting case provides an Archimedean analog of the results of \cite{Sh},
\cite{CS}. This leads to an explicit  realization of the
 Langlands duality pattern for the Archimedean Whittaker functions.

In this note the  approach of \cite{GLO1}, \cite{GLO2} to a
construction of local Archimedean $L$-factors
in terms of topological field theories is generalized
to a class of Whittaker functions introduced below.
Recall that standard $\mathfrak{gl}_{\ell+1}$-Whittaker functions
are common eigenfunctions of  quantum $\mathfrak{gl}_{\ell+1}$-Toda chain
Hamiltonians and according to \cite{Gi3}  appear in a description
of $S^1\times U_{\ell+1}$-equivariant Gromov-Witten invariants of
complete flag spaces $GL_{\ell+1}(\IC)/B$, $B$ a Borel subgroup.
In the first part of this note we introduce a class of generalized
$\mathfrak{gl}_{\ell+m}$-Whittaker functions associated with a
parabolic subgroup $P\subset GL_{\ell+m}(\IC)$.
We conjecture that a $P$-parabolic $\mathfrak{gl}_{\ell+m}$-Whittaker
function describes $S^1\times U_{\ell+m}$-equivariant Gromov-Witten invariants of
the partial flag space $GL_{\ell+m}(\IC)/P$.
The parabolic $\mathfrak{gl}_{\ell+m}$-Whittaker functions are common eigenfunctions
of quantum Hamiltonians of  generalized quantum  Toda chain.
We explicitly check, in the case of $GL_{\ell+m}(\IC)/P=Gr(m,\ell+m)$,
that  the corresponding Toda chain recovers Astashkevich-Sadov-Kim
description \cite{AS}, \cite{K} of quantum cohomology of Grassmannian
${\rm Gr}(m,\ell+m)$.
In the case $m=1$ we explicitly verify that the corresponding
parabolic Whittaker functions describe equivariant quantum cohomology
of ${\rm Gr}(1,\ell+1)=\IP^{\ell}$.

In the second part of this
note we propose two representations of the parabolic Whittaker functions
associated with maximal parabolic subgroup $P$ (i.e. such that
$GL_{\ell+1}(\IC)/P=\IP^{\ell}$) as correlation functions
in two-dimensional topological fields theories on a disk.
The first representation ( see Theorem \ref{Amodel})
 is given in terms of an equivariant  type $A$ topological sigma model
with a target space $\IP^{\ell}$ and provides an
integral representation of the parabolic Whittaker function
of the Mellin-Barnes type. In this representation, the  Whittaker
function arises as an equivariant volume of a space of holomorphic
maps of the disk into projective
space $\IP^{\ell}$ and thus, following \cite{GLO1},  shall be  considered as
an ``arithmetic'' construction  of the parabolic Whittaker function.
This representation provides an Archimedean analog of SCS formula \cite{Sh}, \cite{CS}
for the parabolic Whittaker functions.
The second representation is given in terms of a type $B$ topological Landau-Ginzburg
model on a disk with a target space $\IC^{\ell+2}$ supplied with  appropriate
superpotential (see Theorem \ref{Bmodel}). 
In this description  the parabolic Whittaker function
is given by a holomorphic period \cite{Gi2}.
In contrast with the type $A$ model representation
the correlation functions  in $B$ model can be reduced to
certain finite-dimensional integrals with  simple integrands.
Thus obtained finite-dimensional integral representation naturally arises
in  a matrix element interpretation of the Whittaker function
according to \cite{GKLO}. Hence, in analogy with
\cite{GLO2},  the type $B$ topological field theory
representation  shall  be considered as an ``automorphic''/representation theory
construction of the parabolic Whittaker functions. The discussed
type $A$ and type $B$ topological quantum field
theories are related by mirror transformation.
This leads to interpretation of
the Archimedean Langlands duality between
``arithmetic'' and ``automorphic'' constructions of the parabolic
Whittaker functions  as a mirror duality between underlying topological field theories.

Finally let us  note that the results of this note can be generalized to the case of
general parabolic Whittaker functions. For a general
parabolic subgroup, the corresponding parabolic version of
the quantum $\mathfrak{gl}_{\ell+m}$-Toda chain
provides a new interesting example  of a quantum integrable system 
and will be considered elsewhere. 
One should also stress that   explicit calculations of correlation
functions in topological field theories on non-compact
manifolds is an interesting  subject by itself and undoubtedly deserves further
attention.  According
to the standard lore  boundary conditions in
topological field theories encode  geometry
of topological branes in a target space. It would be interesting
to compare the  choice of the boundary conditions used in this note
with an equivariant generalization of the standard brane boundary conditions.
We are going to clarify this relation elsewhere.

The plan of the paper is as follows.
In Section 1 we introduce parabolic generalizations of the standard
$\mathfrak{gl}_{\ell+m}$-Whittaker function
as  particular matrix elements of infinite-dimensional representations of
Lie algebra $\mathfrak{gl}_{\ell+m}$.
In Section 2 we construct a
representation of the parabolic
$\mathfrak{gl}_{\ell+1}$-Whittaker  functions
associated  with a maximal parabolic subgroup as  correlation
functions of a type $A$ topological sigma model on
a disk with a target space $\IP^{\ell}$. In Section 3 we
 provide a mirror dual representation of the same Whittaker
function as a correlation
function in a type $B$ topological Landau-Ginzburg model on a  disk.
In Section 4 we give a heuristic derivation of the result
of Sections 2 and 3. In  Section 5 we discuss directions of further research.
Finally in Appendixes the proofs and some technical results
are given.

{\em Acknowledgments}: The research was supported by  Grant
RFBR-09-01-93108-NCNIL-a. The research of AG was  also partly
supported by Science Foundation Ireland grant and the research of SO
was partially supported by Deligne's 2004 Balzano prize in
Mathematics.   The authors are thankful to
Max-Planck-Institut f\"ur Mathematik in Bonn for hospitality and
excellent working conditions.

\section{Parabolic Whittaker functions}

In this Section we introduce a generalization of
$\mathfrak{gl}_{\ell+m}$-Whittaker functions associated with a
parabolic subgroup $P\subset GL_{\ell+m}(\IC)$. The standard
$\mathfrak{gl}_{\ell+m}$-Whittaker functions are associated with
Borel subgroups $B\subset GL_{\ell+m}(\IC)$  and are
common  eigenvalues of  quantum $\mathfrak{gl}_{\ell+m}$-Toda
chain Hamiltonians (for standard facts on
quantum Toda chains see e.g. \cite{STS}).
The classical Whittaker functions are relevant to
a description of (equivariant)  Gromov-Witten invariants of flag
spaces $G/B$ \cite{Gi3}. The parabolic generalizations of
Whittaker functions introduced below
are common eigenvalues of generalized quantum Toda chains
defined below. In the next Sections we demonstrate that the parabolic
Whittaker functions describe equivariant
Gromov-Witten invariants of partial flag spaces.
In this Section we restrict considerations to
the case of the  parabolic subgroup $P_{m,\ell+m}$ such that
$GL_{\ell+m}(\IC)/P_{m,\ell+m}=Gr(m,\ell+m)$ and refer to the
corresponding Whittaker functions as $(m,\ell+m)$-Whittaker
functions.  The general case follows basically the same pattern and
will be treated elsewhere.

First we recall the representation theory construction
of the standard $\mathfrak{gl}_{\ell+1}$-Whittaker functions.
Let $E_{ij}$, $i,j=1,\ldots \ell+1$ be the standard basis of the Lie
algebra $\mathfrak{gl}_{\ell+1}$.
Let $\CZ(\CU\mathfrak{gl}_{\ell+1})\subset
\CU\mathfrak{gl}_{\ell+1}$ be the center of the  universal
enveloping algebra $\CU\mathfrak{gl}_{\ell+1}$.
Let  $B_{\pm}\subset GL_{\ell+1}(\IC)$
be  upper-triangular and lower-triangular
Borel subgroups and  $N_{\pm}\subset B_{\pm}$
be  upper-triangular and lower-triangular
unipotent subgroups. We denote $\mathfrak{b}_{\pm}={\rm Lie}(B_{\pm})$ and
$\mathfrak{n}_{\pm}={\rm Lie}(N_{\pm})$ their Lie algebras.
 Let $\mathfrak{h}\subset \mathfrak{gl}_{\ell+1}$
be a diagonal Cartan subalgebra and $S_{\ell+1}$ be the associated Weyl
group of $GL_{\ell+1}$.   Using the Harish-Chandra
isomorphism of $\CZ(\CU\mathfrak{gl}_{\ell+1})$ with the Weyl group
invariant subalgebra of the symmetric algebra of the Cartan subalgebra
$\mathfrak{h}$ we  identify central characters with
homomorphisms $c:\IC[h_1,\cdots,
h_{\ell+1}]^{S_{\ell+1}}\to \IC$ of an algebra of $S_{\ell+1}$-invariant
polynomials of the generators of $\mathfrak{h}$ into complex numbers.
Let $\pi_{\underline{\lambda}}:\CU\mathfrak{gl}_{\ell+1}\to
{\rm End}(\CV_{\underline{\la}})$,
$\CV_{\underline{\la}}={\rm   Ind}_{\CU\frak{b}_-}^{\CU \frak{gl}_{\ell+1}}$
be a family of principal series representations   of
$\CU\frak{gl}_{\ell+1}$  induced from  one-dimensional
representations of $\CU\frak{b}_-$ such that
images of the symmetric polynomials of $h_i$ are symmetric
polynomials of $\lambda_j\in\IC$,
$\underline{\lambda}=(\lambda_1,\cdots ,\lambda_{\ell+1})\in
\IC^{\ell+1}$.  Let $\CV'_{\underline{\lambda}}$ be a dual module
supplied with induced action of $\CU\mathfrak{gl}_{\ell+1}^{opp}$
(universal enveloping algebra of $\mathfrak{gl}_{\ell+1}^{opp}$
obtained by inverting the signs of the structure constants of
$\mathfrak{gl}_{\ell+1}$). Denote $\<\,,\,\>$ the pairing
between $\CV'_{\underline{\lambda}}$ and $\CV_{\underline{\lambda}}$.
We suppose that the action of the Cartan subalgebra $\mathfrak{h}$
 in representation  $\CV_{\underline{\lambda}}$ can be integrated
to the action of the corresponding Cartan subgroup $H\subset GL_{\ell+1}(\IC)$.

According to Kostant the $\mathfrak{gl}_{\ell+1}$-Whittaker function
can be defined as  a matrix element
\be\label{Wf}
\Psi_{\underline{\lambda}}(x_1,\ldots, x_{\ell+1})\,=\,e^{-\rho(x)}
\<\psi_L|\,\pi_{\underline{\lambda}}(e^{-\sum_{i=1}x_iE_{ii}})\,|\psi_R\>,
\ee
where  the  vectors $\<\psi_L|\in \CV'_{\underline{\lambda}}$ and
$|\psi_R\>\in \CV_{\underline{\lambda}}$
provide  one-dimensional representations
of $N_-$ and $N_+$ correspondingly
\be\label{Wclass}
\<\psi_L|E_{i+1,i}=-\<\psi_L|,\qquad
E_{i,i+1}|\psi_R\>=|\psi_R\>,\qquad i=1,\ldots ,\ell
\ee
 and $\rho_r=(1/2)(\ell+2-2k),\,\,\,k=1,\ldots,\ell+1$ are the components of the vector
 $\rho$ in  $\IR^{\ell+1}$.
The standard considerations (see e.g. \cite{STS})
show that  the matrix element \eqref{Wf}  is a common
eigenfunction  of a family of commuting differential operators
descending  from the action of the generators of
$\CZ(\CU\mathfrak{gl}_{\ell+1})$  in $\CV_{\underline{\lambda}}$.
These differential operators  can be identified with quantum Hamiltonians of
$\mathfrak{gl}_{\ell+1}$-Toda chain.

Below we propose a generalization of the $\mathfrak{gl}_{\ell+m}$-Whittaker
functions \eqref{Wf} (for convenience we slightly change notations
replacing $\ell+1$ by $\ell+m$). Let $P_{m,\ell+m}$ be  a
parabolic subgroup of $GL_{\ell+m}(\IC)$ such that the corresponding Levi
factor is $GL_{m}(\IC)\times GL_{\ell}(\IC)$. The corresponding partial flag
space $GL_{\ell+m}(\IC)/P_{m,\ell+m}$ is isomorphic to Grassmannian ${\rm
  Gr}(m,\ell+m)$. The associated  Whittaker
 function is then  defined as the following
matrix element of the principle series representation
$\CV_{\underline{\la}}={\rm   Ind}_{\CU(\frak{b}_-)}^{\CU(\frak{gl}_{\ell+m})}$.
Let us  associate with $P_{m,\ell+m}$  a decomposition of the Borel
subalgebra $\frak{b}_+\subset\frak{gl}_{\ell+m}$
$$
 \frak{b}_+\,\,=\,\,\frak{h}^{(m,\ell+m)}
 \,\,\oplus\,\,
 \frak{n}_+^{(m,\ell+m)}\,,
$$
with the commutative subalgebra
$\frak{h}^{(m,\,\ell+m)}\subset\frak{b}_+$ generated by
\be \label{Cartanml}
  H_1\,=\,E_{11}+\ldots+E_{mm}\,;
  \hspace{1.5cm}
  H_k\,=\,E_{1,k},\,\quad 2\leq k\leq m\,;\\
  H_{m+k}\,=\,E_{m+k,\,\ell+m},\,\quad
  1\leq k\leq\ell-1\,;
  \hspace{1.5cm}
  H_{\ell+m}\,=\,E_{m+1,\,m+1}+\ldots+E_{\ell+m,\,\ell+m}\,,
 \ee
and the subalgebra $\frak{n}^{(m,\,\ell+m)}_+\subset\frak{b}_+$ given  by
 \be
  \frak{n}^{(m,\,\ell+m)}_+\,=\,\bigl\<E_{1,\,\ell+m};\,E_{1,\,m+1};\,
  E_{m,\,\ell+m};\hspace{5cm}\\
  \hspace{2cm}
  E_{kk}\,,2\leq k\leq \ell+m-1\,;\,
  E_{j,\,j+1},\,2\leq j\leq\ell+m-2\bigr\>\,.
 \ee
Note that
$\dim\frak{h}^{(m,\,\ell+m)}=\rk\,\frak{gl}_{\ell+m}=\ell+m$ and
$\dim\frak{n}^{(m,\,\ell+m)}_+=(\ell+m)(\ell+m-1)/2$. Let
$H^{(m,\ell+m)}$ and $N_+^{(m,\ell+m)}$ be the  Lie groups
corresponding to  the Lie algebras $\mathfrak{h}^{(m,\ell+m)}$ and
$\mathfrak{n}_+^{(m,\ell+m)}$. An open part $\dot{GL}_{\ell+m}$
of $GL_{\ell+m}$ allows the following analog of the  Gauss
decomposition: \be\label{modG}
\dot{GL}_{\ell+m}=N_-\,H^{(m,\ell+m)}\,N_+^{(m,\ell+m)}. \ee

\begin{de} The  Whittaker vectors
$\<\psi_L|\in \CV'_{\underline{\lambda}}$,
$|\psi_R\>\in\CV_{\underline{\lambda}}$ are defined by the following
conditions:
 \be\label{LeftWhittEqs}
  \<\psi_L|E_{n+1,\,n}\,=\,\h^{-1}\<\psi_L|\,,
  \hspace{2.5cm}
  1\leq n\leq\ell+m-1\,,
 \ee
 \be\label{RightWhittEqs}
  \left\{
  \begin{array}{lc}
  E_{kk}|\psi_R\>\,=\,0\,,&2\leq k\leq\ell+m-1\,;\\
  E_{k,\,k+1}|\psi_R\>\,=\,0,\,&2\leq k\leq\ell+m-2\,;\\
  E_{1,\,m+1}|\psi_R\>\,=E_{m,\,\ell+m}|\psi_R\>\,=\,0\,;&\\
  E_{1,\,\ell+m}|\psi_R\>\,=(-1)^{\epsilon(\ell,m)}\frac{1}{\h}|\psi_R\>,&
  \end{array}
  \right.
 \ee
where  $\epsilon(\ell,m)$ is an integer number and $\hbar \in \IR$.
\end{de}
Here, in comparison with \eqref{Wclass}, we introduce additional 
parameter $\hbar$ to make a contact with the results of other Sections
( \eqref{Wclass} corresponds to $\hbar=1$).
Note that the equations \eqref{LeftWhittEqs} define a one-dimensional
representation $\<\psi_L|$ of the Lie algebra $\frak{n}_-$ of strictly
lower-triangular matrices and
the equations \eqref{RightWhittEqs} define a one-dimensional
representation of $\frak{n}^{(m,\,\ell+m)}_+$.

\begin{de} The  $(m,\ell+m)$-Whittaker function associated
with  the principal series representation
$\bigl(\pi_{\underline{\lambda}},\,\CV_{\underline{\lambda}}\bigr)$
is defined as the following matrix element:
 \be\label{parTod}
  \Psi^{(m,\,\ell+m)}_{\underline{\lambda}}(x)\,
  =\,e^{-\rho_1(x_1-x_{\ell+m})}\<\psi_L|\pi_{\underline{\lambda}}\bigl(g(x)\bigr)
|\psi_R\>\,,
 \ee
where  the left and right  vectors  solve the equations \eqref{LeftWhittEqs} and
\eqref{RightWhittEqs} respectively and $\rho_1=(\ell+m-1)/2$. 
Here $g(x)$ is a Cartan group valued function given by
 \be\label{GrGroupElement}
  g(x)\,=\,\exp\Big\{-\sum_{i=1}^{\ell+m}\,x_iH_i\Big\}\,,
 \ee
where the generators $H_i$, $i=1,\ldots , (\ell+m)$  are defined by  \eqref{Cartanml}.
\end{de}
 Similar to the classical Whittaker functions $(m,\ell+m)$-parabolic
 Whittaker functions are naturally
common eigenfunctions of a family of commuting differential
operators. Let us define a set of mutually commuting
differential operators $\CH^{(m,\ell+m)}_k$, $k=1,\ldots, (\ell+m)$
by the following conditions
 \be\label{GrTodaHamiltonians}
  \CH^{(m,\ell+m)}_k(x,\pr_x)\cdot\Psi_{\underline{\la}}^{(m,\,\ell+m)}(x)\,
  =\,\hbar^k\,e^{-\rho_1(x_1-x_{\ell+m})}\bigl
\<\psi_L\bigl|\,\pi_{\underline{\la}}\bigl(C_k\,g(x)\bigr)
  \bigr|\psi_R\bigr\>\,,
 \ee
where $C_k\in\CZ(U\frak{gl}_{\ell+m})$ is a  Casimir element of the
center   acting in $\CV_{\underline{\lambda}}$ by multiplication on $k$-th elementary
symmetric polynomial $\sigma_k(\underline{\lambda})$ of the variables $\lambda_i$,
$i=1,\ldots, (\ell+m)$. Thus,
Whittaker function $\Psi_{\underline{\la}}^{(m,\,\ell+m)}(x)$
tautologically satisfy the
following system of differential equations
\be\label{GrTodaeq}
  \CH^{(m,\ell+m)}_k(x,\pr_x)\cdot\Psi_{\underline{\la}}^{(m,\,\ell+m)}(x)\,
  =\,\sigma_k(\underline{\lambda})\,\Psi_{\underline{\la}}^{(m,\,\ell+m)}(x)\,,
\qquad k=1,\ldots, (\ell+m),
 \ee

\begin{rem} The set of mutually commuting
differential operators $\CH^{(m,\ell+m)}_k$,
$k=1,\ldots, (\ell+m)$  defines a quantum integrable
system generalizing  the standard  quantum
$\mathfrak{gl}_{\ell+m}$-Toda chain. The Whittaker functions
$\Psi_{\underline{\la}}^{(m,\,\ell+m)}(x)$ provide a solution of the
corresponding eigenfunction problem.
\end{rem}

For the first
two lowest degree differential operators $\CH^{(m,\ell+m)}_1$,
$\CH^{(m,\ell+m)}_2$ we have
\be\label{Casimirs}
 C_1\,=\,\sum_{j=1}^{\ell+m} E_{jj},\qquad
 C_2\,=\,\sum_{i<j}^{\ell+m}\bigl(
 E_{ii}E_{jj}\,-\,E_{ji}E_{ij}\bigr)\,
 -\,\sum_{j=1}^{\ell+m}\rho_jE_{jj}+\s_2(\rho)\,,
\ee
where $\rho=(\rho_1,\ldots,\rho_{\ell+m})$ with
$\rho_k=\frac{\ell+m+1}{2}-k,\,k=1,\ldots,\ell+m$  and
$\sigma_2(\rho)=\sum_{i<j}\rho_i\rho_j$.
These operators act in the representation
$\CV_{\underline{\lambda}}$ via  multiplication on
$$
\sigma_1(\underline{\lambda})=\sum_{j=1}^{\ell+m}\lambda_j,\qquad
\sigma_2(\underline{\lambda})=\sum_{1\leq i< j\leq \ell+m}\lambda_i\lambda_j\,.
$$
The explicit form of the differential operators $\CH_1$, $\CH_2$
corresponding to $C_1$ and $C_2$ is as follows.

\begin{prop}\label{propone} The following differential
operators satisfy the equations \eqref{GrTodaHamiltonians}
 \be\label{GrQuadraticHamiltonian}
  \CH^{(m,\ell+m)}_1(x,\pr_x)\,=\,-\hbar\frac{\pr}{\pr x_1}-
\hbar\frac{\pr}{\pr x_{\ell+m}},\\
  {\CH}_2^{(m,\,\ell+m)}\,=\,\h^2\Big\{
  \frac{\pr^2}{\pr x_1\pr x_{\ell+m}}\,
  +\,\sum_{k=2}^m x_k\frac{\pr^2}{\pr x_1\pr x_k}\,
  -\,\sum_{2\leq k\leq a}^mx_kx_a\frac{\pr^2}{\pr x_k\pr x_a}\\
  -\,\sum_{n=1}^{\ell-1}
  x_{m+n}\frac{\pr^2}{\pr x_{m+n}\pr x_{\ell+m}}\,
  -\,\sum_{1\leq n\leq b}^{\ell-1}
  x_{m+n}x_{m+b}\frac{\pr^2}{\pr x_{m+n}\pr x_{m+b}}\\
  -\,\sum_{k=2}^m\,(1-\rho_k)x_k\frac{\pr}{\pr x_k}\,
  -\,\sum_{n=1}^{\ell-1}(1+\rho_{m+n})x_{m+n}\frac{\pr}{\pr x_{m+n}}
  \Big\}\\
  +\,\h\Big\{(1-\delta_{m,\,1})\frac{\pr}{\pr x_2}\,
  -\,(1-\delta_{\ell,\,1})\frac{\pr}{\pr x_{\ell+m-1}}\,
  +\,\sum_{k=2}^{m-1}x_k\frac{\pr}{\pr x_{k+1}}\,
  -\,\sum_{n=1}^{\ell-2}x_{m+n+1}\frac{\pr}{\pr x_{m+n}}\Big\}\\
  +\,(-1)^{\delta_{\ell,\,1}+\epsilon(\ell,m)}
  (x_m)^{1-\delta_{m,\,1}}(x_{m+1})^{1-\delta_{\ell,\,1}}e^{x_1-x_{\ell+m}}\,
  -\,\frac{\h^2}{24}(\ell+m-1)(\ell+m-2)(\ell+m-3)\,.
 \ee

\end{prop}
\noindent {\it Proof}.
The case of $\CH^{(m,\ell+m)}_1$ is trivial and the proof of the
expression for  $\CH^{(m,\ell+m)}_2$  is given in Appendix A. $\Box$

\begin{cor}  The $(m,\ell+m)$-parabolic Whittaker function
  \eqref{parTod} 
satisfies the following  equations:
$$
\CH^{(m,\ell+m)}_1(x,\pr_x)\,
 \Psi^{(m,\ell+m)}_{\underline{\lambda}}(x)=\sum_{j=1}^{\ell+m}\lambda_j\,\,\,
\Psi^{(m,\ell+m)}_{\underline{\lambda}}(x),
$$
\be\label{FirstTwo}
\CH^{(m,\ell+m)}_2(x,\pr_x)\, \Psi^{(m,\ell+m)}_{\underline{\lambda}}(x)
=\sum_{i<j}^{\ell+m}\lambda_i\lambda_j\,\,\,
\Psi^{(m,\ell+m)}_{\underline{\lambda}}(x).
\ee
where $\CH^{(m,\ell+m)}_1(x,\pr_x)$ and $\CH^{(m,\ell+m)}_2(x,\pr_x)$
are given by \eqref{GrQuadraticHamiltonian}.
\end{cor}

\begin{ex}
For  $m=1$ the quadratic Hamiltonian
\eqref{GrQuadraticHamiltonian} has the following form:
 \be\label{PellQuadraticHamiltonian}
  \hspace{-2cm}
  {\CH}^{(1,\,\ell+1)}_2\,=\,
  \hbar^2\frac{\pr^2}{\pr x_1\pr x_{\ell+1}}\,
  -\,\sum_{k=2}^{\ell}
  x_k\hbar^2\frac{\pr^2}{\pr x_k\pr x_{\ell+1}}\,
  -\,\sum_{2\leq k\leq a}^{\ell}
  x_kx_a\hbar^2\frac{\pr^2}{\pr x_k\pr x_a}
  -\,\sum_{k=2}^{\ell}(\rho_k+1)x_k \hbar^2\frac{\pr}{\pr x_k}\\
  +\,\h\frac{\pr}{\pr x_{\ell}}\,
  -\,\h\sum_{j=2}^{\ell-1}x_{j+1}\frac{\pr}{\pr x_j}
  +\,(-1)^{\epsilon(\ell,1)}
  x_2e^{x_1-x_{\ell+1}}\,-\,\frac{\h^2}{24}\ell(\ell-1)(\ell-2)\,.
 \ee
For  $\ell=1$ the quadratic Hamiltonian
\eqref{GrQuadraticHamiltonian} reads as follows:
 \be\label{PmQuadraticHamiltonian}
  \hspace{-1.5cm}
  {\CH}^{(m,\,m+1)}_2\,=\,\hbar^2\frac{\pr^2}{\pr x_1\pr x_{m+1}}\,
  +\,\sum_{k=2}^mx_k\hbar^2\frac{\pr^2}{\pr x_1\pr x_k}\,
  -\,\sum_{2\leq k\leq a}^mx_kx_a\hbar^2\frac{\pr^2}{\pr x_k\pr x_a}\,
  -\,\sum_{k=2}^m(1-\rho_k)x_k\h^2\frac{\pr}{\pr x_k}\\
  +\,\h\frac{\pr}{\pr x_2}\,
  +\,\h\sum_{k=2}^{m-1}x_k\frac{\pr}{\pr x_{k+1}}\,
  -(-1)^{\e(m,1)}\,x_me^{x_{1}-x_{m+1}}\,-\,\frac{\h^2}{24}m(m-1)(m-2)\,.
 \ee
\end{ex}

We conjecture that $(m,\ell+m)$-parabolic Whittaker functions describe
equivariant Gromov-Witten invariants of the Grassmannian
${\rm Gr}(m,\ell+m)=GL_{\ell+1}(\IC)/P_{m,\ell+m}$
thus generalizing the Givental description
\cite{Gi3} of $S^1\times U_{\ell+1}$-equivariant Gromov-Witten
invariants of the complete flag spaces $GL_{\ell+1}(\IC)/B$.
We support our  conjecture  by matching it with
a description of  quantum cohomology of ${\rm Gr}(m,\ell+m)$ due to
Astashkevich-Sadov-Kim \cite{AS}, \cite{K}.
To establish a relation with
\cite{AS}, \cite{K} let us define  a quantum  $\CL$-operator
associated with the quantum integrable system
\eqref{GrTodaHamiltonians} as  a  matrix-valued differential
operator satisfying the relation \be\label{Loper}
\CL(x,\pr_x)\,e^{-\rho_1(x_1-x_{\ell+m})}
\<\psi_L|\pi_{\underline{\lambda}}(g(x))|\psi_R\>:=
\hbar\sum_{i,j=1}^{\ell+1}\,e_{ij}e^{-\rho_1(x_1-x_{\ell+m})}\,
\<\psi_L|\pi_{\underline{\lambda}}(E_{ij}\,
g(x))|\psi_R\>, \ee where
$(e_{ij})_{kn}=\delta_{ik}\delta_{jn},\,\,$ $i,j,k,n=1,\ldots
,(\ell+m)$   are matrix unites.

\begin{prop}\label{PropTwo}
The matrix $\CL=\|\CL_{ij}\|$ of the quantum $L$-operator
\eqref{Loper} is given by
 \be\label{GrLaxOperator}
  \CL_{j+1,\,j}\,=\,1\,,
  \hspace{2.5cm}
  1\leq j\leq\ell+m-1\,;\\
  \CL_{j+s,\,j}\,=\,0\,,
  \hspace{1.5cm}
  1\leq j\leq\ell+m-1,
  \quad
  2\leq s\leq\ell+m-j\,;\\
  \CL_{11}\,=\,-\hbar\pr_{x_1}\,-\,\h\frac{\ell+m-1}{2}\,
  +\,\sum_{k=2}^mx_k\hbar\pr_{x_k}\,;
  \hspace{2.5cm}
  \CL_{1,\,k}\,=\,-\hbar\pr_{x_k}\,,
  \hspace{1cm}
  2\leq k\leq m\,;\\
  \CL_{k,\,k+j}\,=\,-x_k\hbar\pr_{x_{k+j}}\,,
  \hspace{1.5cm}
  2\leq k\leq m,\,\,
  0\leq j\leq m-k\,;\\
  \CL_{1,\,m+j}\,=\,-(-1)^{\epsilon(\ell,m)}x_{m+j}e^{x_1-x_{\ell+m}}\,,
  \hspace{1.5cm}
  1\leq j\leq\ell-1\,;\\
  \CL_{1,\,\ell+m}\,=\,(-1)^{\epsilon(\ell,m)}e^{x_1-x_{\ell+m}}\,;\\
  \CL_{k,\,m+j}\,=\,-(-1)^{\epsilon(\ell,m)}x_kx_{m+j}e^{x_{x_1-\ell+m}}\,,
  \hspace{1.5cm}
  2\leq k\leq m,
  \quad
  1\leq j\leq\ell-1\,;\\
  \CL_{k,\,\ell+m}\,=\,(-1)^{\epsilon(\ell,m)}x_ke^{x_1-x_{\ell+m}}\,,
  \hspace{1.5cm}
  2\leq k\leq m\,;\\
  \CL_{m+n,\,\ell+m}\,=\,-\hbar\pr_{x_{m+n}}\,,
  \hspace{1.5cm}
  1\leq n\leq \ell-1\,;\\
  \CL_{m+n,\,m+n+j}\,=\,x_{m+n+j}\hbar\pr_{x_{m+n}}\,,
  \hspace{1.5cm}
  1\leq n\leq\ell-2,
  \quad
  0\leq j\leq\ell-n-1\,;\\
  \CL_{\ell+m,\,\ell+m}\,=\,-\hbar\pr_{x_{\ell+m}}\,
  +\,\h\frac{\ell+m-1}{2}\,
  -\,\sum_{k=1}^{\ell-1}x_{m+k}\hbar\pr_{x_{m+k}}\,.
 \ee
\end{prop}
{\it Proof}. The proof is given in Appendix B.  $\Box$

The classical limit $L$ of the operator \eqref{Loper}
is defined by replacing derivatives by the classical
momenta $-\h\frac{\pr}{\pr x_j}\to p_j$ and taking the limit $\hbar\to
0$.   Let us specialize the resulting
matrix function $L(x_1,\ldots
\,x_{\ell+m};\,p_1,\ldots,p_{\ell+m})$ by
taking $x_2=\ldots=x_{\ell+m-1}=0$. This way  we obtain the
matrix  $L=\|L_{ij}\|$ with the following entries
 \be\label{GrLaxMatrix}
  L_{j+1,\,j}\,=\,1\,,
  \hspace{2.5cm}
  1\leq j\leq\ell+m-1\,;\\
  L_{1,\,k}\,=\,p_k\,,
  \hspace{1.5cm}
  1\leq k\leq m\,;\\
  L_{m+n,\,\ell+m}\,=\,p_{m+n}\,,
  \hspace{1.5cm}
  1\leq n\leq\ell\,;\\
  L_{1,\,\ell+m}\,=\,(-1)^{\epsilon(\ell,m)}e^{x_1 -x_{\ell+m}}\,.
 \ee
It is easy to verify that thus defined matrix $L=\|L_{ij}\|$
coincides (up to a conjugation by a simple matrix) with the
matrix entering a description of small quantum cohomology
of ${\rm Gr}(m,\ell+m)$  \cite{AS}, \cite{K}. This supports the conjecture
that $(m,\ell+m)$-Whittaker functions are relevant to a description of
$S^1\times U_{\ell+m}$-equivariant quantum cohomology of
${\rm Gr}(m,\ell+m)$.

 In the rest of the note  we will consider only the case of $m=1$ and
arbitrary $\ell$. In this case the conjectural relation between
solutions of the generalized Toda chain given by
$(1,\ell+1)$-Whittaker functions and $S^1\times
U_{\ell+m}$-equivariant Gromov-Witten invariants of
$\IP^{\ell}=GL_{\ell+1}/P_{1,\ell+1}$ can be proved
as follows. For  $m=1$ there is a well-known   description of
$S^1\times U_{\ell+1}$-equivariant quantum cohomology of
${\rm Gr}(1,\ell+1)=\IP^{\ell}$ in  terms of the functions given by
the following integral expressions  (see e.g. \cite{Gi2}). Let  us introduce
a modified $\Gamma$-function
\be \Gamma_1(z|\omega)=\omega^{\frac{z}{\omega}}\Gamma(\frac{z}{\omega}).
\ee
 Then
 \be\label{x=0one}
  \Phi_{\underline{\lambda}}(x) =\int_{\CC}
  \prod_{j=1}^{\ell}dt_j\,\, \exp \left(-\frac{1}{\h}\left(\sum_{j=1}^{\ell}\lambda_j
  t_j +\lambda_{\ell+1}(x-\sum_{j=1}^{\ell}t_j)+
  \sum_{j=1}^{\ell}e^{t_j}+
  e^{x-\sum_{j=1}^{\ell}t_j}\right)\right)=
 \ee
\be\label{x=0two}
=\frac{1}{2\pi\hbar}\int_{\RR-\imath\epsilon} \,\,\,d H \,\,
e^{-\frac{\imath}{\h} x H}\,\prod_{j=1}^{\ell+1}
\Gamma_1\Big(\imath H-\lambda_j)|\hbar\Big),
\ee
satisfies  the differential equation
\be\label{Deq}
\left(\prod_{j=1}^{\ell+1}(-\h\pr_{x}-\lambda_j)-e^{x}\right)
\Phi_{\underline{\lambda}}(x)=0.
\ee
Here $\epsilon > max(\lambda_j)\,\,\,j=1,\ldots,\ell+1$ 
 and ${\cal C} $ is a slightly deformed subspace $\RR^{\ell}\subset
 \IC^{\ell}$   making the integral \eqref{x=0one} convergent. 
For $\ell=1$  the function \eqref{x=0one} is the 
classical $\mathfrak{gl}_2$-Whittaker function.

\begin{te}\label{propthree} The $(1,\ell+1)$- parabolic
Whittaker function specialized to $x=x_1$ and $x_i=0$, $i\neq 1$
and for $\epsilon(1,\ell+1)=\frac{\ell(\ell-1)}{2}+1$
coincides with the generating function \eqref{x=0one} of
$S^1\times U_{\ell+1}$-equivariant quantum cohomology of
$\IP^{\ell}$. That is
\be\label{identone}
\Psi^{(1,\ell+1)}_{\underline{\lambda}}(x,0,\cdots ,0)\,=
\,\frac{1}{2\pi\hbar}\int_{\RR-\imath\epsilon} \,\,\,d H \,\,
e^{-\frac{\imath}{\h} x H}\,\prod_{j=1}^{\ell+1}
\Gamma_1\Big(\imath H-\lambda_j|\hbar\Big),
\ee
where $\underline{\lambda}\,\in\RR^{\ell+1}$
and $\epsilon > max (\lambda_j),\,\,\,j=1,\ldots,\ell+1$.

\end{te}
{\it Proof}. The proof is given in Appendix C. $\Box$

The integral representation \eqref{x=0two}
arises naturally when the matrix element \eqref{parTod} for $m=1$
is represented by using the Gelfand-Zetlin
realizations of the infinite-dimensional representations of
$\CU\mathfrak{gl}_{\ell+1}$ \cite{GKL} (similar relation holds
for the integral representation \eqref{x=0one} and the
representation of $\CU\mathfrak{gl}_{\ell+1}$ constructed in
\cite{GKLO}).
Note that one has an obvious symmetry ${\rm Gr}(m,\ell+m)={\rm
  Gr}(\ell,\ell+m)$. The compatibility of our conjecture with this
isomorphism  is explicitly  checked for $m=1$, $\ell=2$  in Appendix D.

In the next Sections we propose an identification of the
generating functions given by the integral representations
\eqref{x=0two}, \eqref{x=0one} with particular correlation functions
 in type $A$ and type $B$ equivariant
two-dimensional topological sigma models on a disk.

\section{Type A topological sigma model  with a target space
  $\IP^{\ell}$}

In this Section we define a class of correlation functions in
   $S^1\times U_{\ell+1}$-equivariant type $A$ topological
 sigma model on a disk  with a target space
  $\IP^{\ell}$ and calculate the correlation functions explicitly.
The resulting expressions coincide with the integrals
\eqref{x=0one}. This provides an infinite-dimensional integral
representations of $(1,\ell+1)$-Whittaker functions
\eqref{parTod} in terms of topological type $A$  equivariant
sigma models on a disk.

\subsection{Topological  gauged  linear sigma model}

We start recalling a gauge linear sigma model realization of the
sigma-model with the target space $\IP^{\ell}$ (see e.g. \cite{W3},
\cite{MP}). Consider a type $A$ topological linear sigma model on a
Riemann surface $\Sigma$ with the target space $\IC^{\ell+1}$. Let
$(z,\zb)$ be local complex coordinates on  $\Sigma$. We pick a
Hermitian metric $h$ on $\Sigma$ and denote $\sqrt{h}\,d^2z$ the
corresponding measure on $\Sigma$. The complex structure defines a
decomposition $d=\pr+\apr$, $\pr=dz\,\pr_z$, $\apr=d\zb\,\pr_{\zb}$
of the de Rham differential $d$ acting on differential forms on $\Sigma$.
Let $K$ and $\bar{K}$ be canonical and anti-canonical bundles over
$\Sigma$. Let $T_{\IC}\IC^{\ell+1}=T^{1,0}\IC^{\ell+1} \oplus
T^{0,1}\IC^{\ell+1}$ be a decomposition of the (trivial)
complexified tangent bundle to $\IC^{\ell+1}$ induced by a standard
complex structure on $\IC^{\ell+1}$.  We denote linear
complex coordinates
 on $\IC^{\ell+1}$ by $(\varphi^j, \bar{\varphi}^{j})$.
Consider a  two-dimensional topological quantum filed theory
based in the maps $\Phi:\,\Sigma\to \IC^{\ell+1}$  with 
the  action functional
\be\label{linsigmact} S= \int_{\Sigma}\,\,d^2z
\sqrt{h}\,\delta_0\CV= \int_{\Sigma}\,d^2z\,h^{z\zb}
\sqrt{h}\,\sum_{j=1}^{\ell+1}\Big( t\bar{F}_z^j F_{\zb}^j+ \imath
\bar{F}_{z}^j\pr_{\zb} \varphi^j-\imath F_{\zb}^j\pr_z
\bar{\varphi}^j+\imath \psi_{\zb}^j\,\pr_z\bar{\chi}^j-\imath
\bar{\psi}_z^j\,\pr_{\zb}\chi^j\Big), \ee where
$$\CV=h^{z\zb}
\sum_{j=1}^{\ell+1}(\bar{\psi}_z^j\,(\frac{t}{2}F_{\zb}^j+\imath
\pr_{\zb} \varphi^j) + \psi_{\zb}^j(\frac{t}{2}\bar{F}_z^{j}-\imath
\pr_z\bar{\varphi}^j)),
$$
and BRST transformations are given by \be\label{transformeqonea}
\delta_0 \varphi^j=\chi^j,\qquad \delta_0 \chi^j=0,\qquad \delta_0
\psi_{\zb}^j=F_{\zb}^j,\qquad \delta_0 F_{\zb}^j=0, \ee \be\nonumber
\delta_0 \bar{\varphi}^j=\bar{\chi}^j,\qquad \delta_0
\bar{\chi}^j=0, \qquad \delta_0 \bar{\psi}_z^j=\bar{F}_z^j,\qquad
\delta_0 \bar{F}_z^j=0. \ee Here the  commuting fields $F$ and
$\bar{F}$ are sections of $K\otimes \Phi^*(T^{0,1}\IC^{\ell+1})$ and
of $\bar{K}\otimes \Phi^*(T^{1,0}\IC^{\ell+1})$ correspondingly. The
anticommuting  fields $\chi$, $\bar{\chi}$ are sections  of the
bundles $\Phi^*(\Pi T^{1,0}\IC^{\ell+1})$, $\Phi^*(\Pi
T^{0,1}\IC^{\ell+1})$ and anticommuting fields  $\psi$, $\bar{\psi}$
are sections of the bundles $K\otimes \Phi^*(\Pi
T^{0,1}\IC^{\ell+1})$, $\bar{K}\otimes \Phi^*(\Pi
T^{1,0}\IC^{\ell+1})$. By $\Pi \CE$ we denote a vector bundle $\CE$
with a reversed parity of the fibres. The action \eqref{linsigmact}
is  $\delta_0$-invariant.

A gauged linear sigma model description of the  sigma model with the
target space $\IP^{\ell}$ is based on the representation of the
projective space $\IP^{\ell}$ as a Hamiltonian quotient of
$\IC^{\ell+1}$.  Let us
supply $\IC^{\ell+1}$ with a symplectic structure \be
\Omega=\frac{\imath}{2}\sum_{j=1}^{\ell+1}\,d\varphi^j\wedge
d\bar{\varphi}^j. \ee The following action of $U_1$:
\be\label{gsym} \varphi^j\longrightarrow e^{\imath
\alpha}\varphi^j,\qquad e^{\imath
  \alpha}\in U(1)
\ee is Hamiltonian and the corresponding momentum i.e. a solution
of the equation 
$$
\iota_v\Omega=d\mu
$$
is given by
\be
\mu(\varphi,\bar{\varphi})=-\frac{1}{2}\sum_{j=1}^{\ell+1}\,|\varphi^j|^2.
\ee
The projective space $\IP^{\ell}$ has the following
representation as the Hamiltonian quotient:
$$
\IP^\ell=\{\mu (\varphi,\bar{\varphi})+r^2/2=0\}/U(1).
$$
where the value  $r^2/2$ of the momentum $\mu$ defines a K\"{a}hler
class of $\IP^{\ell}$.

The interpretation of the projective space as a Hamiltonian
reduction allows to describe topological non-linear
two-dimensional sigma model with the target space $\IP^{\ell}$  in
terms of a linear sigma model  with the target space $\IC^{\ell+1}$
and gauged $U_1$-symmetry \eqref{gsym}. Topological
$U_1$ gauge theory can be constructed using the
following three  sets  of fields
$(A,\lambda,\sigma)$, $(b,\eta)$, $(\xi,H)$ where $A$ is a connection
in $U_1$-bundle,  $\psi$ is an odd one-form, $\sigma$, $b$, $H$ are
real even zero forms and $\xi$,$\eta$  are real odd zero forms. Define the
topological BRST transformations as follows
\be\label{equivmult}
\delta_{\CG}\,A= \lambda,\qquad \delta_{\CG}\,\lambda=-\imath
d\sigma, \qquad \delta_{\CG}\,\sigma=0, \ee
$$
\delta_{\CG}\,b=\eta,\qquad \delta_{\CG} \eta=0,\qquad
\delta_{\CG}\,\xi= H,\qquad \delta_{\CG} H=0.
$$
Note that
$\delta^2_{\CG}$ acts as a gauge transformation with the gauge
parameter  $\sigma$ e.g. \be \delta_{\CG}^2\,A= -\imath
d\sigma,\qquad \delta_{\CG}^2\,\lambda=0,\qquad
\delta_{\CG}^2\,\sigma=0. \ee
The space of functionals  of the
fields  $(A,\lambda,\sigma)$ supplied with the action of
$\delta_{\CG}$ can be considered as  a  model for the
$\CG$-equivariant
 de Rham complex on the space of connections on a Riemann  surface $\Sigma$.
In the following we consider the gauge multiplet
 $(A,\lambda,\sigma)$ interacting with the fields $(b,\eta)$,
 $(\xi,H)$ and with the fields entering the action
 \eqref{linsigmact}.
The gauge theory generalization $\delta_{\CG}$ of the BRST
transformation  \eqref{transformeqonea} acts as follows
\be\label{transformeqone}
\delta_{\CG} \varphi^j=\chi^j,\qquad \delta_{\CG} \chi^j=-\imath
\sigma \varphi^j,\qquad \delta_{\CG} \psi_{\zb}^j=F_{\zb}^j,\qquad
\delta_{\CG} F_{\zb}^j=-\imath \sigma \psi^j_{\zb}, \ee \be\nonumber
\delta_{\CG} \bar{\varphi}^j=\bar{\chi}^j,\qquad \delta_{\CG}
\bar{\chi}^j=\imath \sigma \bar{\varphi}^j, \qquad \delta_{\CG}
\bar{\psi}_z^j=\bar{F}_z^j,\qquad \delta_{\CG} \bar{F}_z^j=\imath
\sigma \,\bar{\psi}^j_{z}.
\ee
Let $*dz=\imath dz $ and $*d \bar{z}=-\imath d\bar{z}$ be the Hodge
star operators.  A topological  $U_1$-gauged linear
sigma model  is described  by the following action
 \be\label{actionS1}
S=\frac{1}{2\pi}\int_{\Sigma} d^2z\,\sqrt{h}\delta_{\CG} \CV
=\\ \frac{1}{2\pi e^2}\int_{\Sigma}\,\,d^2z\,\,\sqrt{h}h^{z\zb}(- HF_{z\zb}(A)
+\pr_{z}\lambda_{\zb} \,(\xi-\imath \eta)\, -
\,\pr_{\zb}\lambda_{z}\,(\xi+\imath \eta)\, - 2 b\,\pr_z\pr_{\zb}\sigma) \ee
$$
+\frac{1}{2\pi}\int_{\Sigma}\,d^2z\,\sqrt{h}h^{z\zb}
( t\bar{F}^j_zF^j_{\zb}+\imath \bar{F}_z^{j} (\pr_{\zb}-
A_{\zb})\varphi^j - \imath F_{\zb}^{j}(\pr_{z}-
A_{z})\bar{\varphi}^{j} + \imath t\psi_{\bar{z}}^j\sigma\bar{\psi}^j_z
$$
$$
-\imath\bar{\psi}_z^{j}(\pr_{\zb}\chi^j- A_{\zb}\chi^j)
+\imath\psi_{\zb}^{j}(\pr_{z}\bar{\chi}^{j}-
 A_{\zb}\bar{\chi}^{j})-
 \imath\psi_{\zb}^j\lambda_z \bar{\varphi}^{j}+\imath
\bar{\psi}_{z}^{j}\lambda_{\zb}
 \varphi^{j}+\eta \sum_{j=1}^{\ell+1}
(\varphi^j\bar{\chi}^j-\bar{\varphi}^j\chi^j)
$$
$$
+2b\sum_{j=1}^{\ell+1}
(\chi^j\bar{\chi}^j+\imath\sigma|\varphi^j|^2)+
\frac{\imath   H}{2}(\sum_{j=1}^{\ell+1}|\varphi|^2-r^2)-\frac{\imath\xi}{2}
(\sum_{j=1}^{\ell+1}\chi^j\bar{\varphi}^j+\bar{\chi}^j\varphi^j),
$$
where $F_{z\zb}(A)=\pr_zA_{\zb}-\pr_{\zb}A_z$  and
$$
\CV=\frac{1}{e^2}h^{z\zb}(-\xi F_{z\zb}(A)-\,\imath b\,
(\pr_z\lambda_{\bar{z}}+\pr_{\bar{z}}\lambda_z))
+\psi^j_{\zb}(\frac{t}{2}\bar{F}^j_z-\imath(\pr_z-A_z)\bar{\varphi}^{j})+
\bar{\psi}^{j}_{z}(\frac{t}{2}F^j_{\zb}+\imath(\pr_{\zb}-A_{\zb})\varphi^j)$$
$$+\frac{\imath\xi}{2} (\sum_{j=1}^{\ell+1}|\varphi|^2-r^2)+
b\sum_{j=1}^{\ell+1} (\varphi^j\bar{\chi}^j-\bar{\varphi}^j\chi^j).
$$
A relation of \eqref{actionS1} (for a generic coupling constant $e^2$)
 with the standard $\IP^{\ell}$ sigma model is illustrated in Appendix F.
Let us note that a variation of the coupling constants $e^2$ and $t$ leads to a
change of the action \eqref{actionS1} on a $\delta_{\CG}$-exact
term. Thus, according to the standard considerations, the
dependence  of  correlation functions of $\delta_{\CG}$-closed
operators  on $e^2$ and $t$ is given only by the boundary
contributions of the space of fields. In many cases these boundary
contributions are zero  and the correlation functions are $e^2$- and
$t$-independent. In the following we make  calculations at
 particular values of the coupling constants $e^2$ and $t$ and will
not consider the issue of the  coupling constant independence.

Now we consider $S^1\times U_{\ell+1}$-equivariant version of the
theory \eqref{actionS1} on a Riemann surface $\Sigma$ allowing
isometry $S^1$. The action of $U_{\ell+1}$ is induced from a
 linear action on the target space $\IC^{\ell+1}$ and the action of
 $S^1$ is induced by $S^1$-isometry of $\Sigma$.  To simplify
notations we consider equivariance with respect to a maximal  abelian
subgroup $U_1^{\ell+1}\subset U_{\ell+1}$ and will work  with
functions invariant with respect to the permutation group
$S_{\ell+1}$ (i.e. Weyl group of $U_{\ell+1}$). Let $\hbar$ and
$\sigma_j$, $j=1,\ldots, (\ell+1)$ be
 generators of abelian Lie algebras ${\rm Lie}(S^1)$ and
${\rm Lie}(U_1^{\ell+1})$. Correlation functions of $S^1\times
U_{\ell+1}$-equivariant extension of the topological field theory
\eqref{actionS1} now  take values in the space of functions of
$\hbar$ and $\sigma_j$, $j=1,\ldots (\ell+1)$ invariant with respect
to permutations  of $\sigma_j$.

To construct an  $S^1\times U_{\ell+1}$-equivariant version of type
$A$ topological linear sigma model define $S^1\times U_1^{\ell+1}$-equivariant
generalization of the BRST transformations  \eqref{transformeqone},
\eqref{equivmult} as follows:
 \be\nonumber \delta \varphi^j=\chi^j,\quad \delta
\chi^j=-(\imath (\sigma_j+\sigma)\varphi^j+
\hbar\,\CL_{v_0}\varphi^j),\quad \delta \psi^j=F^j,\quad
 \delta F^j=-(\imath(\sigma_j+\sigma)\psi^j+\hbar\,\CL_{v_0}\psi^j),
\ee \be\nonumber \delta \bar{\varphi}^j=\bar{\chi}^j,\quad \delta
\bar{\chi}^j=-(-\imath(\sigma_j+\sigma)
\bar{\varphi}^j+\hbar\,\CL_{v_0}\bar{\varphi}^j), \quad \delta
\bar{\psi}^j=\bar{F}^j,\quad \delta \bar{F}^j=-(-\imath
(\sigma_j+\sigma)\bar{\psi}^j+ \hbar\,\CL_{v_0}\bar{\psi}^j), \ee
\be\label{transformeq} \delta\,A= \lambda,\qquad
\delta\,\lambda=-\imath d\sigma+\hbar d(\iota_{v_0} A)+\hbar
\iota_{v_0}F(A), \qquad \delta \sigma=0, \ee \be\nonumber
\delta\,b=\eta,\qquad \delta \eta=\hbar \iota_{v_0}db, \qquad
\delta\,\xi= H,\qquad \delta H=\hbar \iota_{v_0}d\xi, \ee
where $v_0=\imath \left( z\frac{\pr}{\pr z}-\zb\frac{\pr}{\pr
\zb}\right)$ and $\CL_{v_0}=\iota_{v_0}d+d\iota_{v_0}$.
 Note  that the transformation rules \eqref{transformeq} are not explicitly
gauge-invariant (this is related with the fact that the gauge group
$\CG$ and $S^1$ do not  commute).

A $S^1\times U_1^{\ell+1}$-equivariant version of the topological gauged
linear sigma model \eqref{actionS1} (for $t=0$) is then given by
\be\nonumber S_{eq}=\frac{1}{2\pi}\int_{\Sigma} d^2z\,\sqrt{h}\delta_2 \CV= \ee
\be\label{actionS1eq}
=\frac{1}{2\pi e^2}\int_{\Sigma}\,\,d^2z\,\,\sqrt{h}h^{z\zb}
((- HF_{z\zb}(A)\,+\,
(\lambda\wedge d\xi)_{z\zb}\, +\, (\lambda\wedge *d\eta)_{z\zb}
\,-\imath\,  (\,d b\,\wedge*d\sigma )_{z\zb} -
\ee
$$\hbar b d*(\iota_{v_0} A+\iota_{v_0}F(A))_{z\zb})\, +\,
\frac{1}{2\pi}\int_{\Sigma}\,\,d^2z\,\,\sqrt{h}h^{z\zb}(\imath \bar{F}_z^{j} (\pr_{\zb}-
A_{\zb})\varphi^j- \imath F_{\zb}^{j}(\pr_{z}-
A_{z})\bar{\varphi}^{j}
$$
$$
-\imath\bar{\psi}_z^{j}(\pr_{\zb}\chi^j- A_{\zb}\chi^j)
+\imath\psi_{\zb}^{j}(\pr_{z}\bar{\chi}^{j}-
 A_{z}\bar{\chi}^{j})-
 \imath\psi_{\zb}^j\lambda_z \bar{\varphi}^{j}+\imath
\bar{\psi}_{z}^{j}\lambda_{\zb}
 \varphi^{j}+\eta \sum_{j=1}^{\ell+1}
(\varphi^j\bar{\chi}^j-\bar{\varphi}^j\chi^j)
$$
$$
+b\sum_{j=1}^{\ell+1}(2 \chi^j\bar{\chi}^j +2\imath({\sigma}+\sigma_j)
|\varphi^j|^2 + \hbar
\bar{\varphi}^j\iota_{v_0}(d-A){\varphi}^j -\hbar
\varphi^j\iota_{v_0}(d-A)\bar{\varphi}^j)+$$$$
\frac{\imath H}{2}(\sum_{j=1}^{\ell+1}|\varphi|^2-r^2)-
\frac{\imath\xi}{2}(\sum_{j=1}^{\ell+1}\chi^j\bar{\varphi}^j+
\bar{\chi}^j\varphi^j)),
$$
where $F_{z\zb}(A)=\pr_zA_{\zb}-\pr_{\zb}A_z$  and
$$
\CV=\frac{1}{e^2}h^{z\zb}(- \xi F_{z\zb}(A)+ b\,(d*\lambda)_{z\zb})
-\imath\psi^j_{\zb}(\pr_z-
A_z)\bar{\varphi}^{j}+\imath
\bar{\psi}^{j}_{z}(\pr_{\zb}-
A_{\zb})\varphi^j$$
$$+\frac{\imath\xi}{2} (\sum_{j=1}^{\ell+1}|\varphi|^2-r^2)+
b\sum_{j=1}^{\ell+1} (\varphi^j\bar{\chi}^j-\bar{\varphi}^j\chi^j).
$$

\subsection{Calculations in type A topological sigma model on  a disk}

Now  we calculate a particular correlation function in
a topological field theory with the action \eqref{actionS1eq} on a
disk $\Sigma=D$, $D=\{z\in \IC\,|\,|z|^2\leq 1\}$ with
the coupling constant $e^2\to 0$.  We chose a flat metric
$d^2s=|dz|^2$ on the  disk $D$ invariant with respect to the
standard  action of the rotation group $S^1$  generated by the
vector field
$$
v_0=\imath \left( z\frac{\pr}{\pr z}-\zb\frac{\pr}{\pr
\zb}\right),\qquad z=re^{\imath \theta}.
$$
 The following boundary conditions on $S^1=\pr D$ are imposed:
\be\label{bc1} A_\theta|_{S^1}=0, \qquad \qquad
\lambda_{\theta}|_{S^1}=0,\qquad \pr_{\theta}\sigma|_{S^1}=0, \ee
\be\label{bc2} \pr_{\theta}b|_{S^1}=0,\qquad \pr_{\theta}\xi|_{S^1}=0,
\qquad
\pr_{\theta}\eta|_{S^1}=0. \ee  We
also  consider a restricted  gauge group \be\label{restrggroup}
\CG_0=\{g\in {\rm Map}(D,U_1)|\,\,g|_{S^1}=const\}. \ee
It is easy to see that the deformed action \eqref{actionS1eq}
and the boundary conditions are compatible with the gauge symmetry $\CG_0$  and BRST
transformations \eqref{transformeq}. For example,  we have
$$
\delta\,A_{\theta}= \lambda_{\theta},\qquad
\delta\,\lambda_{\theta}=\pr_{\theta}\sigma+ \hbar
\pr_{\theta}A_{\theta},
$$
and restriction of the variations to  the boundary $S^1=\pr D$ gives
$$
\delta\,A_{\theta}|_{S^1}= \lambda_{\theta}|_{S^1}=0,\qquad
\delta\,\lambda_{\theta}|_{S^1}=\pr_{\theta}\sigma|_{S^1}+ \hbar
\pr_{\theta}A_{\theta}|_{S^1}=0.
$$
This verifies the compatibility of the boundary conditions \eqref{bc1}
and BRST transformations for the gauge multiplet.
 Similarly, under infinitesimal gauge transformations  we have
$$
\delta_{\alpha}A_{\theta}|_{S^1}=\pr_{\theta}\alpha|_{S^1}+A_{\theta}|_{S^1}=0,\qquad
\alpha\in {\rm Lie}(\CG_0),
$$
and thus the boundary conditions \eqref{bc1} are compatible with gauge
invariance with respect to group \eqref{restrggroup}. We define a
metric on the space of fields using the standard quadratic form on the
tensor fields on $D$.

\begin{prop} \label{Propone} A functional
integral over fields on the disk with the action \eqref{actionS1eq}
at $e^2\to 0$ has an asymptotic  given by a functional
integral with the effective action on the boundary $\pr D=S^1$ 

\be\label{Mainprop}
  S_{\rm eff}\,=\,\frac{\i\,H_0}{4\pi}\,\int\limits_0^{2\pi}\!d\theta
  \Big(\sum_{j=1}^{\ell+1}|\varphi^j(\theta)|^2-r^2\Big)\,
  -\,\frac{\i\xi_0}{4\pi}\int\limits_0^{2\pi}\!\!d\theta\,\,
  \sum_{j=1}^{\ell+1}\Big(\bar{\varphi}^j(\theta)\chi^j(\theta)
  +\varphi^j(\theta)\bar{\chi}^j(\theta)\Big)\\
  +\,\frac{\eta_0}{2\pi}\int\limits_0^{2\pi}\!\!d\theta\,\,
  \sum_{j=1}^{\ell+1}\Big(\varphi^j(\theta)\bar{\chi}^j(\theta)
  -\bar{\varphi}^j(\theta)\chi^j(\theta)\Big)\\
  +\,2 b_0\sum_{j=1}^{\ell+1}\frac{1}{2\pi}
  \int\limits_0^{2\pi}\!d\theta\Big[\chi^j(\theta)\bar{\chi}^j(\theta)\,
  +\,\i(\sigma_0+\sigma_j)|\varphi^j(\theta)|^2\,
  +\,\h\,\bar{\varphi}^j(\theta)\pr_{\theta}\varphi^j(\theta)\,\Big]\,.
 \ee
Here in the functional integral over $\varphi^j(z)$, $\chi^j(z)$ being
 even and odd holomorphic
functions on $D$, $\sigma_0$ and $b_0$ being even variables, $\xi_0$ and
$\eta_0$ being odd variables. The measure on the space of $\varphi^j(z)$ and
 $\chi^j(z)$  is induced from the standard pairing of functions on
 $D$ with the metric $d^2s=|dz|^2$.
\end{prop}

\noindent {\it Proof}. We will need  an asymptotic at $e^2\to 0$ of the
various integrals. Given a function $F(e,y)$ of the coupling
constant $e$ and a variable $y$ such that at $e^2\to 0$ the leading
asymptotic is given by $F(y,e)=Q(e)f_0(y)+\cdots$ we call $f_0(y)$
a leading term and denote it by $[F(e,y)]_0$.

Consider first an asymptotic of an  integral over $H$. We
have the following identity:
$$
\left(\int DH\,e^{\int_D(\frac{1}{e^2} HF(A)+\CF_1(H)}\right)_0=\int
  d H_0\,\delta(F(A))\,e^{\int_D \CF_1(H_0)},
$$
where $H_0$ is a constant mode of $H$ and
$\CF_1(H)$ is an  arbitrary $e^2$-independent function of $H$ such that
the left hand side of the identity is defined.
 Now let us find an asymptotic of the following integral
 \be\label{intone}
Z_{\CF_2}=\int Db \,D\sigma\, e^{\int_D (\frac{\imath}{e^2} db\wedge
*d\sigma +b\CF_2(\sigma))},
\ee
where  ${\CF}(\sigma)$ is a $e^2$-independent function
of $\sigma$. Consider a quadratic form
\be\label{qform} \<
f,f\>=\int_D d f\wedge *d f, 
\ee
 on the space of functions such that
$\pr_{\theta}f|_{S^1=\pr D}=0$. We have the following obvious exact
sequence
$$
0\rightarrow Fun_0(D)\rightarrow Fun(D) \rightarrow
Fun(S^1)\rightarrow 0,
$$
where $Fun(D)$ is the space of functions on the disk, $Fun_0(D)$ is the subspace
of functions taking zero values on the boundary $S^1=\pr D$ and
$Fun(S^1)$ is   a space of functions on the boundary.  Then the
orthogonal complement $Fun^{\vee}_0(D)$ to the space $Fun_0(D)$ with
respect to the  quadratic form \eqref{qform} is given by the space
of harmonic  functions $\Delta f=0$. Due to the  constraint
$\pr_{\theta}f|_{S^1=\pr D}=0$ the space  $Fun^{\vee}_0(D)$ consists
of constant functions. Taking into account  that the quadratic form
\eqref{qform} is non-degenerate on $Fun_0(D)$ we infer  that the
following asymptotic for \eqref{intone} holds
$$
(Z_{\CF_2})_0=\frac{1}{\det' (d*d)}\int db_0 \,d\sigma_0\, e^{\int_D
b_0\CF_2(\sigma_0))},
$$
where $b_0$ and $\sigma_0$ are constant modes of $b$ and $\sigma$.

By construction the functional integral with the action
\eqref{actionS1eq} is invariant with respect to the gauge group
$\CG_0$ defined  by  \eqref{restrggroup}.
 We fix the gauge symmetry using a gauge fixing condition $d*A=0$.
In the case of abelian gauge  group the introduction of Faddeev-Popov ghosts is not
necessary. Integration over
$H$ implies a constraint $F(A)=dA=0$. The gauge fixing condition for
the gauge group \eqref{restrggroup} implies that the residual gauge
symmetry group is a subgroup of constant gauge transformations.
Therefore one can take $A=0$ and the effective action functional is
given by
 \be
  S_{\rm eff}\,=\,\frac{1}{2\pi}\int\!d^2z\,\Big\{\,\frac{1}{e^2}\,\,\Big[\,
  (\lambda\wedge d\xi)_{z\zb} \,+\, (\lambda\wedge *d\eta)_{z\zb}\Big]\\
  +\sum_{j=1}^{\ell+1}\,\Big[\,
  -\imath F^j_{\zb}(\pr_z\bar{\varphi}^j)
  +\imath\,\bar{F}^j_z(\pr_{\zb}\varphi^j)
  -\,\imath\psi^j_{\zb}\lambda_z\bar{\varphi}^j\,
  +\,\imath\bar{\psi}^j_z\lambda_{\zb}\varphi^j\,
  +\,\imath\psi^j_{\zb}(\pr_z\bar{\chi}^j)
  -\,\imath\bar{\psi}^j_z(\pr_{\zb}\chi^j)\Big]\\
  +\,\frac{\imath H_0}{2}\Big(\sum_{j=1}^{\ell+1}|\varphi^j|^2\,-\,r^2\Big)\,
  -\,\frac{\imath \xi}{2}\sum_{j=1}^{\ell+1}\bigl(\varphi^j\bar{\chi}^j
  +\bar{\varphi}^j\chi^j\bigr)\,
  +\,\eta\sum_{j=1}^{\ell+1}
  (\varphi^j\bar{\chi}^j-\bar{\varphi}^j\chi^j)\\
  +\,b_0\sum_{j=1}^{\ell+1}\Big[2\chi^j\bar{\chi}^j\,
  +\,2\i(\sigma_0+\sigma_j)|\varphi^j|^2\,
  +\,\h\bigl(\bar{\varphi}^j\iota_{v_0}d\varphi^j
  -\varphi^j\iota_{v_0}d\bar{\varphi}^j\bigr)\Big]\,
  \Big\}\,.
 \ee
The integration over $F$ and $\bar{F}$ gives the constraints
$$
\apr\varphi^j=0,\qquad \pr \bar{\varphi}^j=0.
$$
The solutions of the constraints are given by holomorphic functions
$$
\varphi^j=\varphi^j(z), \qquad
\bar{\varphi}^j=\bar{\varphi}^j(\zb).
$$
From now on $\varphi^j$ will denote holomorphic functions.
Now consider an asymptotic of the integrals over odd variables
\be\label{fermint} 
\tilde{Z}_{\CF_3} =\int D\lambda\,D\xi\,D\eta\,
 e^{\int_D( -\frac{1}{e^2}\lambda \wedge d\xi
-\frac{1}{e^2}\lambda \wedge *d\eta +\lambda \CF_3(\eta,\xi))}, 
\ee
where $\CF_3(\eta,\xi)$ is independent of $e$. Let us combine two
zero forms $\xi$ and $\eta$ into a one form
$$
\rho=d\xi+*d\eta.
$$
Taking integral over $\lambda$ in \eqref{fermint} we obtain the
following asymptotic  in the  limit $e^2\to 0$:
\be\label{fermintone} \left(\int D\lambda\,D\xi\,  D\eta \,e^{\int_D
-\frac{1}{e^2}\lambda \wedge d\xi -\frac{1}{e^2}\lambda \wedge * d\eta
+\lambda B}\right)_0\,=\, \det{}'(d*d)\delta(\rho). \ee
We have
$$
 \pr_z(\xi+\imath \eta)=0,\qquad \pr_{\zb}(\xi-\imath \eta)=0\,,
$$
and thus solutions of the equation $\rho=0$ are given by holomorphic
functions $F(z)=\xi-\imath\eta$.
This implies that the functions $\xi,\eta$ are
harmonic. Taking into account the boundary conditions
$\pr_{\theta}\xi|_{S^1}=\pr_{\theta}\eta|_{S^1}=0$, the functional
integration over $\xi$ and  $\eta$ reduces to the integration over
constant modes $\xi_0$ and $\eta_0$.

Now integration over $\psi_z^{\jb}$ and $\psi_{\zb}^j$ provides constraints
$$
\pr_{\zb}\chi^j=0,\qquad \pr_z\bar{\chi}^{\jb}=0,
$$
solved by holomorphic functions $\chi^j(z)$:
$$
\chi^j=\chi^j(z),\qquad \bar{\chi}^j=\overline{\chi^j(z)}.
$$
From now on  $\chi^j$ will denote holomorphic functions.
Combining all ingredients together  we obtain the following functional
integral:
$$
 \int\,d\xi_0d\eta_0\,dH_0\,db_0d\sigma_0\,
 [D^2\varphi^j]\,[D^2\chi^j]\,
 \,\,e^{-S_{\rm eff}}
$$
with
\be\label{effAction}
  S_{\rm eff}\,=\,\frac{\i\,H_0}{4\pi}\,\int\limits_0^{2\pi}\!d\theta
  \Big(\sum_{j=1}^{\ell+1}|\varphi^j(\theta)|^2-r^2\Big)\,
  -\,\frac{\i\xi_0}{4\pi}\int\limits_0^{2\pi}\!\!d\theta\,\,
  \sum_{j=1}^{\ell+1}\Big(\bar{\varphi}^j(\theta)\chi^j(\theta)
  +\varphi^j(\theta)\bar{\chi}^j(\theta)\Big)\\
  +\,\frac{\eta_0}{2\pi}\int\limits_0^{2\pi}\!\!d\theta\,\,
  \sum_{j=1}^{\ell+1}\Big(\varphi^j(\theta)\bar{\chi}^j(\theta)
  -\bar{\varphi}^j(\theta)\chi^j(\theta)\Big)\\
  +\,2 b_0\sum_{j=1}^{\ell+1}\frac{1}{2\pi}
  \int\limits_0^{2\pi}\!d\theta\Big[\chi^j(\theta)\bar{\chi}^j(\theta)\,
  +\,\i(\sigma_0+\sigma_j)|\varphi^j(\theta)|^2\,
  +\,\h\,\bar{\varphi}^j(\theta)\pr_{\theta}\varphi^j(\theta)\,\Big]\,,
 \ee
where  $\varphi^j(\theta)$, $\chi^j(\theta)$  are restrictions on
$S^1=\pr D$ of
even and odd holomorphic functions  $\varphi^j(z)$, $\chi^j(z)$. Here we make a change
of variables
\be\label{chvar}
\varphi_n^j\rightarrow\frac{\varphi_n^j}{n+1} \qquad
\chi_n^j\rightarrow\frac{\chi_n^j}{n+1},
\ee
in the expansions of $\varphi^j(z)$ and $\chi^j(z)$
$$
\varphi^j(z)=\sum_{n=0}^{\infty}\varphi^j_nz^n,\qquad
\chi^j(z)=\sum_{n=0}^{\infty}\chi_n^j z^n.
$$
This change of variables converts the integrals of the product of the
holomorphic and antiholomorphic functions over the disk $D$ into integrals
over the boundary $S^1$. By the standard properties of the
canonical integration measure on super-manifold $\IR^{N|N}$  the
Jacobian of the change of variables \eqref{chvar}  is trivial.
The resulting functional integral with the action \eqref{effAction}
coincides with the one defined by \eqref{Mainprop}. $\Box$

\begin{te}\label{Amodel}  The functional integral in
$S^1\times U_{\ell+1}$-equivariant topological sigma model with the
deformed action  \eqref{Mainprop}  has the following
finite-dimensional integral representation \be\label{finalint}
(2\pi\hbar)^{\frac{\ell-1}{2}}\<e^{\frac{r^2}{2}\CO^{(0)}}\>=
\frac{1}{2\pi\hbar}\int_{\IR-\imath\epsilon}
d H_0\,e^{\imath H_0\frac{r^2}{2}} \prod_{j=1}^{\ell+1}
\Gamma_1\left(\imath H_0+\sigma_j|\hbar\right)\,,
\qquad \CO^{(0)}=\int_{S^1=\pr
D}d\theta \sigma(\theta), \ee
where $\epsilon >\max(-\sigma_j),\,j=1,\ldots,\ell+1$.
\end{te}

\proof  We prove \eqref{finalint} using the representation obtained
in Proposition \ref{Propone}. We shall calculate the following functional
integral
$$
 \frac{1}{(2\pi)^2}\int\,d\xi_0d\eta_0\,dH_0\,db_0d\sigma_0\,
 [D^2\varphi^j]\,[D^2\chi^j]\,
 \,\,e^{-S_{\rm eff} +\frac{r^2}{2}\sigma_0}
$$
Integrating over $b_0$ we obtain the delta-function factor
\be\label{constrP} \delta\left(\frac{1}{\pi}\int_0^{2\pi}d\theta
\left(\sum_{j=1}^{\ell+1}\imath\chi^j(\theta)\bar{\chi}^j(\theta)-
\sum_{j=1}^{\ell+1}(\sigma_0+\sigma_j)|\varphi^j(\theta)|^2+\imath
\hbar\bar{\varphi}^j(\theta)\pr_{\theta}\varphi^j(\theta)\right)\right).
 \ee
Further integration over $\sigma_0$ gives
$$
\sigma_0(\varphi,\chi)=\frac{\frac{1}{2\pi}\int_0^{2\pi}d\theta
  \left(\sum_{j=1}^{\ell+1}\imath\chi^j(\theta)
\bar{\chi}^j(\theta)- \sum_{j=1}^{\ell+1}\sigma_j|\varphi^j(\theta)|^2+
\imath\hbar\bar{\varphi}^j(\theta)\pr_{\theta}\varphi^j(z)\right)}
{ \frac{1}{2\pi}\int_0^{2\pi}d\theta
\sum_{j=1}^{\ell+1}|\varphi^j(\theta)|^2}\,\,.
$$
Taking into account that the integral over $H_0$ implies the
constraint
$$
\frac{1}{2\pi}\int_0^{2\pi}d\theta \sum_{j=1}^{\ell+1}|\varphi^j(\theta)|^2=r^2,
$$
we obtain
$$
\sigma_0(\varphi,\chi)=\frac{1}{2\pi r^2}\int_0^{2\pi}d\theta
  \left(\sum_{j=1}^{\ell+1}\imath\chi^j(\theta)
\bar{\chi}^j(\theta) -\sum_{j=1}^{\ell+1}\sigma_j|\varphi^j(z)|^2+\imath
\hbar\bar{\varphi}^j(\zb)\pr_{\theta}\varphi^j(z)\right)\,\,.
$$
After integration over $H_0$ the integrand reduces to
\be\nn
\frac{2\pi}{2r^2}
\delta\left(-\frac{1}{4\pi}\,\int_{0}^{2\pi}\,d\theta
\left(\sum_{j=1}^{\ell+1}|\varphi^j(\theta)|^2-r^2\right)\right)\times 
\ee
\be\label{BoundaryActionOne}
\exp\Big\{(\frac{\imath\xi_0}{2}+\eta_0)
\frac{1}{2\pi} \int_0^{2\pi}d\theta
 \sum_{j=1}^{\ell+1}\bar{\varphi}^j(\theta)\chi^j(\theta)
+(\frac{\imath \xi_0}{2}-\eta_0)\frac{1}{2\pi} \int_0^{2\pi}d\theta
 \sum_{j=1}^{\ell+1}\varphi^j(\theta)\bar{\chi}^j(\theta)+\\
\frac{1}{4\pi}\int_0^{2\pi}d\theta
  \left(\sum_{j=1}^{\ell+1}\imath\chi^j(\theta)
\bar{\chi}^j(\theta)- \sum_{j=1}^{\ell+1}\sigma_j|\varphi^j(\theta)|^2+\imath
\hbar\bar{\varphi}^j(\theta)\pr_{\theta}\varphi^j(\theta)\right)\Big\},
\ee
where  $\varphi(z)$, $\chi(z)$  are even and odd holomorphic functions.
We normalize the measure as follows:
\be D^2\varphi D^2\chi=\prod_{j=1}^{\ell+1}\frac{\imath}{2}D\varphi^j D\bar{\varphi}^j
\prod_{j=1}^{\ell+1}\frac{2}{\imath}D\chi^j D\bar{\chi}^j.
\ee
In particular
\be \int D^2\chi \exp\{\frac{\imath}{4\pi}
\int_{0}^{2\pi}d\theta\chi^j(\theta)\bar{\chi}^j(\theta)\}\,=\,1.
\ee
It is useful to reintroduce the variable $H_0$ by writing the first
delta-function in \eqref{BoundaryActionOne} in the integral form. Then
integrating over odd variables  $d\xi_0,\,d\eta_0$ and $d^2\chi$
we reduce the  functional integral to the following one
$$
\<e^{\frac{r^2}{2}\CO^{(0)}}\>=\int_{\IR-\imath\epsilon}  dH_0 D^2
\varphi 
e^{\frac{\imath r^2 H_0}{2}}
  \exp\left(-\frac{1}{4\pi}\int_0^{2\pi}
 d\theta\sum_j\left((\imath H_0+\sigma_j)|\varphi^j(\theta)|^2-
\imath\hbar\bar{\varphi}^j(\theta)
  \pr_{\theta}\varphi^j(\theta)\right)\right),
$$
where $\epsilon > max(-\sigma_j),\,j=1,\ldots,\ell+1$.
We can rewrite the  expression for  correlator in equivalent form
 $$ \<e^{\frac{r^2}{2}\CO^{(0)}}\>=\int_{\IR-\imath\epsilon}  dH_0 e^{\frac{\imath r^2
     H_0}{2}}Z_{LSM}(\imath H_0+ \sigma_j,\hbar),
 $$
where
\be\label{LSM}
Z_{LSM}(\sigma_j,\hbar)=\int [D^2\chi]\,[D^2\varphi] e^{-S_{LSM}(\sigma_j,\hbar)}
\ee
 is a functional integral with the action
$$
S_{LSM}(\sigma_j,\hbar)=\frac{1}{4\pi}\int_{0}^{2\pi} d\theta \left(\sum_{j=1}^{\ell+1}
\imath\chi^j(\theta)\bar{\chi}^j(\theta)- \sum_{j=1}^{\ell+1}\sigma_j|\varphi^j(\theta)|^2+
\imath\hbar\bar{\varphi}^j(\theta)\pr_{\theta}\varphi^j(\theta)\right).
$$
The functional integral \eqref{LSM}  is a correlation function  in $S^1\times
U_{\ell+1}$-equivariant  type $A$ topological linear sigma model on
$D$ with the target space $V=\IC^{\ell+1}$. This functional  integral was
calculated in \cite{GLO1} with the following result: 
\be\label{LSMa}
Z_{LSM}(\sigma_j,\hbar)=
\prod_{j=1}^{\ell+1}(2\pi\hbar)^{-\frac{1}{2}}\hbar^{\frac{\sigma_j}{\hbar}}\Gamma
\left(\frac{\sigma_j}{\hbar}\right)=
(2\pi\hbar)^{-\frac{\ell+1}{2}}\prod_{j=1}^{\ell+1}\Gamma_1(\sigma_j|\hbar),
\ee
provided
$Re (\imath H_0+\sigma_j)>0,\,\,j=1,\ldots,\ell+1$.

Using the expression \eqref{LSMa} for $Z_{LSM}$ 
and noting that the integration contour $(\IR-\imath\epsilon$ for 
$\epsilon > max( -\sigma_j),\,j=1,\ldots,\ell+1)$
is compatible with the condition  $Re (\imath
H_0+\sigma_j),\,\,j=1,\ldots,\ell+1$ 
 we obtain the integral representation for $\<e^{\frac{r^2}{2}\CO^{(0)}}\>$
\be\<e^{\frac{r^2}{2}\CO^{(0)}}\>\,=\,
(2\pi\hbar)^{-\frac{\ell+1}{2}}\int_{\IR-\imath\epsilon}
d H_0\,e^{\imath H_0\frac{r^2}{2}} \prod_{j=1}^{\ell+1}
\Gamma_1\left(\imath H_0+\sigma_j|\hbar\right).
\ee
This complete the proof of the theorem.
$\Box$

\begin{cor} The parabolic Whittaker function  \eqref{x=0one}  associated with
$\IP^{\ell}$ has an infinite-dimensional integral representation as
a correlation function in type A topological $S^1\times
U_{\ell+1}$-equivariant sigma model with the target space
$\IP^{\ell}$
$$
\Psi_{\lambda}(x)=(2\pi\hbar)^{\frac{\ell-1}{2}}\<e^{\frac{r^2}{2}\CO^{(0)}}\>,
$$
where $\sigma_j=-\lambda_j$ and $x=-\hbar r^2/2$. \end{cor}

\section{Mirror dual type B topological Landau-Ginzburg model}

In this Section we consider  a type $B$ topological Landau-Ginzburg
model which is a mirror dual to the topological type $A$ sigma model
with the target space $\IP^{\ell}$ considered in the previous
Section. We calculate a correlation function  in the Landau-Ginzburg
model which is mirror dual to the correlation function
\eqref{finalint}. This calculation naturally provides another
integral representation \eqref{x=0one} of $(1,\ell+1)$-Whittaker function
associated with $\IP^{\ell}$.

The mirror duals of the topological gauged linear sigma models were
constructed in \cite{HV}. In the following we apply this
construction to a particular  sigma model considered in the previous
Section. To provide a heuristic explanation of the construction
 let us recall that the topological gauge field multiplets  can be
obtained in a simple way from  $\CN=2$ SUSY gauge multiplets
described by twisted chiral superfields. Under mirror symmetry
 twisted chiral superfields are transformed into chiral multiplets.
Thus a mirror dual of the $U_1$-gauged type $A$ topological linear
sigma model with the target space $\IC^{\ell+1}$ should be described
by a mirror dual of a type $A$ topological linear sigma model with
the target space $\IC^{\ell+1}$ interacting  with an additional
topologically twisted chiral  multiplet $\Xi$. The mirror dual of
the type $A$ twisted linear sigma model was considered in \cite{HV}
(see also \cite{GLO2}) and is  described by a Landau-Ginzburg sigma
model. A coupling with the additional topological multiplet $\Xi$ can be guessed
from simple duality considerations in quadratic theories \cite{HV}.
Actually we are interested in the dual to $S^1\times
U_{\ell+1}$-equivariant  $U_1$-gauged type $A$ linear sigma-model.
It is useful to replace $U_{\ell+1}$-equivariance by
$U_1^{\ell+1}$-equivariance supplied with the overall invariance of
the correlation functions with respect to the permutation group
$S_{\ell+1}$ (the Weyl group of $U_{\ell+1}$). The resulting
Landau-Ginzburg theory, dual to $S^1\times U_1^{\ell+1}$-equivariant
$U_1$-gauged type $A$ linear sigma-model, has the following
superpotential:
\be\label{gaugedSP} W(\phi,\sigma)=-\imath\hbar\sigma
\frac{r^2}{2}+\sum_{j=1}^{\ell+1}
(-(\imath\sigma+\sigma_j)\phi^j+e^{\phi^j})+\hbar\log (2\pi\hbar), \ee
written in terms of the lowest components $\phi_j$, $j=1\ldots
(\ell+1)$ and $\sigma$  of chiral superfields. In \eqref{gaugedSP} 
$\sigma_j$, $j=1,\ldots, \ell+1$  are
$U_1^{\ell+1}$-equivariant parameters  and  $x=\frac{r^2}{2}$ is a parameter
of the K\"{a}hler structure of $\IP^{\ell}$ in the dual type $A$
model.  Note that usually one considers
the Landau-Ginzburg  theory  on $(\IC^*)^{\ell+2}$ as a mirror dual to
$U_1$-gauged linear sigma  model associated with non-linear
$\IP^{\ell}$ sigma model (this implies in particular that one uses  the variables
$\Phi^j=e^{\phi_j}$ as  the  correct description of the target space
of the theory). Taking into account that the mirror dual of $U_{\ell+1}$-equivariant
$\IP^{\ell}$ sigma model has superpotential \eqref{gaugedSP} which is
single-valued on the universal covering space $\IC^{\ell+2}$ we use
the coordinates $\phi_j$ below.

We would like to calculate a particular correlation function on a
disk $D$ in the Landau-Ginzburg topological theory with the
superpotential \eqref{gaugedSP}. Happily this calculation was
already done in \cite{GLO2} for an  arbitrary superpotential
$W(\phi)$. Let us recall briefly this derivation.  We consider a
set of fields with  the following $S^1$-equivariant BRST
transformations
 \be\label{linearS1} \delta_{S^1}
\phi_-^{i}=\eta^{i},\qquad \delta_{S^1} \eta^{i}=-\hbar
\iota_{v_0}d\phi_-^i, \qquad \delta_{S^1}\theta^{i}=G_-^{i},\qquad
\delta_{S^1} G_-^{i}=-\hbar \iota_{v_0}d\theta^{i}, \ee
$$
\delta_{S^1} \rho^i=-d\phi_+^i+\hbar \iota_{v_0}G_+^i, \qquad
\delta_{S^1}
 \phi_+^i=-\hbar \iota_{v_0}\rho^i,
 \qquad \delta_{S^1} G_+^i=d\,\rho^{i},
$$
where   $\phi_+$ and $\phi_-$ are even real zero-form valued fields,  
$\eta^i$ and $\theta^i$ are odd real  zero-form
valued fields, $\rho^i$ are odd real one-form valued  fields,
$G_-^i$ are even real zero-form valued fields  and $G_+^i$ are even
real two-form valued fields. The action of the topological sigma model is  given
by
\be\label{linearTFTRone} S=-\imath \sum_{j=1}^{N}\int_{D}\,
\left((d\phi_+^j-\hbar
\iota_{v_0}G_+^j)\wedge  *d\phi_-^{j}+\rho^j\wedge *d\eta^{j}
- \theta_jd\rho^j+ G_+^jG_-^{j}\right)
\ee
$$
+\sum_{i,j=1}^{N} \int_{D}d^2z \sqrt{h}\left(-\frac{\pr^2
W_-(\phi_-)}{\pr
    \phi_-^i\pr\phi_-^j}
\eta^{i}\theta^{j}-\imath \frac{\pr
W_-(\phi_-)}{\pr\phi_-^i}\,G_-^{i}\right)
$$
$$
+\sum_{i,j=1}^{N}\int_{D}\left(-\frac{1}{2}
  \frac{\pr^2 W_+(\phi_+)}{\pr\phi_+^i\pr \phi_+^j}\rho^i\wedge \rho^j+
\frac{\pr W_+(\phi_+)}{\pr \phi_+^i}\,G_+^i\right)
+\frac{1}{\hbar}\int_{S^1=\pr D}\,d\sigma W_+(\phi_+).
$$
Here $W_+$ and $W_-$ are arbitrary independent regular functions on
$\IR^{N}$. Thus defined action is $\delta_{S^1}$-closed. Below
we consider the case of $W_-(\phi_-)=0$ and $W_+(\phi_+)=W(\phi_+)$.
 Thus we have
\be\label{linearTFTR} S=-\imath \sum_{j=1}^{N}\int_{D}\, \left(
(d\phi_+^j-\hbar \iota_{v_0}G_+^j)\wedge *d\phi_-^{j}+\rho^j\wedge
*d\eta^{j} - \theta_jd\rho^j+ G_+^jG_-^{j}\right) \ee
$$
+\sum_{i,j=1}^{N}\int_{D}\left(-\frac{1}{2}
  \frac{\pr^2 W(\phi_+)}{\pr\phi_+^i\pr \phi_+^j}\rho^i\wedge \rho^j+
\frac{\pr W(\phi_+)}{\pr \phi_+^i}\,G_+^i\right)
+\frac{1}{\hbar}\int_{S^1=\pr D}\,d\sigma W(\phi_+).
$$
Given an observable $\CO$ (i.e. a functional of the fields),
its correlation function is defined as  the following functional integral
 \be\label{fint}
  \bigl\<\CO\,\bigr\>_{W}\,:=\,
  \int\!D\mu\,\,\CO\,e^{-S}\,,\\
  D\mu\,=\,\prod_{i=1}^{N}\,
  [D\phi_+^i][D\phi_-^i][D\eta^i][D\theta^i][D\rho^i][DG_+^i][DG_-^i]\,.
 \ee
We consider a local  $\delta_{S^1}$-invariant observable
 \be\label{soakup}
\CO=\,\prod_{i=1}^{N}\,
  \delta(\phi_-^i(z,\zb))\,\eta^i(z,\zb)|_{z=0},
 \ee
inserted at the center of the disk $D$ .
The calculation of the integral \eqref{fint} with
the observable \eqref{soakup} was done  in \cite{GLO2}.

\begin{te}\label{Bmodel} A correlation function of the observable
\eqref{soakup} in the type $B$ topological $S^1$-equivariant linear
sigma model \eqref{linearTFTR} is given by
\be\label{finalex}
\<\CO\>_{W}=\int_{\IR^{N}}\,\prod_{j=1}^{N}\,\,
dt^j\,\,\,e^{-\frac{1}{\hbar}\,W(t)},
\ee
where $t^j$ are the constant modes of the fields $\phi_+^j$
\end{te}
 Now we can  apply this general result  to
a particular case of the superpotential \eqref{gaugedSP}.

\begin{prop}\label{Bcor}
The correlation function of the observable \eqref{soakup} in the type
$B$ topological $S^1$-equivariant linear sigma model
\eqref{linearTFTR}    with the  superpotential
\be\label{superpot} W(\phi_+,\sigma)= -\imath\hbar\sigma \frac{r^2}{2}
+\sum_{j=1}^{\ell+1}(-(\imath\sigma+\sigma_j)\phi_+^j+e^{\phi^j_+}) + 
\hbar\log(2\pi\hbar), \ee is
given by \be\label{typeBint} \<\CO\>_{W}=\frac{1}{2\pi\hbar}
\int_{\IR-\imath\epsilon}\,\,d\sigma_0
\,e^{\imath\sigma_0\frac{r^2}{2}} \int_{\IR^{\ell+1}}\,\prod_{j=1}^{\ell+1}
dt^j\,\,\,\,e^{\frac{1}{\hbar}\sum_{j=1}^{\ell+1}
((\imath\sigma_0+\sigma_j)t^j-e^{t^j})},\,\,\,\,\epsilon>-\sigma_j,\,j=1,\ldots,\ell+1,
 \ee
where $\sigma_0$ and $t^j$ 
are the constant modes of the real fields $\sigma$ and $\phi_+^j$ respectively.
\end{prop}

The expression \eqref{typeBint} coincides with  \eqref{finalint} and is
equivalent to \eqref{x=0one}  \eqref{x=0two} obtained  in the Section 1. Thus
type $A$ and type $B$ topological sigma model representations related by
mirror symmetry
 give rise to two different integral representations
\eqref{x=0one} \eqref{x=0two} of the same parabolic Whittaker
function \eqref{identone}.

\section{Equivariant  symplectic volume and its mirror}

In this Section, following  \cite{GLO2},
 we provide a simple heuristic derivation of the correlation
function \eqref{finalint}. This calculation directly relates
type $A$  and type $B$ mirror dual
integral representations. First let us demonstrate that \eqref{x=0two}
can be understood as a limit of equivariant volumes of spaces
of holomorphic maps of $\IP^1\to \IP^{\ell}$ when a degree of the maps
goes to infinity.
The  compactified space of holomorphic maps $\IP^1\to \IP^{\ell}$ of a
degree $d$ can be identified with $\IP^{(\ell+1)(d+1)-1}$ as follows.
Holomorphic maps of $\IP^1$ into $\IP^{\ell}$ of degree $d$ can  be described as
a collection  of $\ell+1$
mutually prime polynomials of the  degree $d$
\be\label{polynom}
\varphi^j(z)=\sum_{m=0}^d\varphi^j_{m}\,z^m, \qquad j=1,\ldots ,\ell+1,
\ee
 modulo multiplication by rational functions
$$
\varphi^j(z)\longrightarrow \tilde{\varphi}^j(z)=g(z)\varphi^j(z),
$$
such that the resulting functions $\tilde{f}^j(z)$ are again
polynomials of the degree $d$. In the case of mutually prime
polynomials the function   $g(z)$ is  necessary constant and thus the
space $\CM_d(\IP^1,\IP^{\ell})$ of
holomorphic maps $\IP^1\to \IP^{\ell}$ is given by a projectivization of
the space of mutually prime $\ell+1$ polynomials. This space is
non-compact and its compactification $\CQ\CM_d(\IP^1,\IP^{\ell})$
obtained  by omitting condition to
be mutually prime  is a projectivization $\IP^{(\ell+1)(d+1)-1}$
of the   vector space of  $(\ell+1)$-tuples polynomials of the degree $d$. The space
 $\CQ\CM_d(\IP^1,\IP^{\ell})=\IP^{(\ell+1)(d+1)-1}$ can be obtained as
the  Hamiltonian reduction of the space of $(\ell+1)$-tuples polynomials of the degree
$d$  supplied with the symplectic structure
\be\label{initss}
\Omega=\frac{\imath}{2} \sum_{j=1}^{\ell+1}\sum_{m=0}^d
\delta \varphi^j_{m}\wedge \delta \bar{\varphi}^j_{m},
\ee
with respect to  a diagonal action of $U_1$ generated by the vector
field
$$
 v_{U_1}\,=\,\i\sum_{j=1}^{\ell+1}\sum_{m=0}^d\Big (
 \varphi^j_m\frac{\delta}
{\delta \varphi^j_m}\,
 -\,\bar{\varphi}^j_m\frac{\delta}{\delta\bar{\varphi}^j_m}\Big)\,.
$$
The action of $U_1$ is Hamiltonian and
the corresponding momentum (i.e. solution of the equation
$\iota_{v_{U_1}}\Omega=\delta\mu$)  is given by
$$
 \mu\,=\,-\frac{1}{2} \sum_{j=1}^{\ell+1}\sum_{m=0}^d|\varphi^j_m|^2\,
 =\,-\frac{1}{4\pi}\sum_{j=1}^{\ell+1}\int_0^{2\pi}d\theta\,|\varphi^j(\theta)|^2,
$$
Here  $\varphi^j(\theta)$ denotes a restriction of
the polynomial  $\varphi^j(z)$ on the circle $z=e^{\imath \theta}$,
$\theta\in [0,2\pi)$. The reduced space $\IP^{(\ell+1)(d+1)-1}$ is naturally
supplied with a symplectic structure $\Omega_{ind}$.

The symplectic space $\CQ\CM_d(\IP^1,\IP^{\ell})$ allows a Hamiltonian action
of the larger Lie group $S^1\times U_1^{\ell+1}$. In homogeneous coordinates
it is given by
$$
 \varphi^j_{m}\longrightarrow \varphi^j_{m}e^{\imath \alpha_j}\,e^{\imath
  m\beta},\qquad (e^{\imath \beta}, e^{\imath \alpha_1},\cdots ,e^{\imath
  \alpha_{\ell+1}})\in S^1\times U_1\times\cdots\times U_1\,.
$$
The action of the corresponding Lie algebra generators is realized  by
the  vector fields
 \be
  v_{S^1}\,=\,\i\sum_{j=1}^{\ell+1}\sum_{m=0}^d\,m\Big (
  \varphi^j_m\frac{\delta}{\delta \varphi^j_m}\,
  -\,\bar{\varphi}^j_m\frac{\delta}{\delta\bar{\varphi}^j_m}\Big )\,,\\
  v_{U(1)_j}\,=\,\i\sum_{m=0}^d\,\Big (
  \varphi^j_m\frac{\delta}{\delta \varphi^j_m}\,
  -\,\bar{\varphi}^j_m\frac{\delta}{\delta\bar{\varphi}^j_m}\Big )\,,
  \hspace{1.5cm}
  j=1,\ldots,\ell+1\,.
 \ee
The corresponding momenta are
$$
\mu_{S^1}\,=\,-\frac{1}{2}\sum_{j=1}^{\ell+1}\sum_{m=0}^d\,m |\varphi^j_m|^2\,
 =\,-\frac{1}{4\pi\imath}\sum_{j=1}^{\ell+1}\int_0^{2\pi}\!\!d\theta\,\,
 \bar{\varphi}^j(\theta)\pr_{\theta}\, \varphi^j(\theta)\,,
$$
$$
\mu_j\,=\,-\frac{1}{2}\sum_{m=0}^d\,|\varphi^j_m|^2\,
 =\,-\frac{1}{4\pi}\int_0^{2\pi}\!\!d\theta\,\,|\varphi^j(\theta)|^2.
$$
The $S^1\times U_1^{\ell+1}$-equivariant volume of
$\CQ\CM_d(\IP^1,\IP^{\ell})$ is defined as the following integral
\be\label{Eqvol}
Z_d=\int_{\CQ\CM_d(\IP^1,\IP^{\ell})} \,e^{ \Omega_{ind}^{S^1\times U_1^{\ell+1}}},
\ee
where $ \Omega_{ind}^{S^1\times U_1^{\ell+1}}$ is an $S^1\times
U_1^{\ell+1}$-equivariant extension of the  symplectic
structure $\Omega_{ind}$ on $\CQ\CM_d(\IP^1,\IP^{\ell})$
induced from \eqref{initss} by the Hamiltonian reduction.
Let us identify $H^*_{S^1\times U_1^{\ell+1}}({\rm pt})$ with $\IC[\hbar,
\sigma_1,\ldots, \sigma_{\ell+1}]$ where $\hbar$ corresponds to a
generator of $S^1$ and $\sigma_j$ to a generator of $j$-th factor in
$U_1^{\ell+1}$.

\begin{prop} \label{Eqform} The equivariant volume \eqref{Eqvol} has the following
  integral representation
$$
Z_d=\frac{1}{(2\pi)^2}\int\,
\frac{D^2\varphi\,D^2\chi\,DH_0\,D\xi_0\,D\eta_0\,D\lambda\,D\sigma_0\,}
{\rm Vol(U_1)} e^{\frac{r^2}{2}\phi}e^{-S},
$$
where
$$
S=
\imath H_0\frac{1}{4\pi}\int_0^{2\pi}
d\theta (\sum_{j=1}^{\ell+1}|\varphi^j(\theta)|^2-r^2)-
\frac{\imath \xi_0}{4\pi} \sum_{j=1}^{\ell+1} \int_{0}^{2\pi} d\theta
(\varphi^j(\theta)\bar{\chi}^j(\theta)+
\bar{\varphi}^j(\theta)\chi^j(\theta))
+\frac{\eta_0}{2\pi} \sum_{j=1}^{\ell+1} \int_{0}^{2\pi} d\theta
(\varphi^j(\theta)\bar{\chi}^j(\theta)
$$
$$
-\bar{\varphi}^j(\theta)\chi^j(\theta))+2
 \lambda \sum_{j=1}^{\ell+1}\frac{1}{2\pi} \int_{0}^{2\pi}( \imath(\sigma_0+\sigma_j)
|\varphi^j(\theta)|^2+\hbar \bar{\varphi}^j(\theta)\pr_{\theta}\varphi^j(\theta))+
 \chi^j(\theta)\bar{\chi}^j(\theta))d\theta,
$$
where $\xi_0$, $\eta_0$ and $\chi^j$, $j=1,\ldots, (\ell+1)$ are considered
to be  Grassmann variables.  Also the  functions $\varphi^j(\theta)$ and
$\chi^j(\theta)$ are restricted to be  degree $d$ polynomials of
$z=e^{\imath \theta}$ with the integration measure
 given by
$$
D^2\varphi=\prod_{j=1}^{\ell+1}\prod_{m=0}^{d}\frac{\imath}{2}d\varphi_m^j\wedge
d\bar{\varphi}_m^j,\,\,\,\,
D^2\chi=\prod_{j=1}^{\ell+1}
\prod_{m=0}^{d}\frac{2}{\imath}d\chi_m^j\wedge d\bar{\chi}_m^j.
$$
\end{prop}

\noindent {\it Proof}. The proof reduces to application of the
standard technique (see e.g. \cite {CMR}) and is given in Appendix F. $\Box$

\begin{prop} The equivariant symplectic volume \eqref{Eqvol} has
  the following integral representation
\be\label{inrepone}
Z_d\,=\,\left(2\pi\right)^{(\ell+1)(d+1)-2}\int_{\IR-\imath\epsilon}
dH_0 \frac{e^{\frac{\imath r^2}{2}H_0}}
{\prod_{j=1}^{\ell+1}\prod_{m=0}^{d}\left(\imath H_0+\sigma_j+\hbar
    m\right)},
\ee
where $\epsilon>max(-\sigma_j),\,\,\,j=1\ldots,\ell+1$.
\end{prop}

\noindent {\it Proof}. Using Proposition \ref{Eqform} we have
$$
-S+\frac{r^2}{2}\sigma_0=\frac{-\imath
H_0}{2}\sum_{j=1}^{\ell+1}(\sum_{m=0}^{d}\varphi_m^j\bar{\varphi}^j_m-
\frac{1}{2}r^2)
+\frac{\imath
  \xi_0}{2}\sum_{j=1}^{\ell+1}\sum_{m=0}^{d}\left(\varphi^j_m\bar{\chi}^j_m +
 \bar{\varphi}^j_m\chi_m^{j}\right)$$$$
-\eta_0\sum_{j=1}^{\ell+1}\sum_{m=0}^{d}
\left(\varphi^j_m\bar{\chi}^j-\bar{\varphi}^j_m\chi^j_m\right)
 -2\lambda
\left(\imath\sum_{j=1}^{\ell+1}\sum_{m=0}^{d}(\sigma_0+\sigma_j +\hbar
  m)|\varphi_m^j|^2+ \chi_m^j\bar{\chi}^j_m\right)+\frac{r^2}{2}\sigma_0.
$$
Applying  Proposition \ref{projvol}  in Appendix F
to $\CQ\CM_d(\IP^1,\IP^{\ell})=\IP^{(\ell+1)(d+1)-1}$ with the
action of $S^1\times U_1^{\ell+1}$ we obtain
$$Z_d\,=\,
\left(2\pi\right)^{(\ell+1)(d+1)-2}\int_{\IR-\imath \epsilon} 
dH_0\frac{e^{\frac{\imath r^2}{2}H_0}}
{\prod_{j=1}^{\ell+1}\prod_{m=0}^{d}\left(\imath H_0+\sigma_j+\hbar m\right)},
$$
where
$\epsilon>\max(-\sigma_j),\,\,\, j=1,\ldots,\ell+1$.
 $\Box$

Now let us consider the limit $d\to\infty$ of \eqref{inrepone}. We
use $\zeta$-function regularization to define infinite products 
(see e.g. \cite{Vor} and Appendix in \cite{GLO1}).
More precisely, define logarithm of the regularized infinite product
as a derivative of the zeta-function
$$
\ln \Big[\prod_{n=0}^{\infty} \,(\rho n+\lambda)\Big]_{reg}:=
-\pr_s \zeta_\rho(s,\lambda)|_{s=0},
$$
where $\zeta_\rho(s,\lambda)$ is an analytic continuation of the
infinite  sum
$$
\zeta_\rho(s,\lambda)=\sum_{n=0}^{\infty}\frac{1}{(\rho n+\lambda)^s},\qquad
  -\pi<{\rm arg}(\rho n+\lambda)\leq \pi,\quad {\rm Re}(s)>1.
$$
We have
$$
\zeta_\rho(0,\lambda)=\frac{1}{2}-\frac{\lambda}{\rho},\qquad
\pr_s\zeta_\rho(0,\lambda)=-\left(\frac{1}{2}-\frac{\lambda}{\rho}\right)\ln \rho+
\ln \frac{1}{\sqrt{2\pi}}\Gamma\left(\frac{\lambda}{\rho}\right),
$$
and thus  for the regularized infinite product  we obtain
\be\label{regOne}
\Big[\prod_{n=0}^{\infty} \,(\rho n+\lambda)\Big]_{reg}=\rho ^{1/2-\lambda/\rho}
\frac{(2\pi)^{1/2}}{\Gamma(\lambda/\rho)}.
\ee
Taking $\lambda=\imath H_0+\sigma_j$ and $\rho=\hbar$ we have 
\be
\Big[\frac{1}{\prod_{n=0}^{\infty} \,(\imath H_0+\sigma_j+ \hbar n)}\Big]_{reg}\,=\,
\frac{1}{\sqrt{2\pi\hbar}}\hbar ^{(\imath H_0+\sigma_j)/\hbar}
\Gamma(\frac{\imath H_0+\sigma_j}{\hbar}).
\ee
Now applying this regularization to the products in \eqref{inrepone}
for $d\to +\infty$ we obtain
$$
\lim_{d\rightarrow\infty}\frac{(2\pi\hbar)^{(\ell-1)/2}}
{(2\pi)^{(\ell+1)(d+1)-2}}Z_d\,=\,\frac{1}{2\pi\hbar}
\int_{\IR-\imath\epsilon}dH_0 e^{\frac{\imath r^2}{2} H_0}
\prod_{j=1}^{\ell+1}\hbar^{(\imath H_0+\sigma_j)/\hbar}
\Gamma\left(\frac{\imath H_0+\sigma_j}{\hbar}\right),
$$
where we imply that $-\pi<\arg(\imath H+\sigma_j+ \hbar n)\leq n\pi$
and $\epsilon>max(-\sigma_j),\,\,\,j=1,\ldots,\ell+1$. 
 Thus taking in appropriate way the
 limit $d\to \infty$ we recover the  integral representation
\eqref{finalint} for the functional integral
in topological field theory  on the disk  $D$.
 Note that the interpretation of  the limit $d\to \infty$ as a
replacement of $\IP^1$ by $D$ seems quite natural as in the non-compact
case there is no  notion of a finite  degree map.

To relate the type $A$ model calculations given above with the
integral representation \eqref{typeBint} arising in type $B$ model we
follow the strategy used in \cite{GLO2} i.e. we calculate
the equivariant  volume of the holomorphic maps
using the Duistermaat-Heckman formula \cite{DH}.
We would like to calculate the integral over the infinite-dimensional projective space
$\IP\CM(D,\IC^{\ell+1})$  assuming $\hbar>0$ 
and $\sigma_j>0,\,\,\,j=1,\ldots,\ell+1$. Let 
\be\label{infvolg}
Z(\hbar,\sigma)\,=
\,\int_{\IP\CM(D,\IC^{\ell+1})}e^{\hbar
 \tilde{\mu}_{S^1}+\sum_{j=1}^{\ell+1} \sigma_j\tilde{\mu}_j+\Omega(x)},
\ee where $\Omega$ is a symplectic form on
$\IP\CM(D,\IC^{\ell+1})$ defined by the Hamiltonian reduction
with momentum $x$, $\tilde{\mu}_{S^1}$ is a  momentum corresponding
to the $S^1$-action on $\IP\CM(D,\IC^{\ell+1})$ and $\tilde{\mu}_j$,
$j=1,\ldots, \ell+1$ are  momentum corresponding
to the action of $U_1^{\ell+1}$ on $\IP\CM(D,\IC^{\ell+1})$.
Note that the integral in  \eqref{infvolg} is an infinite-dimensional one  and
thus requires a proper regularization. One can formally apply the
Corollary \ref{usecor} Appendix E to rewrite the integral as follows
\be\label{intrepaux}
Z(\hbar,\sigma)\,=\,\frac{1}{2\pi}
\int dx_1\cdots dx_{\ell+1}\,\delta(x-\sum_{j=1}^{\ell+1}x_j)\,\,
\prod_{j=1}^{\ell+1}2\pi \int_{\IP\CM(D,\IC)}\,\,e^{ x_j\omega_{\IP\CM(D,\IC)}+
\sigma_j \mu_j^{\IP\CM(D,\IC)}+\hbar\tilde{\mu}_{S^1}}\, .
\ee
Thus the calculation reduces to the calculation of the following
integral
\be \label{infvol}
Z_1(\hbar,\sigma)\,=\,2\pi
\,\int_{\IP\CM(D,\IC)}e^{\hbar
 \tilde{\mu}_{S^1}+\sigma_j \tilde{\mu}_j+x_j\omega_{\IP\CM(D,\IC)}}\,=\,
2\pi\,e^{- \sigma_j x_j}\,\int_{\IP\CM(D,\IC)}e^{\hbar
 \tilde{\mu}_{S^1}+x_j\omega_{\IP\CM(D,\IC)}}.
\ee
This integral was already calculated in \cite{GLO2}. Below, for
completeness,  we recall the main steps of the calculation.
To calculate the integral \eqref{infvol} we use
an infinite-dimensional version of the  Duistermaat-Heckman formula
\cite{DH} (for a detailed introduction into the subject see e.g. \cite{Au}).
 Let $M$ be a $2N$-dimensional  symplectic manifold with
the  Hamiltonian action of  $S^1$ having only isolated fixed points.
Let $\mu$ be the corresponding momentum.
The tangent space $T_{p_k}M$ to a fixed point $p_k\in M^{S^1}$ has a
natural action of $S^1$. Let $v$  be  a generator
of ${\rm Lie} (S^1)$ and let $\hat{v}$ be  its action on
$T_{p_k}M$. Then the following identity holds:
 \be\label{DH}
  \int_M\,e^{\hbar \mu+\omega}=\sum_{p_k\in M^{S^1}}\,
  \frac{e^{\hbar \mu(p_k)}}{\det_{T_{p_k}M}\hbar\hat{v}/2\pi}\,\,.
 \ee
Let us formally apply \eqref{DH} to the integral \eqref{infvol}.
A set of fixed points of $S^1$ acting on $\IP\CM(D,\IC)$   can be  easily
found using linear coordinates on $\CM(D,\IC)$ (considered as
homogeneous coordinates on $\IP\CM(D,\IC)$). Let $\varphi(z)$ be a
holomorphic map of $D$ to $\IC$. It represents an $S^1$-fixed point
on $\IP\CM(D,\IC)$ if rotations by $S^1$ can be
compensated  by an action of  $U_1$
\be\label{fpcond}
e^{\imath \alpha(\beta)}\varphi(e^{\imath \beta}z)=\varphi(z), \qquad
\beta\in [0,2\pi].
\ee
It is easy to see that solutions of \eqref{fpcond}
are enumerated by non-negative integers and are given by
\be\label{fixedpt}
 \varphi^{(n)}(z)=\varphi_n z^n,\qquad \varphi_n\in \IC^*\, \quad n\in \IZ_{\geq 0}.
\ee
The tangent space to $\CM(D,\IC)$ at an  $S^1$-fixed point $\varphi^{(n)}$
has natural linear coordinates $\varphi_m/\varphi_n$, $m\in \IZ_{\geq 0},
m\neq  n$ where coordinates $\varphi_k$, $k\in \IZ_{\geq 0}$  are defined by the series
expansion  of $\varphi\in \CM(D,\IC)$
$$
\varphi(z)=\sum_{k=0}^\infty\varphi_kz^k.
$$
After identification of $\hbar$ in \eqref{infvol} with a
generator of ${\rm Lie}(S^1)$ its action on  the tangent space
at the fixed point
is given by a  multiplication of each  $\varphi_m/\varphi_n$  on $(m-n)$.
Thus to define an analog of the  denominator in the right hand side of
the  Duistermaat-Heckman formula \eqref{DH} one should provide a
meaning to the infinite product $\prod_{m=0,m\neq n}^{\infty}\hbar(m-n)/2\pi$.
We use a $\zeta$-function regularization
 \be\label{regprod}
  \ln \Big[\prod_{{m\in\IZ_{\geq 0},},m\neq n}\,
  \frac{\hbar}{2\pi}(m-n)\Big]_a\,:=\,-\frac{\pr}{\pr s}\left.\left(
  \sum_{m=1}^n\frac{e^{-\imath \pi s}}{(a\hbar m/2\pi )^s}
  +\sum_{m=1}^{\infty}\frac{1}{(a \hbar m/2\pi )^s}\right)\right|_{s\to0},
\ee
where $a$ is a normalization multiplier.
The introduction of $a$ is to take into account a multiplicative
anomaly $\det (A B)\neq \det A\cdot \det B$ appearing for generic
operators $A$ and $B$. We specify $a$ at the final step of the calculation of
\eqref{infvol}.

\begin{lem} The regularized product \eqref{regprod} is given by
 \be\label{prodd}
  \frac{1}{\left[\prod_{m\in\IZ_{\geq 0}, m\neq n}\hbar (m-n)/2\pi \right]_a}
  =(-1)^n\frac{(a \hbar/2\pi )^{-n}}{n!}\frac{\sqrt{a\hbar}}{2\pi}.
 \ee
\end{lem}

{\it Proof}.   Using the Riemann $\zeta$-function
$$
\zeta(s)=\sum_{n=1}^{\infty}\frac{1}{n^s},
$$
one can express the right hand side of \eqref{regprod}  as follows:
$$
 \ln \Big[\prod_{{m\in\IZ_{\geq 0},},m\neq n}\,
 \frac{\hbar}{2\pi}(m-n)\Big]_a=
 (\zeta(0)+n)\ln a \hbar/2\pi  +\ln n!-\zeta'(0)+\imath \pi n.
$$
Taking into account  $\zeta(0)=-\frac{1}{2}$ and
$\zeta(0)'=-\frac{1}{2}\ln 2\pi$ we obtain \eqref{prodd}. $\Box$

Let us now calculate the difference of the values of  $S^1$-momentum map
$\tilde{\mu}_{S^1}$ at two $S^1$-fixed points $\varphi^{(n)}, \varphi^{(0)}\in
  \IP\CM(D,\IC)$. Consider an embedded  projective line $\IP^1\subset\IP\CM(D,\IC)$,
containing $\varphi^{(n)}$ and $\varphi^{(0)}$. Let us choose homogeneous
coordinates $[z_0:z_1]$ on $\IP^1$ such that $\varphi^{(0)}=[1:0]$ and
$\vp^{(n)}=[0:1]$. The action of $S^1$ on $\IP\CM(D,\IC)$
descends to the embedded  $\IP^1$ via the  vector field
\be\label{descentvec}
 V\,=\,\i n\,\Big\{w\frac{\pr}{\pr
   w}\,-\,\wb\frac{\pr}{\pr\wb}\Big\}\,,\qquad w=z_1/z_0.
\ee
The pull back of the symplectic form $\Omega(t)$ is
given by
$$
 \omega_{\IP^1}\,=\,\imath t\frac{dw\wedge d\wb}{(1+|w|^2)^2}\,.
$$
The action of the vector field \eqref{descentvec} on $\IP^1$  is the
Hamiltonian one. Let
$\mu^{(n)}_{S^1}$ be the corresponding momentum   given by a
restriction of the momentum $\tilde{\mu}_{S^1}$ for $S^1$-action
$\IP\CM(D,\IC)$. From the definition of the momentum map  we have
 \be
  \mu^{(n)}_{S^1}(\vp^{(n)})\,-\,\mu^{(n)}_{S^1}(\vp^{(0)})\,
  =\,\int_{[1:0]}^{[0:1]}d\mu^{(n)}_{S^1}\,
  =\,\int_{[1:0]}^{[0:1]}\iota_{V}\omega_{\IP^1}\,.
 \ee
A momentum defined as a solution of the equation $i_V\omega=d\mu$ is
unique up an   additive constant. To fix this constant
we normalize the momentum $\tilde{\mu}_{S^1}(\varphi)$ so that
$\mu_{S^1}(\varphi^{(0)})=0$. Thus  we obtain the following:
 \be\label{CriticalValues}
  \mu^{(n)}_{S^1}(\vp^{(n)})\, =\,-nt\int_{[1:0]}^{[0:1]}
  \frac{wd\wb+\wb dw}{(1+|w|^2)^2}\, =\,
nt\Big[\frac{1}{(1+|w|^2)}\Big]_0^{\infty}\,
  =\,-nt\,.
\ee
Substituting \eqref{CriticalValues} and \eqref{prodd}
into \eqref{DH} for $M=\IP\CM(D,\IC)$ we obtain
 \be
 2\pi\,\int_{\IP\CM(D,\IC)}e^{\hbar
 \tilde{\mu}_{S^1}+t\omega_{\IP\CM(D,\IC)}} \,=\,2\pi\sqrt{\frac{a \hbar}{(2\pi)^2}}\,
  \,\sum_{n=0}^{\infty}\,(-1)^n\,\frac{e^{-nt\h}}{(a \hbar/2\pi )^n\,n!}\,\\
  =\,\sqrt{a\hbar}\,
  \exp\Big\{-\frac{2\pi }{a \hbar}\,e^{-\h t}\Big\}\,,
 \ee
where the dependence on the normalization constant $a$ reflects an ambiguity of the
regularized infinite-dimensional integral.

Taking into account \eqref{intrepaux},  the regularized $S^1\times U_{\ell+1}$-equivariant
symplectic  volume of the projectivization of the space of holomorphic maps of $D$
into $\IC^{\ell+1}$  can be written as follows:
$$
Z_{reg}\,=\,\frac{1}{2\pi}
\int_{0}^{\infty}\ldots\int_{0}^{\infty} dx_1\cdots dx_{\ell+1}\,
\delta(x-\sum_{j=1}^{\ell+1}x_j)\,\,
\prod_{j=1}^{\ell+1}2\pi e^{-\sigma_j x_j}\int_{\IP\CM(D,\IC)}\,\,e^{ x_j\omega_{\IP\CM(D,\IC)}
+\hbar\tilde{\mu}_{S^1}}\,=\,$$
$$
(a\hbar)^{\frac{\ell+1}{2}}\int_{\IR-\imath\epsilon}e^{\imath H_0 x}
\prod_{j=1}^{\ell+1}\int_{0}^{\infty} 
dx_j e^{-(\imath H_0+\sigma_j)x_j-\frac{2\pi}{a\hbar}e^{-\hbar x_j}}.
$$
Changing the variables 
$$
u_j=-\hbar x_j-\ln\left(\frac{a\hbar}{2\pi}\right),\,\,\,\,
 u=x+\frac{1}{\hbar}\ln\left(\frac{a\hbar}{2\pi}\right)$$  we obtain
$$
Z_{reg}\,=\,\frac{1}{4\pi^2} \left(\frac{a}{\hbar}
\right)^{\frac{\ell+1}{2}}\left(\frac{a}{2\pi}\right)^{\frac{1}{\hbar}\sum\sigma_j}
\int_{\IR-\imath\epsilon} d H_0 \,
e^{\imath H_0 u}\prod_{j=1}^{\ell+1}\hbar^{\frac{\imath H_0+\sigma_j}{\hbar}}
\int_{-\infty}^{-\frac{1}{\hbar}\ln\left(\frac{a\hbar}{2\pi}\right)}du_j
e^{\frac{1}{\hbar}(\imath H_0+\sigma_j) u_j-e^{u_j}}.
$$
To get rid of the renormalization ambiguity we take 
the limit $a\rightarrow 0$ 
$$
Z(\sigma,\hbar)\,=\,\lim_{a\rightarrow 0}\frac{2\pi}{\hbar}
 \left(\frac{\hbar}{a}\right)^{\frac{\ell+1}{2}}\left(\frac{2\pi}{a}
\right)^{\frac{1}{\hbar}\sum\sigma_j}Z_{reg}\,=
$$
$$\frac{1}{2\pi\hbar}\int_{\IR-\imath\epsilon} d H_0 \,
e^{\imath H_0 u}\prod_{j=1}^{\ell+1}\Gamma_1(\imath
H_0+\sigma_j|\hbar),\qquad \epsilon>{\rm max}(-\sigma_j).
$$
Integrating over $H_0$ we obtain precisely the mirror dual
integral representation \eqref{x=0one}
for the  Whittaker function associated with $\IP^{\ell}$
$$\int_{\CC}
  \prod_{j=1}^{\ell}dt_j\,\, \exp \left(-\frac{1}{\h}\left(\sum_{j=1}^{\ell}\lambda_j
  t_j +\lambda_{\ell+1}(x-\sum_{j=1}^{\ell}t_j)+
  \sum_{j=1}^{\ell}e^{t_j}+
  e^{x-\sum_{j=1}^{\ell}t_j}\right)\right),$$
  where $\sigma_j=\lambda_j,\,\,\,\,x=-\hbar u$.

\section{Conclusions}

To conclude  we briefly outline  some directions for
future research. The most obvious one is a generalization to the
case of an arbitrary partial flag manifold $G/P$. Recall that
partial flag spaces allow a description in terms of the non-abelian
Hamiltonian reduction of a symplectic vector spaces and thus the corresponding
type $A$ sigma models can be represented as  gauged  linear sigma
models (of quiver type) with non-abelian gauge groups.
For  $G/P={\rm Gr}(m,\ell+m)$ the corresponding
parabolic $\mathfrak{gl}_{\ell+1}$-Whittaker function \eqref{parTod}
has the following integral representation generalizing \eqref{x=0two}
(see Appendix D for $m=2$, $\ell=1$):
\be\label{partW}
 \Psi_{\lambda_1,\ldots,
\lambda_{j+1}}(x)=\int_{\CC} d\gamma_1,\cdots
d\gamma_m\,e^{-\frac{x}{\hbar}(\gamma_1+\gamma_2+\cdots +\gamma_m)}
\frac{\prod_{j=1}^m\prod_{a=1}^{\ell+1}
\Gamma_1({\gamma_j-\lambda_a}|{\hbar}
)} {\prod_{i,j=1,\,\,i\neq
j}^{m}\Gamma_1({\gamma_i-\gamma_j}|{\hbar})}.
\ee
A detailed discussion of the (topological) gauged linear sigma
models describing  holomorphic maps into ${\rm   Gr}(m,\ell+m)$
 can be found in \cite{W3}, \cite{MP}.
For a mirror description of the corresponding
gauged linear sigma models see  \cite{HV}.
We expect that ${\rm Gr}(m,\ell+m)$ analogs of the correlation functions
of the topological quantum field theories considered in the previous
Sections are given by the integral expressions \eqref{partW}. Note
also that the Givental type integral representation for Whittaker
functions associated with classical groups was constructed in 
\cite{GLO}. This  provides  a Landau-Ginzburg model description 
of the mirror dual to a type $A$ topological sigma models on the flag
manifolds associated with the classical groups. We are going to
explicitly derive this mirror duality following the line of this note
elsewhere.

Another direction to pursue  is a  higher-dimensional generalizations
of the constructions proposed in this note. One of the motivations
is  a higher-dimensional generalization of the connection between
Archimedean Langlands correspondence and mirror symmetry.
This includes in particular an instanton counting in higher
dimensions.  Note also that the higher dimensional examples
considered in \cite{GLO1} provide also additional insights into the
conjectured relation between local Archimedean Langlands
correspondence and the  mirror symmetry  in two dimensions. We are
going to discuss various higher-dimensional generalization of the
results of \cite{GLO1}, \cite{GLO2} and of this note elsewhere.

\section*{Appendix A: Proof of Proposition \ref{propone}}

We start with the following auxiliary result.

 \begin{lem}\label{Auxlem} The adjoint action of the group element
\eqref{GrGroupElement} on the algebra $\frak{b}_+=\<E_{ij},\,i\leq
j\>$ is given by the following:
 \be\label{mBlockCommutators}
  g^{-1}E_{11}g\,=\,E_{11}\,-\,\sum_{k=2}^mx_kE_{1,\,k}\,,\\
  g^{-1}E_{kk}g\,=\,E_{kk}\,+\,x_kE_{1,\,k}\,,
  \hspace{2.5cm}
  2\leq k\leq m\,,\\
  g^{-1}E_{k,\,k+j}g\,=\,E_{k,k+j}\,+\,x_kE_{1,\,k+j}\,,
  \hspace{1.5cm}
  2\leq k\leq m-1\,,
  \quad
  1\leq j\leq m-k\,;
 \ee
 \be\label{lBlockCommutators}
  g^{-1}E_{m+k,\,m+k}g\,=\,E_{m+k,\,m+k}\,
  -\,x_{m+k}E_{m+k,\,\ell+m}\,,
  \hspace{1.5cm}
  1\leq k\leq \ell-1\,,\\
  g^{-1}E_{m+k,\,m+k+j}g\,=\,E_{m+k,\,m+k+j}\,
  -\,x_{m+k+j}E_{m+k,\,\ell+m}\,,\\
  1\leq k\leq\ell-2\,,
  \quad
  1\leq j\leq\ell-1-k\,,\\
  g^{-1}E_{\ell+m,\,\ell+m}g\,=\,E_{\ell+m,\,\ell+m}\,
  +\,\sum_{k=1}^{\ell-1}x_{m+k}E_{m+k,\,\ell+m}\,.
 \ee
 \be\label{mlBlockCommutators}
 g^{-1}E_{1,\,m+j}g\,=\,e^{x_1-x_{\ell+m}}\bigl(\,E_{1,\,m+j}\,
 -\,x_{m+j}E_{1,\,\ell+m}\bigr)\,,
 \hspace{1.5cm}
 1\leq j\leq\ell-1\,,\\
 g^{-1}E_{k,\,m+j}g\,=\,e^{x_1-x_{\ell+m}}\Big(\,
 E_{k,\,m+j}+x_kE_{1,\,m+j}\,
 -\,x_{m+j}\bigl(E_{k,\,\ell+m}+x_kE_{1,\,\ell+m}\bigr)\Big)\,,\\
 2\leq k\leq m\,,
 \quad
 1\leq j\leq\ell-1\,,\\
 g^{-1}E_{k,\,\ell+m}g\,=\,e^{x_1-x_{\ell+m}}\bigl(
 E_{k,\,\ell+m}\,+\,x_kE_{1,\,\ell+m}\bigr)\,,
 \hspace{1.5cm}
 2\leq k\leq m\,,\\
 g^{-1}E_{1,\,\ell+m}g\,=\,e^{x_1-x_{\ell+m}}\,E_{1,\,\ell+m}\,.
 \ee
\end{lem}
\noindent \emph{Proof.} Consider  functions
$F_{ij}(x):=g(x)^{-1}E_{ij}g(x)$. The commutation relations above then
can be derived by writing  down and
 solving  differential equations for
$F_{ij}(x)$ with the initial  condition $F_{ij}(0)=E_{ij}$. $\Box$

Let us introduce a notation
$$
\< A\>:=\<\psi_L|\pi_{\underline{\lambda}}(A)|\psi_R\>, \qquad A\in
\CU\mathfrak{gl}_{\ell+m}.
$$
Consider $\<C_2 g\>$ where $C_2$ is a quadratic Casimir operator
\eqref{Casimirs}. Taking into account  $\<\psi_L|E_{ji}=0$ for $j-i>1$, we have
$\<C_2g\>=\<\tilde{C}_2g\>$ with
 \be\label{QuadraticCasimir}
  \tilde{C}_2\,=\,\sum_{i<j}^{\ell+m}E_{ii}E_{jj}\,
  -\,\sum_{j=1}^{\ell+m-1}E_{j+1,\,j}E_{j,\,j+1}\,
  -\,\sum_{i=1}^{\ell+m}\rho_iE_{ii}+\s_2(\rho)\,.
 \ee
Let us derive contributions of each term in
\eqref{QuadraticCasimir}. First we observe that
  \be\label{GrRhoPart}
 - \sum_{i=1}^{\ell+m}\rho_i\<E_{ii}g\>\,
  =\,-\frac{\ell+m-1}{2}\Big[\<E_{11}g\>\,
  -\,\<E_{\ell+m,\,\ell+m}g\>\Big]\,
  -\,\sum_{j=2}^{\ell+m-1}\rho_j\<E_{jj}g\>\\
  =\,-\frac{\ell+m-1}{2}\Big[
  \<(E_{11}+\ldots+E_{mm})g\>\,
  -\,\<(E_{m+1,\,m+1}+\ldots+E_{\ell+m,\ell+m})g\>\\
  -\,\sum_{k=2}^m\<E_{kk}g\>\,
  +\,\sum_{n=1}^{\ell-1}\<E_{m+n,\,m+n}g\>\Big]\,
  -\,\sum_{j=2}^{\ell+m-1}\rho_j\<E_{jj}g\>\\
  =\,-\frac{\ell+m-1}{2}\Big[
  \<(E_{11}+\ldots+E_{mm})g\>\,
  -\,\<(E_{m+1,\,m+1}+\ldots+E_{\ell+m,\ell+m})g\>\Big]\\
  -\,\sum_{k=2}^m\bigl(\rho_k-\frac{\ell+m-1}{2}\bigr)
  \<E_{kk}g\>\,
  -\,\sum_{n=1}^{\ell-1}\bigl(\rho_{m+n}+\frac{\ell+m-1}{2}\bigr)
  \<E_{m+n,\,m+n}g\>\\
  =\,\frac{\ell+m-1}{2}\Big[\frac{\pr}{\pr x_1}\,
  -\,\frac{\pr}{\pr x_{\ell+m}}\Big]\,
  -\,\sum_{k=2}^m(k-1)x_k\frac{\pr}{\pr x_k}\,
  -\,\sum_{n=1}^{\ell-1}(\ell-n)
  x_{m+n}\frac{\pr}{\pr x_{m+n}}\,,
 \ee
 since $\rho_k-\rho_1\,=\,1-k$ with
$$
 \<E_{kk}g\>\,=\,\bigl\<g\bigl(E_{kk}-x_kE_{1k}\bigr)\bigr\>\,
 =\,-x_k\frac{\pr}{\pr x_k}\<g\>\,,
 \hspace{1.5cm}
 k=2,\ldots,m\,,
$$
and $\rho_{m+n}-\rho_{\ell+m}\,=\,\ell-n$ with
$$
 \<E_{m+n,\,m+n}g\>\,
 =\,\bigl\<g\bigl(E_{m+n,\,m+n}+x_{m+n}E_{m+n,\,\ell+m}\bigr)\bigr\>\,
 =\,x_{m+n}\frac{\pr}{\pr x_{m+n}}\<g\>\,,
 \hspace{1.5cm}
 n=1,\ldots,\ell-1\,.
$$

Next, by using
$\<\psi_L|E_{j+1,\,j}\,=\,\h^{-1}\<\psi_L|$ for
$j=1,\ldots,\ell+m-1$, we find the following:
 \be\nn
  \hspace{-2cm}
  -\sum_{j=1}^{\ell+m-1}\<E_{j+1,\,j}E_{j,\,j+1}g\>\,
  =\,-\frac{1}{\h}\sum_{j=1}^{\ell+m-1}\<E_{j,\,j+1}g\>\,
  =\,-\frac{1}{\h}\Big(\<E_{12}g\>\,+\,\<E_{\ell+m-1,\,\ell+m}g\>\,
  +\,\sum_{j=2}^{\ell+m-2}\<E_{j,\,j+1}g\>\Big)\\
  \hspace{-2cm}
  =\,\frac{1}{\h}\bigl\{\pr_{x_2}+\pr_{x_{\ell+m-1}}\bigr\}\<g\>\,
  -\,\frac{1}{\h}\sum_{k=2}^{m-1}\<g(E_{k,\,k+1}+x_kE_{1,\,k+1})\>\,
  -\,\frac{1}{\h}\<E_{m,\,m+1}g\>\\
  -\,\frac{1}{\h}\sum_{n=1}^{\ell-2}\<g(E_{m+n,\,m+n+1}-x_{m+n+1}E_{m+n,\,\ell+m})\>
 \ee
In generic case when $m>1,\,\ell>1$ we have
$$
 -\frac{1}{\h}\<E_{m,\,m+1}g\>\,\\
 =\,-\frac{1}{\h}e^{x_1-x_{\ell+m}}\bigl\<g\bigl(E_{m,\,m+1}+x_mE_{1,\,m+1}
 -x_{m+1}E_{m,\,\ell+m}-x_mx_{m+1}E_{1,\,\ell+m}\bigr)\bigr\>
$$
and therefore
 \be\label{GrRootPart}
  \hspace{-2cm}
  -\frac{1}{\h}\sum_{j=1}^{\ell+m-1}\<E_{j,\,j+1}g\>\,\\
  =\,\frac{1}{\h}\Big\{\frac{\pr}{\pr x_2}+\frac{\pr}{\pr x_{\ell+m-1}}\,
  +\,\sum_{k=2}^{m-1}x_k\frac{\pr}{\pr x_{k+1}}
  -\,\sum_{n=1}^{\ell-2}x_{m+n+1}\frac{\pr}{\pr x_{m+n}}\,
  +\,\frac{1}{\h}(-1)^{\epsilon(\ell,m)}x_mx_{m+1}e^{x_1-x_{\ell+m}}\Big\}\<g\>\,.
 \ee
In the special case $m=1$ the following holds:

 \be
 -\frac{1}{\h}
 \sum_{k=1}^{\ell}\<E_{k,\,k+1}g\>\,=\,-\frac{1}{\h}\sum_{k=1}^{\ell-1}
\<E_{k,\,k+1}g\>\,
 -\frac{1}{\h}\<E_{\ell,\ell+1}g\>\,=\,\\
 -\frac{1}{\h}\sum_{k=1}^{\ell-1}
\<g E_{k,\,k+1}-x_{k+1}E_{k,\ell+1}\>\,
 -\frac{1}{\h}\<E_{\ell,\ell+1}g\>
  =\,\Big\{\frac{(-1)^{\epsilon(\ell,1)}}{\h^2}x_2e^{x_{\ell+1}-x_1}\,
  +\,\frac{1}{\h}\frac{\pr}{\pr x_{\ell}}\,
  -\frac{1}{\h}\sum_{j=2}^{\ell-1}x_{j+1}\frac{\pr}{\pr x_j}
  \Big\}\<g\>\,.
 \ee
In the other special case $\ell=1$ we have
 \be
  -\frac{1}{\h}\sum_{j=1}^m\<E_{j,\,j+1}g\>\,
  =\,\frac{1}{\h}\Big\{\frac{\pr}{\pr x_2}\,
  -\,\frac{(-1)^{\epsilon(1,m)}}{\h}x_me^{x_{m+1}-x_1}\,
  +\,\sum_{k=2}^{m-1}x_k\frac{\pr}{\pr x_{k+1}}\Big\}\<g\>\,.
 \ee
Finally, we calculate the contribution of the quadratic part of the
Casimir element \eqref{QuadraticCasimir}.  We have
 \be\label{QuadraticPart}
  \sum_{i<j}\<E_{ii}E_{jj}g\>\,=\,\<E_{11}E_{\ell+m,\,\ell+m}g\>\,
  +\,\sum_{k=2}^m\<E_{11}E_{kk}g\>\,
  +\,\sum_{n=1}^{\ell-1}\<E_{11}E_{m+n,\,m+n}g\>\\
  +\,\sum_{k=2}^m\<E_{kk}E_{\ell+m,\,\ell+m}g\>\,
  +\,\sum_{n=1}^{\ell-1}\<E_{m+n,\,m+n}E_{\ell+m,\,\ell+m}g\>\\
  +\,\sum_{2\leq k<a}^m\<E_{kk}E_{aa}g\>\,
  +\,\sum_{1\leq n<b}^{\ell-1}\<E_{m+n,\,m+n}E_{m+b,\,m+b}g\>\,
  +\,\sum_{k=2}^m\sum_{m=1}^{\ell-1}\<E_{kk}E_{m+n,\,m+n}g\>\,.
 \ee
Then for each term in the above decomposition we derive the
following:
 \be\label{Term1}
  \<E_{11}E_{\ell+m,\,\ell+m}g\>\,
  =\,\<(E_{11}+\ldots+E_{mm})
  (E_{m+1,\,m+1}+\ldots+E_{\ell+m,\,\ell+m})g\>\\
  \hspace{-1cm}
  -\,\sum_{k=2}^m\<E_{kk}E_{\ell+m,\,\ell+m}g\>\,
  -\,\sum_{n=1}^{\ell-1}\<(E_{11}E_{m+n,\,m+n}g\>\,
  -\,\sum_{k=2}^m\sum_{n=1}^{\ell-1}\<E_{kk}E_{m+n,\,m+n}g\>\,;
 \ee
 \be\label{Term2}
  \sum_{k=2}^m\<E_{11}E_{kk}g\>\,
  =\,\sum_{k=2}^m\<(E_{11}+\ldots+E_{mm})E_{kk}g\>\,
  -\,\sum_{k,a=2}^m\<E_{aa}E_{kk}g\>\,;
 \ee
 \be\label{Term5}
  \sum_{n=1}^{\ell-1}\<E_{m+n,\,m+n}E_{\ell+m,\,\ell+m}g\>\,
  =\,\sum_{n=1}^{\ell-1}
  \<E_{m+n,\,m+n}(E_{m+1,\,m+1}+\ldots+E_{\ell+m,\,\ell+m})g\>\\
  -\,\sum_{n,b=1}^{\ell-1}\<E_{m+n,\,m+n}E_{m+b,\,m+b}g\>\,.
 \ee
Substituting \eqref{Term1}-\eqref{Term5} into \eqref{QuadraticPart},
and making evident cancelations, we arrive at
 \be\label{GrQuadraticPart}
  \sum_{i<j}\<E_{ii}E_{jj}g\>\,
  =\,\<(E_{11}+\ldots+E_{mm})(E_{m+1,\,m+1}+\ldots+E_{\ell+m,\,\ell+m})g\>\\
  -\,\sum_{k=2}^m\<E_{kk}^2g\>\,
  -\,\sum_{n=1}^{\ell-1}\<E_{m+n,\,m+n}^2g\>\,
  +\,\sum_{k=2}^m\<(E_{11}+\ldots+E_{mm})E_{kk}g\>\\
  -\,\sum_{k<a}^m\<E_{kk}E_{aa}g\>\,
  -\,\sum_{n<b}^m\<E_{m+n,\,m+n}E_{m+b,\,m+b}g\>\\
  +\,\sum_{n=1}^{\ell-1}
  \<E_{m+n,\,m+n}(E_{m+1,\,m+1}+\ldots+E_{\ell+m,\,\ell+m})g\>\\
  \hspace{-2cm}
  =\,\Big\{\frac{\pr^2}{\pr x_1\pr x_{\ell+m}}\,
  -\,\sum_{k=2}^m\Big(x_k^2\frac{\pr^2}{\pr x_k^2}
  +x_k\frac{\pr}{\pr x_k}\Big)\,
  -\,\sum_{n=1}^{\ell-1}\Big(x_{m+n}^2\frac{\pr^2}{\pr x_{m+n}^2}
  +x_{m+n}\frac{\pr}{\pr x_{m+n}}\Big)\\
  \hspace{-2cm}
  -\,\sum_{2\leq k<a}^mx_kx_a\frac{\pr^2}{\pr x_k\pr x_a}\,
  -\,\sum_{1\leq n<b}^{\ell-1}x_{m+n}x_{m+b}\frac{\pr^2}{\pr x_{m+n}\pr x_{m+b}}\,
  +\,\sum_{k=2}^mx_k\frac{\pr^2}{\pr x_1\pr x_k}\,
  -\,\sum_{n=1}^{\ell-1}
  x_{m+n}\frac{\pr^2}{\pr x_{m+n}\pr x_{\ell+m}}
  \Big\}\<g\>\,.
 \ee
since
 \be\nn
  \<E_{kk}^2g\>\,=\,(-x_k\pr_{x_k})(-x_k\pr_{x_k})\<g\>\,
  =\,\bigl\{x_k^2\pr_{x_k}^2\,+\,x_k\pr_{x_k}\bigr\}\<g\>\,,
  \qquad
  2\leq k\leq m\,;\\
  \hspace{-2.5cm}
  \<E_{m+n,\,m+n}^2g\>\,
  =\,(x_{m+n}\pr_{x_{m+n}})(x_{m+n}\pr_{x_{m+n}})\<g\>\\
  =\,\bigl\{x_{m+n}^2\pr_{x_{m+n}}^2\,
  +\,x_{m+n}\pr_{x_{m+n}}\bigr\}\<g\>\,,
  \hspace{2.5cm}
  1\leq n\leq\ell-1\,.
 \ee
At last we collect \eqref{GrRhoPart}, \eqref{GrRootPart},
\eqref{GrQuadraticPart}, multiply by $\hbar^2$ 
and conjugate them by $e^{-\rho_1(x_1-x_{\ell+m})}$. We obtain
\eqref{GrQuadraticHamiltonian}.
This completes the proof of Proposition \ref{propone}.

\section*{Appendix B: Proof of Proposition \ref{PropTwo}}

 The lower-triangular part of the Lax-operator $\CL=\|\CL_{ij}\|$ easily
follows from \eqref{LeftWhittEqs}:
 \be
  \CL_{n+1,\,n}\,=\,1\,,
  \hspace{1.5cm}
  \CL_{n+j,\,n}\,=\,0\,,
  \quad
  2\leq j\leq\ell+m-n\,,
 \ee
for $1\leq n\leq\ell+m-1$.

The calculation of the upper-triangular part of $\CL$ can be done
using  Lemma \ref{Auxlem}. Namely, if $\CL_{ij}$ is in diagonal
$(m\times m)$-block the following holds:
 \be
  \CL_{1,\,k}\<g\>\,=\,\hbar\<E_{1,\,k}g\>\,=\,-\hbar\pr_{x_k}\<g\>\,,
  \hspace{3.5cm}
  2\leq k\leq m\,;\\
  \hspace{-1cm}
  \CL_{kk}\<g\>\,=\,\hbar\<E_{kk}g\>\,=\,\hbar\<g(E_{kk}+x_kE_{1,\,k})\>\,
  =\,-\hbar x_k\pr_{x_k}\<g\>\,,
  \hspace{1.5cm}
  2\leq k\leq m\,;\\
  \CL_{11}\<g\>\,=\,\hbar\<(E_{11}+\ldots+E_{mm})g\>\,
  -\,\hbar\sum_{k=2}^m\<E_{kk}g\>\,=\,\bigl\{-\hbar\pr_{x_1}
  +\hbar\sum_{k=2}^mx_k\pr_{x_k}\bigr\}\<g\>\,;\\
  \CL_{k,\,k+j}\<g\>\,=\,\hbar\<E_{k,\,k+j}g\>\,
  =\,\hbar\<g(E_{k,\,k+j}+x_kE_{1,\,k+j})\>\,
  =\,-\hbar x_k\pr_{x_{k+j}}\,2\leq k\leq m,\quad 1\leq j\leq m-k.
 \ee
For the diagonal $(\ell\times\ell)$-block we have
 \be
  \CL_{m+k,\,\ell+m}\<g\>\,=\,\hbar\<E_{m+k,\,\ell+m}g\>\,
  =\,-\hbar\pr_{x_{m+k}}\<g\>\,,
  \hspace{3.5cm}
  1\leq k\leq\ell-1\,;\\
  \hspace{-2cm}
  \CL_{m+k,\,m+k}\<g\>\,
  =\,\hbar\<g(E_{m+k,\,m+k}-x_{m+k}E_{m+k,\,\ell+m})\>\,
  =\,\hbar x_{m+k}\pr_{x_{m+k}}\<g\>\,,
  \hspace{1.5cm}
  1\leq k\leq\ell-1\,;\\
  \CL_{\ell+m,\,\ell+m}\<g\>\,
  =\,\hbar\<(E_{m+1,\,m+1}+\ldots+E_{\ell+m,\,\ell+m})g\>\,
  -\,\hbar\sum_{k=1}^{\ell-1}\<E_{m+k,\,m+k}g\>\\
  =\,-\hbar\pr_{x_{\ell+m}}\<g\>\,-\,\hbar\sum_{k=1}^{\ell-1}
  \<g(E_{m+k,\,m+k}-x_{m+k}E_{m+k,\,\ell+m})\>\\
  =\,\bigl\{\hbar\pr_{x_{\ell+m}}
  -\hbar\sum_{k=1}^{\ell-1}x_{m+k}\pr_{x_{m+k}}\bigr\}\<g\>\,;\\
  \CL_{m+k,\,m+k+j}\<g\>\,
  =\,\hbar\<g(E_{m+k,\,m+k+j}-x_{m+k+j}E_{m+k,\,\ell+m})\>\,
  =\,\hbar x_{m+k+j}\pr_{x_{m+k}}\<g\>\,,\\
  1\leq k\leq\ell-2\,,
  \quad
  1\leq j\leq\ell-k-1\,;
 \ee
Finally, for the  upper-triangular
$(m\times\ell)$-block the following holds:
 \be
  \hspace{-2cm}
  \CL_{1,\,m+j}\<g\>\,=\,\hbar e^{x_1-x_{\ell+m}}\<g(E_{1,\,m+j}
  -x_{m+j}E_{1,\,\ell+m})\>\,
  =\,-(-1)^{\epsilon(\ell,m)}x_{m+j}e^{x_1-x_{\ell+m}}\<g\>\,,
  \hspace{1.5cm}
  1\leq j\leq\ell-1\,;\\
  \CL_{1,\,\ell+m}\<g\>\,=\,\hbar e^{x_1-x_{\ell+m}}\<gE_{1,\,\ell+m}\>\,
  =\,(-1)^{\epsilon(\ell,m)}e^{x_1-x_{\ell+m}}\<g\>\,;\\
  \hspace{-2cm}
  \CL_{k,\,\ell+m}\<g\>\,=\,\hbar e^{x_1-x_{\ell+m}}\<g(E_{k,\,\ell+m}
  +x_kE_{1,\,\ell+m})\>\,=\,
(-1)^{\epsilon(\ell,m)}x_ke^{x_1-x_{\ell+m}}\<g\>\,,
  \hspace{1.5cm}
  2\leq k\leq m\,;\\
  \hspace{-3cm}
  \CL_{k,\,m+j}\<g\>\,=\,\hbar e^{x_1-x_{\ell+m}}\bigl\<g\bigl(E_{k,\,m+j}+x_kE_{1,\,m+j}
  -x_{m+j}E_{k,\,\ell+m}-x_kx_{m+k}E_{1,\,\ell+m}\bigr)\bigr\>\\
  =\,-(-1)^{\epsilon(\ell,m)}x_kx_{m+k}e^{x_1-x_{\ell+m}}\<g\>\,,
  \hspace{4.5cm}
  2\leq k\leq m,
  \quad
  1\leq j\leq\ell-1\,.
 \ee
After conjugation $\CL\rightarrow e^{-\rho_1(x_1-x_{\ell+m})}\CL
e^{\rho_1(x_1-x_{\ell+m})}$ we  arrive at the proof of Proposition \ref{GrLaxOperator}.

\section*{Appendix C: Proof of Theorem \ref{propthree}}

Consider the following decomposition of
the Borel subalgebra
$\frak{b}_+=\frak{h}^{(1,\ell+1)}+\frak{n}^{(1,\ell+1)}_+$:
 \be\label{Decomp}
  \frak{h}^{(1,\ell+1)}\,
  =\,\Big\<E_{11},\,E_{11}+\ldots+E_{\ell+1,\,\ell+1},\,
  E_{k,\,\ell+1},\,1<k\leq\ell\Big\>\,,\\
  \frak{n}^{(1,\ell+1)}_+\,
  =\,\Big\<E_{12},\,E_{1,\,\ell+1};\,
  E_{kk},\,1<k\leq\ell;\,
  E_{k,\,k+1},\,1<k<\ell\Big\>\,.
 \ee
Recall the construction of a generalized Gelfand-Zetlin representation
of $\frak{gl}_{\ell+1}$ \cite{GKL}.
Namely, let $\underline{\gamma}_1,\ldots,\underline{\gamma}_{\ell+1}$
be a triangular array consisting of $\ell(\ell+1)/2$ variables
$\gamma_n=(\gamma_{n1},\ldots,\gamma_{nn})\in\IC^n,n=1,\ldots,\ell+1$.
The operators
\be\label{GZRep}
  E_{kk}\,=\,\frac{1}{\h}\Big(\sum_{j=1}^n\gamma_{n,j}\,
  -\,\sum_{i=1}^{n-1}\gamma_{n-1,\,i}\Big)\,,
  \hspace{1cm}
  1\leq k\leq\ell+1\,;\\
  E_{n,\,n+1}\,
  =\,-\frac{1}{\h}\sum_{i=1}^n\frac{\prod\limits_{j=1}^{n+1}
  (\gamma_{n,i}-\gamma_{n+1,\,j}-\frac{\h}{2})}
  {\prod\limits_{s\neq i}(\gamma_{n,i}-\gamma_{n,s})}\,
  e^{-\h\pr_{n,i}}\,,
  \hspace{1.5cm}
  1\leq n\leq\ell\,;\\
  E_{n+1,\,n}\,
  =\,\frac{1}{\h}\sum_{i=1}^n\frac{\prod\limits_{j=1}^{n-1}
  (\gamma_{n,i}-\gamma_{n-1,\,j}+\frac{\h}{2})}
  {\prod\limits_{s\neq i}(\gamma_{n,i}-\gamma_{n,s})}\,
  e^{\h\pr_{n,i}}\,,
  \hspace{1.5cm}
  1\leq n\leq\ell\,,
 \ee
form a representation of $\frak{gl}_{\ell+1}$ in
the space $\CM$ of meromorphic functions
 in $\ell(\ell+1)/2$ variables
$(\underline{\gamma}_1,\ldots,\underline{\gamma}_{\ell})$.
The Whittaker vectors $\<\psi_L|\in \CV'_{\underline{\la}}$ and
$|\psi_R\>\in \CV_{\underline{\la}}$  are defined by
 \be\label{WhittakerVecEqs}
  \<\psi_L|E_{n+1,\,n}\,=\,\frac{1}{\h}\<\psi_L|\,,
  \hspace{1.5cm}
  1\leq n\leq\ell\,;\\
  E_{12}|\psi_R\>\,=\,0\,,
  \hspace{2.5cm}
  E_{1,\,\ell+1}|\psi_R\>\,
  =\,\frac{(-1)^{1+\frac{\ell(\ell-1)}{2}}}{\h}|\psi_R\>\,,\\
  E_{kk}|\psi_R\>\,=\,0,\,
  \hspace{1cm}
  2\leq k\leq\ell,\,
  \hspace{1.5cm}
  E_{k,\,k+1}|\psi_R\>\,=\,0,\,
  \quad
  2\leq k\leq\ell-1\,.
 \ee
We identify both $\CV_{\underline{\lambda}}$ and
$\CV'_{\underline{\lambda}}$ with subspaces of the space of functions
of $\gamma_{ij}$, $i=1,\ldots,\ell+1$, $j=1,\ldots i$. The action of
$\CU\mathfrak{gl}_{\ell+1}$ on  $\CV_{\underline{\lambda}}$ is given
by \eqref{GZRep} and the action on  $\CV'_{\underline{\lambda}}$
is given adjoint
generators
$$
E_{ij}^{\dagger}=\mu^{-1}(\gamma)E_{ij}\mu(\gamma),
$$
where
 \be \mu(\gamma)=\prod_{n=2}\prod_{s\neq n}
\Gamma^{-1}\Big(\frac{\gamma_{nj}-\gamma_{ns}}{\h}\Big).
 \ee
 \begin{lem} The equations \eqref{WhittakerVecEqs} admit the solution
\be \<\psi|_L=\frac{1}{2\pi\imath\hbar};
\ee
 \be\label{RightVector}
  |\psi_R\>\,=\,\prod_{k=2}^{\ell}
  \delta\Big(\sum_{j=1}^k\gamma_{k,\,j}\,
  -\,\sum_{i=1}^{k-1}\gamma_{k-1,\,i}\Big)
  \prod_{1\leq i\leq k}^{\ell-1}
  \delta\Big(\gamma_{k,\,i}-\gamma_{k+1,\,i}+\frac{\h}{2}\Big)\\
  \times\prod_{n=2}^{\ell}\prod_{i_n\neq j_n}
  \Gamma\Big(\frac{\gamma_{n,\,i_n}-\gamma_{n,\,j_n}}{\h}\Big)\,
  \prod_{j=1}^{\ell+1}
  \h^{\frac{\gamma_{\ell,\,1}-\gamma_{\ell+1,\,j}}{\h}+\frac{1}{2}}\,
  \Gamma\Big(\frac{\gamma_{\ell,\,1}-\gamma_{\ell+1,\,j}}{\h}+\frac{1}{2}\Big)
 \ee
\end{lem}
{\it Proof}. The equations on the left vector are similar  to those in
\cite{GKL}
and by the same reason  admit the solution $\<\psi|_L=\frac{1}{2\pi\imath\hbar}$.
To find the right Whittaker vector one needs an explicit expression  of the element
$E_{1,\,\ell+1}=[\ldots[[E_{12},\,E_{23}],\,\ldots,\,E_{\ell-1,\,\ell}\,],\,
 E_{\ell,\,\ell+1}]\,$
\be\label{E1ell+1}
 \hspace{-1.5cm}
  E_{1,\,\ell+1}\,=\,-\frac{1}{\h}\Big(
  \sum_{i_1=1}^{\ell}\frac{\prod\limits_{j_1=1}^{\ell+1}
  (\gamma_{\ell,\,i_1}-\gamma_{\ell+1,\,j_1}-\frac{\h}{2})}
  {\prod\limits_{k_1\neq i_1}
  (\gamma_{\ell,\,i_1}-\gamma_{\ell,\,k_1})}
  \cdots
  \sum_{i_{\ell-1}=1}^2\frac{\prod\limits_{j_{\ell-1}\neq i_{\ell-2}}
  (\gamma_{2,\,i_2}-\gamma_{3,j_2}-\frac{\h}{2})}
  {\prod\limits_{k_{\ell-1}\neq i_{\ell-1}}
  (\gamma_{2,\,i_{\ell-1}}-\gamma_{2,\,k_{\ell-1}})}\\
  \cdot\prod_{j_{\ell}\neq i_{\ell-1}}
  \bigl(\gamma_{11}-\gamma_{2,\,j_{\ell}}-\frac{\h}{2}\bigr)\Big)\,
  e^{-\h\sum_{i_1=1}^{\ell}\sum_{i_2=1}^{\ell-1}...
\sum_{i_{\ell=1}}^{1}\sum\limits_{k=1}^{\ell}\pr_{\gamma_{\ell+1-k,\,i_k}}}\,.
 \ee
The constraints $$
 E_{22}|\psi_R\>\,=\,\ldots\,=\,E_{\ell\ell}|\psi_R\>\,=\,0\,,
$$
obviously hold. Similarly due to the presence of the product of delta-functions
$$
 \prod_{1\leq i\leq k}^{\ell-1}
 \delta\Big(\gamma_{k,\,i}-\gamma_{k+1,\,i}+\frac{\h}{2}\Big)
$$
we have $E_{12}|\psi_R\>=\ldots=E_{\ell-1,\,\ell}|\psi_R\>=0$.
Thus we have to check that \eqref{RightVector} satisfies the
relation $E_{1,\,\ell+1}|\psi_R\>\,=\,\h^{-1}|\psi_R\>$ with
\eqref{E1ell+1}. At first we note that due to the factor
$$
 \prod_{1\leq i\leq k}^{\ell-1}
 \delta\Big(\gamma_{k,\,i}-\gamma_{k+1,\,i}+\frac{\h}{2}\Big),
$$
we have
 \be
  E_{1,\,\ell+1}|\psi_R\>\\
  \hspace{-1.5cm}
  =\,-\frac{1}{\h}\frac{\prod\limits_{j=1}^{\ell+1}
  (\gamma_{\ell,\,1}-\gamma_{\ell+1,\,j}-\frac{\h}{2})}
  {\prod\limits_{k=2}^{\ell}
  (\gamma_{\ell,\,1}-\gamma_{\ell,\,k})}
  \cdots\frac{(\gamma_{21}-\gamma_{32}-\frac{\h}{2})
  (\gamma_{21}-\gamma_{33}-\frac{\h}{2})}
  {\gamma_{21}-\gamma_{22}}\\
  \bigl(\gamma_{11}-\gamma_{22}-\frac{\h}{2}\bigr)\,
  e^{-\h\sum_{i_1=1}^{\ell}\sum_{i_2=1}^{\ell-1}...\sum_{i_{\ell=1}}^{1}
  \sum\limits_{k=1}^{\ell}\pr_{\gamma_{\ell+1-k,\,i_k}}}
  \bigl|\psi_R\bigr\>\,.
 \ee
Finally straightforward calculations provide
 \be
  E_{1,\,\ell+1}|\psi_R\>\,=\,(-1)^{1+\frac{\ell(\ell-1)}{2}}
  \prod_{k=1}^{\ell-1}\prod_{i=1}^k
  \frac{\gamma_{k,\,1}-\gamma_{k+1,\,i+1}-\frac{\h}{2}}
  {\gamma_{k+1,\,1}-\gamma_{k+1,\,i+1}-\h}|\psi_R\>\,
  =\,\frac{(-1)^{1+\frac{\ell(\ell-1)}{2}}}{\h}|\psi_R\>.
 \ee
 $\Box$

Now we are ready to prove Proposition \ref{propthree}.
Define the left and right $\cal{U}$-modules as $V'=\<\psi_L|\cal{U}$ and
$V=\cal{U}|\psi_R\>$ respectively. Let $\phi\in V'$ and $\psi\in V$.
Define the paring $\<\cdot,\cdot\>:V'\otimes V\rightarrow \IC$ by
 \be \<\phi,\psi\> =
\int_{\CC}
\mu(\gamma)\phi(\gamma)\psi(\gamma)\prod_{n=1}^{\ell}\prod_{j\leq n} d\gamma_{nj},
 \ee
 where we define the integration domain $\CC$ shortly below. Let $x=x_1$. We have
$$
 \Psi^{(1,\,\ell+1)}_{\underline{\gamma}_{\ell+1}}(x,0,\ldots,0)\,
 =\,e^{-\frac{\ell}{2}x}\<\psi_L,\,
 e^{-x E_{11}}
 \psi_R\>\,,
$$
and thus
\be \label{intconv} 
  \Psi^{(1,\,\ell+1)}_{\underline{\gamma}_{\ell+1}}(x,0,...,0)\,
  =\frac{1}{2\pi\imath\hbar}
e^{-\frac{\ell}{2}x}\int_{\CC}\!\prod_{n=1}^{\ell}
d\underline{\gamma}_n\,\,\mu(\gamma)\,\,
  e^{-\frac{1}{\h}x\gamma_{11}}\\
  \cdot\prod_{k=2}^{\ell}
  \delta\Big(\sum_{j=1}^k\gamma_{k,\,j}\,
  -\,\sum_{i=1}^{k-1}\gamma_{k-1,\,i}\Big)
  \prod_{1\leq i\leq k}^{\ell-1}
  \delta\Big(\gamma_{k,\,i}-\gamma_{k+1,\,i}+\frac{\h}{2}\Big)\\
  \cdot\prod_{n=2}^{\ell}\prod_{i_n\neq j_n}
  \Gamma\Big(\frac{\gamma_{n,\,i_n}-\gamma_{n,\,j_n}}{\h}\Big)\,
  \prod_{j=1}^{\ell+1}
  \h^{\frac{\gamma_{\ell,\,1}-\gamma_{\ell+1,\,j}}{\h}+\frac{1}{2}}\,
  \Gamma\Big(\frac{\gamma_{\ell,\,1}-\gamma_{\ell+1,\,j}}{\h}
  +\frac{1}{2}\Big)\,,
 \ee
for appropriate choice of integration domain $\CC$. 
Taking into account the Stirling formula for gamma function
 $$
\Gamma(c+z)=\sqrt{2\pi}z^{c+z-1/2}e^{-z}(1+O(1/z),
 $$
for $z\rightarrow\infty$, $0<|arg(z)|<\pi$ and $c,z\in\IC$ we infer 
 that the integral \eqref{intconv} converges absolutely.
Making obvious  cancelations and integrating out the
delta-functions one obtains
 \be
  \Psi^{(1,\,\ell+1)}_{\underline{\gamma}_{\ell+1}}(x)\,=\,
  \frac{1}{2\pi\imath\hbar}e^{-\frac{\ell}{2}x}
\int_{\imath\IR+\epsilon-\frac{\hbar}{2}}\!d\gamma_{\ell,1}\,e^{-\frac{1}{\h}
\gamma_{\ell,1}x+\frac{\ell-1}{2}x}
  \prod_{j=1}^{\ell+1}
  \h^{\frac{\gamma_{\ell,\,1}-\gamma_{\ell+1,\,j}}{\h}+\frac{1}{2}}\,
  \Gamma\Big(\frac{\gamma_{\ell,\,1}-\gamma_{\ell+1,\,j}}{\h}
  +\frac{1}{2}\Big)\,,
 \ee
 where $\epsilon>\gamma_{\ell+1,j},\,\,\,j=1,\ldots,\ell+1$.
Finally introducing the 
 variable $\gamma_{\ell,\,1}=\imath H-\frac{\h}{2}$ and setting 
$\lambda_j=\gamma_{\ell+1,j}$ we
obtain \eqref{identone}. $\Box$

\section*{Appendix D: Explicit calculations for  ${\rm Gr}(2,3)$}

In this Appendix we derive, using another version of the
Gelfand-Zetlin realization,  an integral
representation for a specialization of
the matrix elements \eqref{parTod} for $m=2$, $\ell=1$. Note
that due to isomorphism ${\rm Gr}(1,3)={\rm Gr}(2,3)=\IP^2$
the resulting integral expressions should be equal to
\eqref{identone} for $\ell=2$ after appropriate
identification of the parameters. Below we explicitly  verify this
equivalence using an integral identity due to Gustafson \cite{Gu}.

We use the following  version of the Gelfand-Zetlin
realization of the universal enveloping algebra $\CU(\frak{gl}_3)$
( see \cite{GKL}):
 \be\label{GL3Cartans}
  E_{11}=\frac{1}{\hbar}\gamma_{11}\,,
  \hspace{1.5cm}
  E_{22}=\frac{1}{\hbar}(\gamma_{21}+\gamma_{22}-\gamma_{11})\,,\\
  E_{33}=\frac{1}{\hbar}(\gamma_{31}+\gamma_{32}+\gamma_{33}-
  \gamma_{21}-\gamma_{22})
 \ee
 \be\label{GL3Lowering}
  E_{21}\,=\,\frac{1}{\hbar}e^{\h\pr_{11}}\,,\\
  E_{32}\,=\,\frac{1}{\hbar}\Big\{
  \frac{\gamma_{21}-\gamma_{11}+\frac{\h}{2}}
  {\gamma_{21}-\gamma_{22}}e^{\h\pr_{21}}\,
  +\,\frac{\gamma_{22}-\gamma_{11}+\frac{\h}{2}}
  {\gamma_{22}-\gamma_{21}}e^{\h\pr_{22}}\Big\}\,,\\
 \ee
 \be\label{GL3Raising}
  E_{12}\,=\,-\frac{1}{\hbar}\Big(\gamma_{11}-\gamma_{21}-\frac{\h}{2}\Big)
  \Big(\gamma_{11}-\gamma_{22}-\frac{\h}{2}\Big)\,
  e^{-\h\pr_{11}}\,,\\
  E_{23}\,=\,-\frac{1}{\h}\Big\{
  \frac{\prod\limits_{j=1}^3(\gamma_{21}-\gamma_{3j}-\frac{1}{2})}
  {\gamma_{21}-\gamma_{22}}e^{-\h\pr_{21}}\,
  -\,\frac{\prod\limits_{j=1}^3(\gamma_{22}-\gamma_{3j}-\frac{\h}{2})}
  {\gamma_{22}-\gamma_{21}}e^{-\h\pr_{22}}\Big\}\,,\\
  E_{13}=[E_{12},E_{23}]\,
  =\,-\frac{1}{\hbar}\Big\{(\gamma_{11}-\gamma_{22}-\frac{\h}{2})
  \frac{\prod\limits_{j=1}^3(\gamma_{21}-\gamma_{3j}-\frac{\h}{2})}
  {\gamma_{21}-\gamma_{22}}e^{-\h\pr_{21}-\h\pr_{11}}\\
  +\,(\gamma_{11}-\gamma_{21}-\frac{\h}{2})
  \frac{\prod\limits_{j=1}^3(\gamma_{22}-\gamma_{3j}-\frac{\h}{2})}
  {\gamma_{22}-\gamma_{21}}e^{-\h\pr_{22}-\h\pr_{11}}\Big\}
 \ee
The conjugated generators
$$
E_{ij}^{\dagger}=\mu^{-1}(\gamma)E_{ij}\mu(\gamma),\qquad
\mu(\gamma)=\frac{1}{\Gamma(\frac{\gamma_{21}-\gamma_{22}}{\hbar})
\Gamma(\frac{\gamma_{22}-\gamma_{21}}{\hbar})},
$$
are given by
 \be
  E^{\dag}_{21}\,=\,\frac{1}{\h}e^{-\h\pr_{11}}\,,\\
  E^{\dag}_{32}\,=\,\frac{1}{\h}\Big\{
  \frac{\gamma_{21}-\gamma_{11}-\frac{\h}{2}}
  {\gamma_{21}-\gamma_{22}}e^{-\h\pr_{21}}+
  \frac{\gamma_{22}-\gamma_{11}-\frac{\h}{2}}
  {\gamma_{22}-\gamma_{21}}e^{-\h\pr_{22}}\Big\}\,.
 \ee
In the case $\IP^2\simeq{\rm Gr}(2,3)$ we have the following
defining equations on the  Whittaker vectors:
$$
 \<\psi_L|E_{21}^{\dag}\,=\,\frac{1}{\h}\<\psi_L|\,,
 \qquad
 \<\psi_L|E_{32}^{\dag}\,=\,\frac{1}{\h}\<\psi_L|\,;
$$
\be\label{Gr23RightVectorEqs}
  E_{22}|\psi_R\>\,=\,0\,,
  \qquad
  E_{13}|\psi_R\>\,=\,-\frac{1}{\h}|\psi_R\>\,,
  \qquad
  E_{23}|\psi_R\>\,=\,0\,.
\ee

The defining equations for the right vector
\eqref{Gr23RightVectorEqs} also can be solved.

\begin{lem} The Whittaker vectors
defined by \eqref{Gr23RightVectorEqs} read as follows:
 \be
\<\psi_L|\,=\,\frac{1}{2\pi\hbar},
\ee
\be
  |\psi_R\>\,=\,\prod_{k=1}^2\prod_{j=1}^3
  \h^{(\gamma_{2i}-\gamma_{3j})/{\h}+1/2}
  \Gamma\Big(\frac{\gamma_{2i}-\gamma_{3j}}{\h}+\frac{1}{2}\Big)\,\,
  \delta(\gamma_{21}+\gamma_{22}-\gamma_{11})\,.
 \ee
 \end{lem}
{\it Proof}. Direct verification. $\Box$

Finally, we consider the ${\rm Gr}(2,3)$-Whittaker function
$$
 \Psi^{(2,3)}_{\underline{\la}}(x_1,x_2,x_3)\,
 =\,e^{-(x_1-x_3)}\<\psi_L|\,
 \pi_{\underline{\lambda}}\bigl(
 e^{-x_1(E_{11}+E_{22})-x_2E_{12}-x_3E_{33}}\bigr)\,|\psi_R\>\,,
$$
with $\la_j=\gamma_{3,j},\,j=1,2,3$ and  $x_1=x$, $x_2=x_3=0$.
\begin{prop} There is the following integral representation of
  ${\rm Gr}(2,3)$-Whittaker function:
 \be\nn
\Psi_{\underline{\la}}^{(2,3)}(x)\,=\,
\frac{1}{2\pi\hbar}\int_{{\CC}_{\epsilon_1}\times{\CC}_{\epsilon_2}}\!
  d\gamma_{21}d\gamma_{22}\,\,
  e^{-\frac{1}{\h}(\sum_{j=1}^{2}\gamma_{2j})x}
  \frac{1}{\prod_{i\neq j}\Gamma_1(\gamma_{2i}-\gamma_{2j}|{\h})}
  \prod_{j=1}^3\prod_{i=1}^2
  \Gamma_1({\gamma_{2i}-\lambda_{j}}|{\h})\,,
 \ee
 where $\epsilon_i>\lambda_j,\,\,\,i=1,2,\,\,\,j=1,2,3$.
\end{prop}
{\it Proof}. We have
 \be\Psi_{\underline{\la}}^{(2,3)}(x)\,=\,
\frac{e^{-x}}{2\pi\hbar}\int\!d\gamma_{21}d\gamma_{22}d\gamma_{11}\,\,
\delta(\gamma_{11}-\gamma_{21}-\gamma_{22})\times\\
  e^{\frac{1}{\h}(\sum\gamma_{3j}-\sum\gamma_{2k})x}
  \frac{1}{\prod_{i\neq j}\Gamma(\frac{\gamma_{2i}-\gamma_{2j}}{\h})}
  \prod_{j=1}^3\prod_{k=1}^2
  \h^{(\gamma_{2k}-\lambda_{j})/{\h}+1/2}
  \Gamma\Big(\frac{\gamma_{2k}-\lambda_{j}}{\h}+\frac{1}{2}\Big)\,.
 \ee
 After taking integral over $\gamma_{11}$ and shifting the variables
$\gamma_{21}\rightarrow\gamma_{21}-\h/2,\,\gamma_{22}\rightarrow\gamma_{22}-\h/2$
one obtains
 \be
  \Psi_{\underline{\la}}^{(2,3)}(x)\,=\,
\frac{1}{2\pi\hbar}\int_{\CC_{\epsilon_1}\times\CC{\epsilon_2}}\!
  d\gamma_{21}d\gamma_{22}\,\,
  e^{-\frac{1}{\h}(\sum_{j=1}^{2}\gamma_{2j})x}
  \frac{1}{\prod_{i\neq j}\Gamma_1({\gamma_{2i}-\gamma_{2j}}|{\h})}\times\\
  \prod_{j=1}^3\prod_{i=1}^2
  \Gamma_1({\gamma_{2i}-\gamma_{3j}}|{\h})\,.
 \ee

\begin{prop} The following relation between $(1,3)$- and
  $(2,3)$-Whittaker functions holds
$$
\Psi^{(1,3)}_{\lambda_1,\lambda_2,\lambda_3}(x)=\Psi^{(2,3)}_
{\tilde{\lambda}_1,\tilde{\lambda}_2,\tilde{\lambda_3}}(x),
$$
where $\lambda_i=\tilde{\lambda}_j+\tilde{\lambda}_k$, $i\neq j\neq
k$.

\end{prop}

\noindent {\it Proof}.     We have
$$
\Psi_{\lambda_1,\lambda_2,\lambda_3}^{(2,3)}(x)=\frac{1}{2\pi\hbar}
\int_{\CC_{\epsilon_1}\times\CC_{\epsilon_2}} 
d\gamma_1d\gamma_2e^{-\frac{1}{\hbar}(\gamma_1+\gamma_2)x}
\frac{\prod_{j=1}^3\Gamma_1({\gamma_1-\lambda_j}|{\hbar})
\Gamma_1({\gamma_2-\lambda_j}|{\hbar})}
{\Gamma_1({\gamma_1-\gamma_2}|{\hbar})\Gamma_1({\gamma_2-\gamma_1}|{\hbar})},
$$
where $\CC_{\epsilon_i}=\imath\IR+\epsilon_i,\,\,\,
\epsilon_i>\lambda_j,j=1,2,3$. 
Let us introduce new variables $\gamma=\gamma_1+\gamma_2$ and
$\gamma_*=\gamma_1-\gamma_2$  to obtain  
$$
\Psi_{\lambda_1,\lambda_2,\lambda_3}^{(2,3)}(x)=
\frac{1}{2\pi\hbar}\int d\gamma d\gamma_*e^{-\frac{1}{\hbar} \gamma x}
\frac{\prod_{j=1}^3\Gamma_1({\frac{1}{2}(\gamma+\gamma_*)-\lambda_j}|{\hbar})
\Gamma_1(\frac{1}{2}(\gamma-\gamma_*)-\lambda_j|\hbar)}
{\Gamma_1({\gamma_*}|\hbar)\Gamma(-\gamma_*|\hbar)}. 
$$
Thus to establish equivalence with $\Psi^{(1,3)}$ we should prove
$$
\int_{\CC_{(\epsilon_1-\epsilon_2)/2}} d\gamma_*
\frac{\prod_{j=1}^3\Gamma_1(\frac{1}{2}(\gamma+\gamma_*)-\lambda_j|\hbar)
\Gamma_1(\frac{1}{2}(\gamma-\gamma_*)-\lambda_j|\hbar)}
{\Gamma_1(\gamma_*|\hbar)\Gamma_1(-\gamma_*|\hbar)}=\prod_{1\leq i<j\leq 3}
\Gamma_1(\gamma-\lambda_i-\lambda_j|\hbar). 
$$
This follows from the limiting form of the 
identity due to Gustafson \cite{Gu} (eq. (9.4) with n=1 and $a_4=\infty$) 
$$
\int_{\CC_0} dt\,
\frac{\prod_{j=1}^3\Gamma_1(\alpha_j+t|\hbar)\Gamma_1(\alpha_j-t|\hbar)}
{\Gamma_1(2t|\hbar)\Gamma_1(-2t|\hbar)}= \prod_{1\leq i<j\leq 3}
\Gamma_1(\alpha_i+\alpha_j|\hbar),
$$
where the integration contour $\CC_0$ is goes between  
the sets of poles $\alpha_i+n\hbar$, $n\in \IZ_{\geq 0}$ and 
 $-\alpha_i - n\hbar $, $n\in \IZ_{\geq 0}$. $\Box$

\section*{Appendix E: Gauge theory description of non-linear sigma models}

In this Appendix we recall the standard representation of a
bosonic two-dimensional sigma model with the target space $\IP^{\ell}$
in terms of $U_1$-gauged sigma model. To simplify the arguments we
consider the  equivalence  of the classical theories i.e. identifying
the spaces of solutions of the equations of motions in two theories.
Omitting fermionic fields in
the action \eqref{actionS1} we obtain the following action of the
bosonic $U_1$-gauged sigma model
 \be\label{actionS1bos}
S_{bos}=\int_{\Sigma}\,d^2z \sqrt{h}h^{z\zb}\,\,
\left(\frac{1}{t}((\pr_{\zb}- A_{\zb})\varphi^j)(\pr_{z}-
A_{z})\bar{\varphi}^{j} +2\imath
b\,\sum_{j=1}^{\ell+1}\sigma|\varphi^j|^2+
H(\sum_{j=1}^{\ell+1}|\varphi^j|^2-r^2)\right). \ee
To obtain the classically equivalent field theory one may eliminate
some fields using  conditions of zero variations of the action
$S_{bos}$ (substituting instead of an independent  field $\Phi$ a solution of the
equation $\delta S_{bos}/\delta \Phi=0$). Using a shift of
the variable $H\to H-2\imath b\sigma$ and eliminating the fields  $b$ and
$\sigma$ via zero variation condition
we obtain (up to some $r$-dependent additive constant)
\be\label{actionS3bos}
S_{bos}=\int_{\Sigma}\,d^2z
\sqrt{h}h^{z\zb}\,\, \left(\frac{1}{t}((\pr_{\zb}-
A_{\zb})\varphi^j)(\pr_{z}- A_{z})\bar{\varphi}^{j} +
H(\sum_{j=1}^{\ell+1}|\varphi^j|^2-r^2)\right).
\ee
Now it is easy to show that the field theory  with the action
\eqref{actionS3bos} after a  gauge fixing is  equivalent on the
classical level to the  sigma
model  on $\IP^{\ell}$ with the action
\be\label{staction}
S_{\sigma-mod}=\frac{1}{2}\int_{\Sigma} d^2z \sqrt{h}h^{z\zb}\,\,
G_{i\bar{j}}(\xi(z,\zb))\pr_{\zb}\xi^i(z,\zb)\pr_{z}\bar{\xi}^{\jb}(z,\zb),
\ee
where the K\"{a}hler metric $G(\xi)$ is associated with  the
Fubini-Studi two form on $\IP^{\ell}$  locally written as
$$
\omega=\frac{\imath}{2\pi t}\left(
\sum_{j=1}^{\ell}\frac{d\xi_j\wedge d\bar{\xi}_{j}}
{(1+\sum_{j=1}^{\ell}|\xi_j|^2)}-\frac{(\sum_{j=1}^{\ell}\bar{\xi}_jd\xi_j)\wedge
(\sum_{i=1}^{\ell}\xi_id\bar{\xi}_i)}{(1+\sum_{j=1}^{\ell}|\xi_j|^2)^2}\right).
$$
Indeed, eliminating gauge fields $A$  by using zero variation
condition we obtain \be\label{actionnew} S_{bos}=\int_{\Sigma}\,d^2z
\sqrt{h}h^{z\zb}\,\, \left(\frac{1}{t}((\pr_{\zb}-
A_{\zb}(\varphi))\varphi^j)(\pr_{z}-
  A_{z}(\varphi))\bar{\varphi}^{j}
+H(\sum_{j=1}^{\ell+1}|\varphi^j|^2-r^2)\right), \ee with
$$
A_{\zb}=
\left(\sum_{j=1}^{\ell+1}|\varphi^j|^2\right)^{-1}\,\sum_{j=1}^{\ell+1}
\bar{\varphi}^j\pr_{\zb}\varphi^j,\qquad A_{z}=
\left(\sum_{j=1}^{\ell+1}|\varphi^j|^2\right)^{-1}\,\sum_{j=1}^{\ell+1}
\varphi^j\pr_{z}\bar{\varphi}^j.
$$
Note that the action functional \eqref{actionnew} is still invariant
with respect to $U_1$ gauge symmetry
$$
\varphi^i(z)\longrightarrow e^{\imath \alpha(z)}\,\varphi^i(z).
$$
Zero variation condition over $H$ imposes the constraint \be\label{constr}
\sum_{j=1}^{\ell+1}\,|\varphi^j|^2=r^2. \ee The solutions  of
\eqref{constr}  can be  parameterized as follows:
$$
\varphi^j=\frac{\xi^j}{(r^2+\sum_{j=1}^{\ell}|\xi^j|^2)^{1/2}},\quad
j=1,\ldots ,\ell,\qquad \varphi_{\ell+1}=\frac{re^{\imath
\Theta}}{(r^2+\sum_{j=1}^{\ell}|\xi^j|^2)^{1/2}}.
$$
Fixing the gauge freedom by taking $\Theta=0$ we recover the
standard representation \eqref{staction} of $\IP^{\ell}$ sigma
model.

\section*{Appendix F: Intersection theory on $\IP^{\ell}$ via
Hamiltonian reduction}

Let us given a manifold $X$ supplied with an action of a Lie group $G$ and an
$G$-equivariant vector bundle $E$. Let $s$ be a section $E$ such that
$G$ acts freely on the zero locus $s^{-1}(0)$.
There is a universal representation of integrals of  closed differential forms over
factor $s^{-1}(0)/G$ in terms of integrals over $X$ (see
e.g. \cite{CMR}  for detailed exposition and  relations with  quantum field theory
constructions). Below we review a  simple instance of this
construction providing a description of integrals of (equivariantly)
closed differential forms over
$\IP^{\ell}$. In this case
$X=\IC^{\ell+1}$, the bundle $E$ is trivial and the group $G$ is abelian group $U_1$.

Let us supply  complex vector space $\IC^{\ell+1}$  with a symplectic
structure
\be\label{symform}
\Omega=\frac{\imath}{2}\sum_{j=1}^{\ell+1}\,d\varphi^j\wedge d\bar{\varphi}^j.
\ee
The action of $U_1$
$$
\varphi^j\longrightarrow e^{\imath \alpha}\varphi^j, \qquad e^{\imath \alpha}\in U_1,
$$
is Hamiltonian i.e. there exists a momentum
$$
\mu(\varphi)=-\frac{1}{2}\sum_{j=1}^{\ell+1}|\varphi^j|^2.
$$
such that $\iota_v\Omega\,=\,d\mu$ where
$$
v=\imath \sum_{j=1}^{\ell+1}\left(\varphi^j\frac{\pr}{\pr \varphi^j}-
\bvarphi^j\frac{\pr}{\pr \bvarphi^j}\right)
$$
generates the action of $U_1$ on $\IC^{\ell+1}$.
Complex projective space $\IP^{\ell}$
can be constructed via Hamiltonian reduction
as a quotient of a hypersurface of
the fixed level  of the momentum $\mu$ over a free action of $U_1$
\be\label{projspace}
\IP^{\ell}=\mu^{-1}(\frac{1}{2} r^2)/U_1,\qquad r\in \IR.
\ee
Thus constructed $\IP^{\ell}$ is supplied with  a canonical symplectic structure
$\omega_{\IP^{\ell}}$ proportional to the Fubini-Study form.  In terms of
inhomogeneous coordinates $w_j=\varphi_j/\varphi_{\ell+1}$, $\varphi_{\ell+1}\neq 0$ it is
given by
 \be\label{FS}
  \omega_{\IP^{\ell}}=\frac{\imath r^2}{2}
  \frac{(1+\sum_{i=1}^{\ell}|w_i|^2)\sum_{j=1}^{\ell}dw_j\wedge d\wb_j
  -\sum_{i,j}^{\ell}w_i\wb_jdw_j\wedge d\wb_i}{(1+\sum_{i=1}^{\ell}|w_i|^2)^2}.
 \ee
The problem to write down the integral of closed  differential forms over
$\IP^{\ell}$ in terms of  integrals over $\IC^{\ell+1}$ is naturally
divided into two parts. First we shall
write an integral over hypersurface in a complex space in terms of
an integral over the ambient space. Second, we shall write down
an integral over a factor of a space over a free action of a
Lie group in terms of an integral over a space before
factorization. Let us first consider the problem of writing integral
over hypersurface. Thus given a real valued function $s(x)$ on $\IR^N$
let $i:Z\hookrightarrow\IR^N$  be a zero locus subset of $s$ (we consider the
case when $Z$ is compact).  Let $R_Z$ be a de Rham current such that
for a closed differential form  $\omega$ on $\IR^N$ the following holds
\be\label{Current}
\int_{Z}\,i^*\omega=\int_{\IR^N}\,\omega\wedge R_Z.
\ee
To write $R_z$ explicitly let us fix coordinates $(x^1,\ldots, x^N)$
on $\IR^N$.  We  identify algebra of differential forms
$\CA^*(\IR^N)$ on $\IR^N$ with the algebra of functions
${\rm Fun}(\IR^{N|N})$  on the
superspace $\IR^{N|N}=\Pi T\IR^N$ where $\Pi$ is a functor of the
parity change of the fibers of vector bundles.  Thus, we have associated
coordinates $(x^1,\ldots, x^N,\psi^1,\ldots \psi^N)$  in $\IR^{N|N}$
and de Rham differential is given by a vector field
$$
Q=\sum_{j=1}^N\,\psi^i\frac{\pr}{\pr x^j}, \qquad Q^2=0.
$$
Consider an extended space $\IR^{N+1,N+1}=\IR^{N|N}
\times \IR^{1|1}$ with the second factor understood as reversed-parity
tangent bundle to the one-dimensional odd space $\IR^{0|1}$. Let
$(H,\xi)$ be coordinates in $\IR^{1|1}$ and the de Rham differential
on extended space is given by the vector field
\be \label{BRSTvect}
Q=\sum_{j=1}^N\,\psi^j\frac{\pr}{\pr x^j}+H\frac{\pr}{\pr \xi}.
\ee
Now a  one-parameter family of the differential forms $R_Z(t)$ is
given by the following Berezin integral over the superspace $\IR^{1|1}$
\be\label{Current1}
R_Z(t)=\frac{1}{2\pi}\,
\int_{\IR^{1|1}} d\xi \,dH\,\exp\left( Q\,(\xi(\imath s(x)-\frac{t}{2}H))\right),
\ee
satisfies the relation \eqref{Current} (see e.g. \cite{CMR}).

To write down an integral of a closed differential
form over a factor of a space $Y$ over  a free action of  Lie group
$G=U_1$ in terms of an integral over $Y$  we
use $G$-equivariant cohomology of $Y$.
Let us recall that for a free action of a compact Lie  group  $G$ on
$Y$ equivariant  cohomology are
isomorphic to the cohomology of the factor
\be\label{Cohident}
H^*_G(Y)=H^*(Y/G).
\ee
Cartan model of  $U_1$-equivariant de Rham complex on $X$
is given by
\be\label{Cartan}
\Omega_{U_1}^*(Y)=(\Omega^*(Y))^{U_1}\otimes \IC[\sigma],\qquad
d_G=d-\iota_{v(\sigma)},\qquad \sigma \in \mathfrak{u}_1^*=({\rm Lie}(U_1))^*,
\ee
where $\sigma$ is of degree  two. Equivariant cohomology $H^*_{U_1}(Y)$
is a module over the algebra $H^*_{U_1}({\rm pt})=H^*(BU_1)$
isomorphic to  $\IC[\sigma]$.
The algebra $H^*_{U_1}({\rm pt})$ is   generated by
Chern class $c^{univ}_1$ of a universal $U_1$-bundle $EU_1\to
BU_1$.  Given a free
action of $U_1$ on $Y$ one has a principle $U_1$-bundle $\CV$ given by the
projection $\pi: Y\to Y/U_1$. By definition of
the universal bundle  for a principle $U_1$-bundle $\CV$ over
$Y/U_1$ there exist a map $u:Y/U_1\to BU_1$
such that $\CV$ is a pull back of the universal bundle over $BU_1$.
and  $c_1=u^*(c^{univ}_1)$ is a pull back of the first Chern class
$c_1^{univ}$ of the universal bundle $EU_1$.  The structure  of
$H^*_{U_1}(Y)\sim H^*(Y/U_1)$  as
$H^*_{U_1}({\rm pt})$-module is then defined by the condition that
$c_1^{univ}$ acts on $H^*(Y/U_1)$ by a multiplication on $c_1$.
In terms of the algebraic model \eqref{Cartan} the image of the
class $c_1^{univ}$ is represented  by $\sigma$.

Now we would like to relate integration of cohomology classes in $H^*(Y/U_1)$ and
$H^*_{U_1}(Y)$. Equivariant de Rham complex can be represented as a space
of functions on a super-space $\Pi TY \times \mathfrak{u}_1$ with
an odd vector field $Q$ given in  local coordinates $(y^i,\psi^i,\sigma)$
by
$$
Q=\sum_{i=1}^M\psi^i\frac{\pr}{\pr y^i}+\sigma v^i(y)\frac{\pr}{\pr \psi^i}.
$$
To defined an integration over a $U_1$-factor we consider
$S^1$-equivariant cohomology  of the extended space
$\tilde{Y}=Y\times \IR^{0|1}$ where the last factor is understood as a
reversed parity Lie algebra $\mathfrak{u}_1$. Then we have
$$
\Omega^*_{U_1}(\tilde{Y})=(\Omega^*(Y))^{U_1}\otimes \IC[\sigma]\otimes
 \IC[\lambda, \eta],
$$
where the de Rham differential $Q_{U_1}$ acts as follows
$$
Q_{U_1}y^i=\psi^i,\qquad Q_{U_1}\psi^i=-\sigma v^i(y),\qquad
Q_{U_1}\lambda=\eta,\qquad Q_{U_1}\eta=0.
$$
Let $\omega(y,\psi,\sigma)$ be a representative of a
cohomology class in $H^*_{U_1}(Y)$ and
$\widetilde{\omega}$ is a representative of the corresponding class in $H^*(Y/U_1)$
according to identification \eqref{Cohident}.
Then the following identity holds
\be\label{factorrep}
\int_{Y/G}\,\widetilde{\omega}=\frac{1}{2\pi}
 \int_{\Pi TY\times \IR\times \IR^{1|1}}
\,\,\frac{dy\,d\psi\,d\sigma\,d\lambda\,d\eta}{{\rm
  Vol}(U_1)}\,\,\omega(y,\psi,\sigma)\,\,e^{\imath
Q_{U_1}(\lambda v_i(y)\psi^i)},
\ee
where  $(\lambda,\eta)$ are local coordinates on $\IR^{1|1}$. The
identity \eqref{factorrep} can be derived replacing
$\omega(y,\psi,\sigma)$  by a representative of the same class
in $H_{U_1}^*(Y)$ which does not contain $\sigma$. Then
integrating over $\lambda$, $\eta$ and $\sigma$ in the right hand side
of  \eqref{factorrep} one obtains  the left hand side
of \eqref{factorrep}.

Now we can apply \eqref{Current} and \eqref{factorrep}
to the $\IP^{\ell}$ obtained via Hamiltonian reduction of
$\IC^{\ell+1}$. Let us  introduce the following variables
\be\label{coord}
(\varphi^i,\bvarphi^i,\chi^i,\bar{\chi}^i,\sigma),\qquad (\xi,H),\qquad (\lambda,\eta)
\ee
and the action of the $U_1$-equivariant de Rham differential
$Q_{U_1}$ is given by
$$
Q_{U_1} \varphi^i=\chi^i,\,\,\, Q_{U_1} \chi^i=-\imath \sigma \varphi^i,\,\,\,
Q_{U_1} \xi=H,\,\,\, Q_{U_1} H=0,\,\,\, Q_{U_1} \lambda=\eta,\,\,\,
Q_{U_1} \eta=0.
$$
We take as a function $s(\varphi)$ on $X=\IC^{\ell+1}$ the shifted momentum
$$
s(\varphi)=\mu(\varphi)+\frac{1}{2}r^2.
$$
Then we have
\be\label{intOne}
\int_{s^{-1}(0)/U_1}\, \widetilde{\omega}=\,
\int_{\Pi TY\times \IR\times \IR^{2|2}}
\,\,\frac{d^2\varphi\,d^2\chi\,d\sigma\,d\lambda\,d\xi\,d\eta\, d
  H}{(2\pi)^2\rm Vol (U_1)}
 \,\,\omega(\varphi,\bvarphi,\chi,\bar{\chi},\sigma)\,\,
 e^{\imath
Q_{U_1}(\lambda v_i(\varphi)\chi^i+ \xi s(\varphi))},
\ee
where $Y=s^{-1}(0)$, $s^{-1}(0))/U_1=\IP^{\ell}$ according to
\eqref{projspace}  and we take $t=0$ in \eqref{Current}.
We consider the following cohomology class
\be\label{integrand}
\widetilde{\omega}=e^{ac_1}\in H^*(\IP^{\ell}), \qquad a\in \IR,
\ee
where $c_1$ is a first Chern class of the line bundle $\CO(1)$ on
$\IP^{\ell}$. As it was discussed above  $c_1$ can be
represented by $\sigma$ in Cartan model of equivariant cohomology. Thus
we can chose $\omega=e^{a\sigma}$ as a representative of
\eqref{integrand}. Thus we have the following integral representation
\be\label{noneq}
\int_{\IP^{\ell}}\,e^{a c_1}=
\frac{1}{(2\pi)^2}\,
\int_{\Pi TY\times \IR\times \IR^{2|2}}
\,\,\frac{d^2\varphi\,d^2\chi\,d\sigma\,d\lambda\,d\xi\,d\eta\,dH}
{\rm Vol (U_1)}e^{a\sigma}
\ee
$$
\exp \left(\imath H(-\frac{1}{2}\sum_{j=1}^{\ell+1} \varphi^j\bvarphi^j+\frac{1}{2}r^2)+
\frac{\imath \xi}{2} \sum_{i=1}^{\ell+1} (\varphi^i\bar{\chi}^i+\bvarphi^i\chi^i)
-\eta\sum_{i=1}^{\ell+1} (\varphi^i\bar{\chi}^i-\bvarphi^i\chi^i) -
 2\lambda(\imath\sigma \sum_{j=1}^{\ell+1} |\varphi^i|^2+
\sum_{i=1}^{\ell+1}\chi^i\bar{\chi}^i)\right).
$$
Define an action of $U_{\ell+1}$ on $\IC^{\ell+1}$ so that $j$-th
factor in the diagonal subgroup $U_1^{\ell+1}\subset U_{\ell+1}$ acts
as follows
$$
e^{\imath \alpha_j}:\,\,\, \varphi^i
\longrightarrow e^{\imath \alpha_j\delta_{ij}}\,\varphi^j.
$$
This action is Hamiltonian with respect to \eqref{symform} and the
corresponding momenta are given by
$$
\mu_j(\varphi)=-\frac{1}{2}|\varphi^j|^2.
$$
The action of $U_{\ell+1}$ descends to the Hamiltonian action on $\IP^{\ell}$
obtained by the Hamiltonian reduction \eqref{projspace}. Corresponding
momenta are given in inhomogeneous coordinates
$w_j=\varphi_j/\varphi_{\ell+1}$, $\varphi_{\ell+1}\neq 0$ by
 \be\label{PH1}
  \mu^{\IP^\ell}_j=-\frac{r^2}{2}\,\frac{|w_j|^2}{1+\sum_{j=1}^\ell|w_j|^2}\,,
  \qquad j=1,\ldots \ell,
 \ee
\be\label{PH2}
  \mu^{\IP^\ell}_{\ell+1}=-\frac{r^2}{2}\,\frac{1}{1+\sum_{j=1}^\ell|w_j|^2}.
 \ee
Let $c_1^{U_{\ell+1}}$ be
a $U_{\ell+1}$-equivariant extension of the first Chern class $c_1$ of the
line bundle $\CO(1)$ on $\IP^{\ell}$. We would like to express the
integral of $\exp (a c_1^{U_{\ell+1}})$ over $\IP^{\ell}$ in terms  of
an integral over larger space generalizing  \eqref{noneq} to
$U_{\ell+1}$-equivariant setting.  To simplify the presentation we
use the standard properties of equivariant cohomology to replace the
calculations in $U_{\ell+1}$-equivariant cohomology by
equivalent calculations in
equivariant cohomology with respect to a diagonal subgroup
$U_{1}^{\ell+1}\subset U_{\ell+1}$ with additionally imposed condition of
an invariance with respect to the Weyl group $W=S_{\ell+1}$  of
$U_{\ell+1}$.

In the following we need a $U_{\ell+1}$-equivariant
version of \eqref{noneq} which can be obtained straightforwardly
generalizing the previous constructions. Thus we just state the main
identity in this case.  Consider the same set of coordinates as in \eqref{coord}
but with the following equivariant de Rham differential
$Q$  given by
$$
Q_{U_{\ell+1}} \varphi^i=\chi^i,\qquad Q_{U_{\ell+1}} \chi^i=-\imath
(\sigma+\sigma_j) \varphi^i,\qquad
Q_{U_{\ell+1}} \xi=H,
$$
$$
Q_{U_{\ell+1}} H=0,\qquad Q_{U_{\ell+1}}\lambda=\eta,\qquad Q_{U_{\ell+1}} \eta=0,
$$
where $(\sigma_1,\ldots,\sigma_{\ell+1})$  are elements of the Lie algebra of
$U_{1}^{\ell+1}$. The expression for the $U_{1}^{\ell+1}$ equivariant
version \eqref{noneq} is given by
\be\label{noneq1}
\int_{\IP^{\ell}}\,\exp \left(a c_1^{U_{1}^{\ell+1}}\right)\,=\,
\frac{1}{(2\pi)^2}\int_{\Pi TY\times \IR\times \IR^{2|2}}
\,\,\frac{d^2\varphi\,d^2\chi\,d\sigma\,d\lambda\,d\xi\, d\eta\,d H}{\rm Vol( U_1)}
  \,\, e^{a\sigma}
\ee
$$
\exp \left(\imath
Q_{U_{\ell+1}}\left(\xi(-\frac{1}{2}\sum_{j=1}^{\ell+1}
 \varphi^j\bar{\varphi}^j+\frac{1}{2}r^2)+\imath\lambda\sum_{j=1}^{\ell+1}
  (\varphi^j\bar{\chi}^j-\bar{\varphi}^j\chi^j)\right)\right),
$$
where
$$
Q_{U_{\ell+1}}\left(\xi(-\frac{1}{2}\sum_{j=1}^{\ell+1}
 \varphi^j\bar{\varphi}^j+\frac{1}{2}r^2)+\imath\lambda\sum_{j=1}^{\ell+1}
  (\varphi^j\bar{\chi}^j-\bar{\varphi}^j\chi^j)\right)=
H(-\frac{1}{2}\sum_{j=1}^{\ell+1} \varphi^j\bar{\varphi}^j+\frac{1}{2}r^2)
 +\frac{1}{2}\xi \sum_{j=1}^{\ell+1} (\varphi^j\bar{\chi}^j+\bar{\varphi}^j\chi^j)+
$$
$$
\imath\eta\sum_{j=1}^{\ell+1} (\varphi^j\bar{\chi}^j-\bar{\varphi}^j\chi^j) -
2 \lambda  \sum_{j=1}^{\ell+1} (\sigma+\sigma_j)|\varphi^j|^2+
2\imath\lambda\sum_{j=1}^{\ell+1}\chi^j\bar{\chi}^j.
$$
\begin{prop}\label{projvol}
  Let $(\sigma_1,\ldots,\sigma_{\ell+1})$  be an  element of 
the Lie algebra of
$U_{1}^{\ell+1}$.  The $U_1^{\ell+1}$-equivariant symplectic volume
\eqref{noneq1} has the following integral representation
\be\label{projvol1}
\int_{\IP^{\ell}}\,\exp \left(\frac{r^2}{2} c_1^{U_{1}^{\ell+1}}\right)\,=\,
\left(2\pi\right)^{\ell-1}
\int_{\IR-\imath\epsilon} d H e^{\frac{\imath r^2 H}{2}} 
\frac{1}{\prod (\imath H+\sigma_j)},
\ee
where $\epsilon > max (-\sigma_j),\,\,\,j=1,\ldots,\ell+1$.
\end{prop}
{\it Proof.} Integrating  over $H$, $\lambda$ and $\sigma$ we obtain 
the following:
\be\label{Interone}
\<e^{a\sigma}\>\,=\,
\frac{1}{2\pi}\int \,\frac{d^2\varphi\,d^2\sigma\,d\xi\,d\eta}{\rm Vol (U_1)}
e^{a\sigma(\varphi,\chi)}\,\,\delta(-\frac{1}{2}\sum_{j=1}^{\ell+1}
|\varphi_j|^2+\frac{1}{2}r^2)
\frac{1}{2r^2}\times\\
\exp\{(\frac{\imath\xi}{2}-\eta)\sum_{j=1}^{\ell+1}\varphi^j\bar{\chi}^j
+(\frac{\imath\xi}{2}+\eta)\sum_{j=1}^{\ell+1}\bar{\varphi}^j{\chi}^j\}
\ee
where
$$
\sigma(\chi,\varphi)\,=\,
\frac{1}{r^2}\left(
\imath\sum_{j=1}^{\ell+1}\chi^j\bar{\chi}^j-\sum_{j=1}^{\ell+1}\sigma_j|\varphi_j|^2\right).
$$
Here we use the following  normalization of the integration measure:
$$d^2\varphi\,d^2\chi\,=\,\prod_{j=1}^{\ell+1}\frac{\imath}{2}
 d\varphi^j d\bar{\varphi}^j\,\,\,\,\,
\prod_{j=1}^{\ell+1}\frac{2}{\imath} d\chi^j d\bar{\chi}^j.
$$
It is useful to reintroduce the variable $H$ by writing the first
delta-function in \eqref{Interone} in the integral form. Then
integrating over odd variables  $d\xi\,d\eta\,$ and  $d^2\chi$
and taking into account that ${\rm Vol}(U_1)=2\pi$
we arrive at
\be \<e^{a\sigma}\>\,=\,\frac{1}{(2\pi)^2}
\left(\frac{2 a}{r^2}\right)^{\ell} \int dH\int
 \,\prod_{j=1}^{\ell+1}\frac{\imath}{2}d\varphi^jd\bar{\varphi}^j\,
e^{+\frac{\imath r^2}{2}H}\,\,\,
 e^{-\frac{1}{2}\sum\varphi^j(\imath H+\frac{2a}{r^2}\sigma_j)\bar{\varphi}^j}.
\ee
For a finite-dimensional Gaussian integral we have 
\be\label{Gausseven}
\int_{\IC^N}\,\,e^{-\frac{1}{2}\sum_{i,j=1}^N\,\zb_iA_{ij}z_j}
\,\,\prod_{j=1}^N\,\frac{\imath}{2} dz^j\,d\zb^j\,=
\frac{1}{\det\,A/2\pi},
\ee
where the matrix $A$ has positive eigenvalues. More generally,
the Gaussian integral \eqref{Gausseven} for $A$
having complex eigenvalues $a_j$ such that ${\rm Re}(a_j)\geq 0$,
$j=1,\ldots ,N$ is  defined as a limit of the integral
 for $A$ having complex eigenvalues $a_j$ such that ${\rm Re}(a_j)>0$,
$j=1,\ldots ,N$ expressed through \eqref{Gausseven}.
 Now assuming $Re (\imath H +\frac{2a}{r^2}\sigma_j)>0,
\,\,j=1,\ldots,\ell+1$ and taking integral over $\varphi$ we obtain
\be
\<e^{a\phi}\>\,=\,
\left(\frac{2a}{r^2}\right)^{\ell}
\left(2\pi\right)^{\ell-1}\int_{\IR-\imath\epsilon} 
d H e^{\frac{\imath r^2 H}{2}} \frac{1}{
\prod (\imath H+\frac{2a}{r^2}\sigma_j)},
\ee
where  $\epsilon >
max(-\frac{2a}{r^2}\sigma_j),\,\,\,j=1,\ldots,\ell+1$. 
Taking  $a=\frac{r^2}{2}$ we finally obtain
\be
\<e^{\frac{r^2}{2}\phi}\>\,=\,
\left(2\pi\right)^{\ell-1}
\int_{\IR-\imath\epsilon} d H e^{\frac{\imath r^2 H}{2}}
 \frac{1}{\prod (\imath H+\sigma_j)},
\ee
where $\epsilon >max(-\sigma_j),\,\,\,j=1,\ldots,\ell+1$.
 This completes the proof of the proposition. $\Box$

Finally let us provide a reformulation of the integral representation
of the integrals of the equivariantly closed forms over $\IP^{\ell}$
that does not include integrations over odd variables.
\begin{lem}\label{Auxlema}
  Let $\sigma_j>0,\,\,\, j=1,\ldots,\ell+1$. The following identity holds:
 \be\label{PellVolumeForms}
\frac{1}{2\pi}  \int_{\IC^{\ell+1}}\,\,
  \delta\Big(\mu_{U(1)}\,+\,r^2/2\Big)\,e^{(\omega_{\IC^{\ell+1}}+\sum_{j=1}^{\ell+1}
\sigma_j \mu_j)}
 =\,\int_{\IP^\ell}\,\,e^{(\omega_{\IP^\ell}+\sum_{j=1}^{\ell+1}
\sigma_j\mu^{\IP^\ell}_j)}\,,
 \ee
where $\omega_{\IP^\ell}$ is given by \eqref{FS} and
the reduced Hamiltonians $\mu^{\IP^\ell}_j$ are given by \eqref{PH1} and \eqref{PH2}.
\end{lem}
\noindent {\it Proof.}  This can be deduced from the previous
considerations but allows simple direct derivation. Let us introduce new variables
$w_j=\varphi_j\varphi_{\ell+1},\,j=1,\ldots,\ell$ and $t=|\varphi_{\ell+1}|^2$,
$\theta=\frac{1}{2\imath}\ln\frac{\varphi_{\ell+1}}{\bar{\varphi}_{\ell+1}}$, so that
$\varphi_{\ell+1}=\sqrt{t}\,e^{\imath\theta}$. Then we have
 \be
\frac{1}{2\pi}\left(\frac{\imath }{2}\right)^{\ell+1}\,
  \int_{\IC^{\ell+1}}\bigwedge_{i=1}^{\ell+1}
  d\varphi_i\wedge d\bar{\varphi}_i\,\delta\Big(\frac{1}{2}
\sum_{i=1}^{\ell+1}|\varphi_i|^2-\frac{r^2}{2}\Big)e^{\sum_{j=1}^{\ell+1}
\sigma_j \mu_j}\\
  =\,\frac{r^{2\ell}}{2\pi}\left(\frac{\imath }{2}\right)^{\ell}
\int_0^{2\pi} d\theta\int_0^{\infty}dt\,t^{\ell}\int_{\IC^{\ell}}
  \frac{\bigwedge\limits_{n=1}^{\ell}(dw_n\wedge d\wb_n)}{
  1+\sum|w_n|^2}\,
  \delta\Big(t-\frac{r^2}{1+\sum|w_n|^2}\,\Big)e^{\sum_{j=1}^{\ell+1}\sigma_j \mu_j}\\
  = r^{2\ell}\left(\frac{\imath }{2}\right)^{\ell}
\int_{\IC^{\ell}}
  \frac{\bigwedge\limits_{n=1}^{\ell}(dw_n\wedge d\wb_n)}
  {\bigl(1+\sum|w_n|^2\bigr)^{\ell+1}}\,e^{\sum_{j=1}^{\ell+1}\sigma_j
\mu^{\IP^\ell}_j}.
 \ee
Taking into account that
$$
\frac{\omega_{\IP^\ell}^\ell}{\ell!}= r^{2\ell}\left(\frac{\imath }{2}\right)^{\ell}
\frac{\bigwedge\limits_{n=1}^{\ell}(dw_n\wedge d\wb_n)}
  {\bigl(1+\sum|w_n|^2\bigr)^{\ell+1}},
$$
we obtain the identity \eqref{PellVolumeForms}. $\Box$

\begin{cor}\label{usecor}
Let $\IC^{\ell+1}=\IC^{n_1}\oplus \cdots \oplus
  \IC^{n_k}$, $\sum_{a=1}^k \dim (V_a)=\ell+1$
be a decomposition of the symplectic space vector space
$(\IC^{\ell+1},\Omega)$ where symplectic  structure is given by
\eqref{symform} and $(\varphi_{n_1+\cdots n_{a-1}+1},\ldots,$
$\varphi_{n_1+\cdots n_{a}})$ are coordinates on $\IC^{n_a}$.
Let $U_1^{\ell+1}$ act on $\IC^{\ell+1}$ diagonally. The action is
Hamiltonian  and let $\mu_j$, $j=1,\ldots \ell+1$  be the
momenta corresponding to the action of $j$-th factor $(U_1)_j$.  Let
$U_1^k\subset U_1^{\ell+1}$
act on $\IC^{\ell+1}$ so that the $a$-th $(U_1)_a$ acts
non-trivially only on $\IC^{n_a}$ by multiplication on complex
numbers.  Let $U_1$ be diagonally embedded in $U_1^k$.
Let $\IP(\IC^{\ell+1})$ be a Hamiltonian reduction of $\IC^{\ell+1}$
over $U_1$ with momentum $x$ and
$\IP(\IC^{n_a})$ be a Hamiltonian reduction of $\IC^{n_a}$ over
$(U_1)_a$ with the momentum $x_a$. Then the
following relation between $U_k$-equivariant integrals holds
$$
\int_{\IP(\IC^{\ell+1})}\,\,e^{\omega_{\IP(\IC^{\ell+1})}(x)+
\sum_{j=1}^{\ell+1}\sigma_j \mu^{\IP(\IC^{\ell+1})}_j}\,=
$$
$$
\frac{1}{2\pi}\int dx_1\cdots dx_k\,\delta(x-\sum_{j=1}^kx_j)\,\,
\prod_{a=1}^k 2\pi\int_{\IP(\IC^{n_a})}\,\,e^{x_a\omega_{\IP(\IC^{n_a})}+
\sum_{i=n_1+\ldots n_{a-1}+1}^{n_1+\ldots +n_a}\sigma_i \mu^{\IP(\IC^{n_a})}_i}\,,
$$
where $\omega_{\IP(\IC^{n_a})}$ is a standard Fubini-Studi symplectic form
on $\IP(\IC^{n_a})$ multiplied by $x_j$ and $\mu^{\IP(\IC^{n_a})}_j$ is
a momentum for  the action of $(U_1)_j$.
\end{cor}
\noindent{\it Proof}. This identity straightforwardly follows
from Lemma  \ref{Auxlema} and the following identity
$$
\int dx_1\cdots dx_k\,\prod_{j=1}^k\delta(\mu_j+x_j)\delta(x-\sum_{j=1}^kx_j)=
\delta(\sum_{j=1}^k\mu_k+x).
$$
$\Box$

\vskip 1cm

\noindent {\small {\bf A.G.} {\sl Institute for Theoretical and
Experimental Physics, 117259, Moscow,  Russia; \hspace{8 cm}\,
\hphantom{xxx}  \hspace{2 mm} School of Mathematics, Trinity College
Dublin, Dublin 2, Ireland; \hspace{6 cm}\hspace{5 mm}\,
\hphantom{xxx}   \hspace{2 mm} Hamilton Mathematics Institute,
Trinity College Dublin, Dublin 2, Ireland;}}

\noindent{\small {\bf D.L.} {\sl
 Institute for Theoretical and Experimental Physics,
117259, Moscow, Russia};\\
\hphantom{xxxx} {\it E-mail address}: {\tt lebedev@itep.ru}}\\

\noindent{\small {\bf S.O.} {\sl
 Institute for Theoretical and Experimental Physics,
117259, Moscow, Russia};\\
\hphantom{xxxx} {\it E-mail address}: {\tt Sergey.Oblezin@itep.ru}}

\end{document}